\documentclass[journal]{IEEEtran}

\usepackage{xcolor,soul,framed} %,caption
\colorlet{shadecolor}{yellow}
\usepackage[pdftex]{graphicx}
\graphicspath{{../pdf/}{../jpeg/}}
\DeclareGraphicsExtensions{.pdf,.jpeg,.png}
\usepackage[cmex10]{amsmath}
\let\labelindent\relax
\usepackage{enumitem}
\usepackage{amsmath, amsthm, amssymb}
\usepackage{array}
\usepackage{comment}
\usepackage{mdwmath}
\usepackage{mdwtab}
\usepackage{eqparbox}
\usepackage{url}
\usepackage{float}
\usepackage{subfig}
\usepackage{lipsum,multicol}
\usepackage{float}
 \usepackage{pifont}
 \usepackage[T1]{fontenc}
\usepackage{placeins}
\usepackage{hyperref}
\usepackage{bbding}
\usepackage{lipsum}
\usepackage{mwe}
\usepackage[utf8]{inputenc}
\usepackage[version=4]{mhchem}
\usepackage{siunitx}
\usepackage{longtable,tabularx}
\usepackage{soul}
\usepackage{wrapfig}
\usepackage{multirow}
\usepackage{relsize}
\newtheorem{proposition}{Proposition}
\newtheorem{lemma}{Lemma}

\newcommand{\xmark}{\ding{55}}

\begin{document}

%=== TITLE & AUTHORS ====================================================================

%\bstctlcite{IEEEexample:BSTcontrol}
    \title{The Meshing of the Sky: Delivering Ubiquitous Connectivity to Ground Internet of Things}
    
\begin{comment}
    \author{\IEEEauthorblockN{Laila Abouzaid\IEEEauthorrefmark{1}, Essaid Sabir\IEEEauthorrefmark{1}, Ahmed Errami\IEEEauthorrefmark{1}, Halima Elbiaze\IEEEauthorrefmark{2} }
\IEEEauthorblockA{\textit{\IEEEauthorrefmark{1}NEST Research Group, ENSEM, Hassan II University, Casablanca, Morocco.} \\
\textit{\IEEEauthorrefmark{2}Computer Science Department, University of Quebec at Montreal (UQAM), Montreal, QC H2L 2C4, Canada.}\\
{laila.abouzaid@ensem.ac.ma}, {e.sabir}@ensem.ac.ma ,{aerrami}@yahoo.fr}{elbiaze.halima@uqam.ca}}
\end{comment}
  \author{Laila Abouzaid\IEEEauthorrefmark{1},~\IEEEmembership{Student Member,~IEEE,}
      Essaid Sabir\IEEEauthorrefmark{1}\IEEEauthorrefmark{3},~\IEEEmembership{Senior,~IEEE,}
      Halima Elbiaze\IEEEauthorrefmark{3},~\IEEEmembership{Senior,~IEEE,}\\
      Ahmed Errami\IEEEauthorrefmark{1} and
     Othmane Benhmammouch\IEEEauthorrefmark{1}\\
     % <-this % stops a space
\IEEEauthorrefmark{1}NEST Research Group, ENSEM, Hassan II University, Casablanca, Morocco.\\

\IEEEauthorrefmark{3}Department of Computer Science, University of Quebec at Montreal (UQAM), Montreal, QC H2L 2C4, Canada.

  %\thanks{}
  }

% The paper headers
%\markboth{IEEE TRANSACTIONS ON MICROWAVE THEORY AND TECHNIQUES, VOL.~60, NO.~12, DECEMBER~2012
%}{Roberg \MakeLowercase{\textit{et al.}}: High-Efficiency Diode and Transistor Rectifiers}

% ====================================================================
\maketitle

% === ABSTRACT ====================================================================
% =================================================================================
\begin{abstract}
%\boldmath
Nowadays, Unmanned Aerial Vehicles (UAVs) are being used in several novel applications, especially in the telecommunication domain. However, ensuring UAV communication and networking for the purpose of a specific application is still challenging. Indeed, due to the mobility of a UAV in a vast area, permanent connectivity over the back-haul is very sporadic and might be lost.
In this paper, we consider an aerial mesh network where each UAV can serve as a flying base station to boost terrestrial base station in case of damaged infrastructure case for example, or/and provide connectivity for uncovered or poorly covered nodes, and behaves as a relay to establish communication between two components owing to a lack of reliable direct communication link between them. We then detail a case study where a UAV-fleet are used  to collect data from the ground Internet of Things (IoT) devices and forward it to cloud for further processing passing by a remote gateway.
 We aim here to build a queueing framework including network layer, MAC layer, and physical layer and investigate both uplink and downlink communication links. Next, we derive some closed forms allowing us to predict the network performance in terms of traffic intensity at every UAV of the aerial mesh network, end-to-end (E2E) throughput and e2e delay of ongoing streams. Next, we conduct extensive simulations to illustrate the benefit of our framework. Results discussion and numerous insights on parameter setting, target quality of service and design consideration are also drawn. 
\end{abstract}


\section{Introduction}

Over the past few years, the Internet of Things (IoT) has gradually penetrated our daily lives and cities. Smart city, Smart grid, Smart home and connected objects are among the marked IoT applications just to mention a few.
Clearly, IoT is about connectivity and pervasive computing. Yet, the ``Objects'' or ``things'' are assumed to be able to connect anywhere, anytime providing some specific services. However, meeting ubiquitous connectivity cannot be granted anytime due to a lack of sufficient infrastructure, and limited transmit power that prevents communicating over long ranges. Recent unseen progress in developing Unmanned Aerial Vehicles (UAVs), also called drones, allowed their use as flying aerial base stations. This nomadic/flying base stations can offer on-demand and ubiquitous connectivity for remote IoT devices. Owing to their fundamental characteristics such as low cost, adaptive altitude, quick reconfiguration, UAVs can be swiftly employed to dynamically offload legacy mobile network (e.g., when a high traffic is observed), improve network responsiveness, improve network coverage, etc.. For instance, UAVs can substantially extend broadband connection capability between a source/destination pair by establishing a Line of Sight (LOS) opportunity. Exhibiting many flexible deployment features, the service provider can control the coverage area by efficiently adjusting the UAV parameters (e.g., transmit power, altitude, etc.). Here are some promising benefits in wireless environments \cite{SURVEY}:
\begin{itemize}
    \item \textbf{UAV-assisted ubiquitous coverage:} A UAV may provide wireless connectivity for ground devices/sensors deployed in remote areas, allow on-demand flexible infrastructure, offload mobile networks, help healing network failures, etc.;
    \item \textbf{UAV-assisted/enabled  data collection:} Surveying related literature, we find out that deploying single or multiple UAV platforms is a promising approach to sense and collect real-time data efficiently from IoT devices in an area of interest (e.g., precision agriculture sensors, smart meters, etc.). Meanwhile, terrestrial BSs (TBSs) are used for backhaul links or UAV control (UAV-enabled IoT networks) or also jointly use UAV and TBSs to perform data collection (UAV-assisted IoT networks) \cite{UAV-assisted/enabled}.
    \item \textbf{UAV-assisted relaying:} Dispatching drone swarm networks in the sky, permits to build a flying backhaul network and maintain reliable communication. This mainly requires to establish inter-linking communication between UAVs. Several communication architectures for UAV fleet can be used. ad hoc architecture turns out to be an effective solution due to its simplicity, robustness, and flexibility. This new assembling gives rise to the appearance of Flying ad hoc Network (FANET) \cite{FANET}, which communicates through a multi-hop manner;
    \item \textbf{UAV-assisted wireless charging:} On one hand, a UAV could be (re)-charged on the fly by wireless power transfer while flying nearby a charging station. On the other hand, a UAV could transfer energy wirelessly to depleted ground IoT devices. 
\end{itemize}
Including UAVs in network design efficiently helps meeting application requirements as it allows bringing the infrastructure closer to end-users. A full understanding of network performance before and after deployment is required, from network planning and design phase to network survivability. Also, visualizing the impact of each UAV parameter on long-term system functioning, and guaranteeing the ability to dependably run the desired application  based on UAV Platforms is of great importance. Also, ensuring acceptable network connectivity whilst meeting a target QoS is the main tradeoff to expect during network planning and designing. Furthermore, the UAV-network planning process address the problem of determining the optimal cost UAV-Network architecture considering the UAV-fleet size, the antenna type and orientation and other physical parameters (e.g. altitude, transmission power, speed, etc.), such that the QoS requirement is respected. Despite the considerable importance of network planning and design, we are interested in conceiving an advanced framework model of UAV-Network where two links are taking into consideration  "Air-to-Ground channel (GC)" and "Air-to-Air channel (AC)". The objective is to analyze and predict the behavior of performance metrics of the UAV-Network in terms of end-to-end throughput and end-to-end delay as well as to characterize network flow-stability region which bears negative side-effects on the overall end-to-end network performance metrics (e.g. network congestion, high delay, low throughput). This performance analysis offers useful insights into constitutive design parameters, to size the UAV network as well as evaluate its proper functioning before/during its deployment. UAV-assisted data collection use case can also be studied using the developed framework. It elucidates the importance of dynamically/on-demand bringing network connectivity to WSN and IoT  devices {\cite{uaviot}}, overcoming this way the energy restriction of such a system, under different applications area including victim search and rescue, monitoring, emergency networks in disasters and so forth {\cite{datacollection}\cite{IoTreliabl}}. Also, due to the fact that such an application typically incorporates AC/GC communication systems, and variety of traffic nature and cadence.

\subsection{Related work}

Performance analysis of various UAV network metrics such as coverage, connectivity, capacity, reliability, throughput, delay, had recently attracted much attention.
Most of the existing works seek to analyze the UAV coverage in static and mobile scenarios \cite{energycoverage}-\cite{downlinkcoverage}-\cite{coverage1}-\cite{Coveragemozaffari}-\cite{Coveragemobility}. To name a few, reference \cite{coverage1} used a stochastic geometric method to provide an analytic expression for coverage probability where ground users are managed by different drones. The work in \cite{coverage1} analyzed the coverage as a function of UAV parameters (e.g. altitude, antenna bandwidth) and under Nakagami-m fading channel; they demonstrate how the coverage can be maximized for a given value of UAV parameters. The optimal altitude of a drone with minimum energy required for maximizing coverage is investigated in \cite{Coveragemozaffari}. Thus, the problem of optimal deployment of two drones serving the same area was studied for two cases; interference-free and with interference. Recent works highlighted the impact of mobility on a UAV performance system. For example, the authors in \cite{Coveragemobility} analyzed the impact of various mobility patterns (e.g. Tractor, Angular, Square, Circular) on UAV performance such as coverage, Time Efficiency, and Utilization, etc.. Through the finding performance results, the authors present a trade-off between maximizing the covered nodes  while minimizing the time efficiency allowing to choose the suitable mobility pattern. 
The coverage metric for a UAV network-served downlink and D2D users is analyzed in \cite{mozaffarid2d}. First, the authors derived the downlink coverage probability and system rate for downlink users (DUs) and D2D devices. Then the impact of D2D density and UAV altitude on the overall performance system such as coverage, system rate, number of stops, was analyzed. Also, the authors in \cite{akram2016coverage} provided the analytical study for achievable coverage and throughput by an aerial base station. In order to find the optimal altitude that maximizes the coverage, the proposed framework was implemented in a  realistic urban environment with architectural statistics based on the recommended parameters by the International Telecommunication Union (ITU).

Nevertheless, to successfully deploy a UAV network it is important to guarantee the reliability of the UAV link. For example, the authors in \cite{azarireliability} derived the outage probability depending on altitude and elevation angle in both communication modes: direct Air-to-ground channel (GC) and Air-to-Air channel (AC) communication using a relaying. Besides, the throughput maximization problem in UAV networks has been studied in several works. In \cite{throughput6*}, the authors designed a suitable trajectory for a UAV mobile relay while maximizing its throughput and optimizing the source/destination nodes transmit power. The authors in \cite{cyclical9} proposed an algorithm that aims to maximize the minimum throughput of ground devices which takes as input the position of UAV. This algorithm allows allocating the time between nodes for the best use of the channel.
 
 \begin{table*}[h!]
\caption{Comparison between our proposal and related work - UAV networks analysis}
\begin{center}
\begin{tabular}{|c|l|c|c|c|c|c|c|c|c|}
\hline
\multirow{2}{*}{\textbf{\begin{tabular}[c]{@{}c@{}} \\ \\ Ref \end{tabular}}} & \multicolumn{1}{c|}{\multirow{2}{*}{\textbf{{\begin{tabular}[c]{@{}c@{}} \\ \\ Performance \\ Metrics  \end{tabular}}}}}        & \multicolumn{5}{c|}{\textbf{Parameters taken into consideration}}                                             & \multirow{2}{*}{\textbf{\begin{tabular}[c]{@{}c@{}} \\ \\ Link\end{tabular}}} & \multicolumn{1}{l|}{\multirow{2}{*}{\textbf{\begin{tabular}[c]{@{}l@{}} \\ \\Single/\\ Multi-UAV\end{tabular}}}} & \multirow{2}{*}{\textbf{\begin{tabular}[c]{@{}c@{}} \\ \\Relevants\\  OSI layers\end{tabular}}} \\ \cline{3-7}   & \multicolumn{1}{c|}{}                                     & \multicolumn{1}{l|}{\textbf{\begin{tabular}[c]{@{}l@{}}Mobility/\\ Trajectory\end{tabular}}} & \multicolumn{1}{l|}{\textbf{Altitude}}     & \multicolumn{1}{l|}{\textbf{\begin{tabular}[c]{@{}l@{}}Density\\ (UAV,\\  ground \\ node)\end{tabular}}} & \multicolumn{1}{l|}{\textbf{\begin{tabular}[c]{@{}l@{}}UAV \\ Antenna\end{tabular}}} & \multicolumn{1}{l|}{\textbf{Speed}}                                       &                                                                              & \multicolumn{1}{l|}{}                                                                                      &            \\
                                                                                                                          \hline                                    \multicolumn{1}{|c|}{\cite{coverage1} }  
                &
\multirow{6}{*}{Coverage}   & \xmark  & \checkmark                  & \begin{tabular}[c]{@{}c@{}}UAV\\  density\end{tabular}  & \checkmark& \xmark   & GC     & UAVs Network         & Physical                 \\
\cline{1-1} \cline{3-10}       

\cite{Coveragemozaffari}  &         & \xmark        & \checkmark      & \xmark     & \xmark      & \xmark  & GC        & \begin{tabular}[c]{@{}c@{}}Case 1: One UAV\\ Case 2: Two UAVs\end{tabular}         & Physical                         \\ 
\cline{1-1} \cline{3-10}     

\cite{Coveragemobility}  &        & \checkmark       & \checkmark                  & \begin{tabular}[c]{@{}c@{}}Density\\  of \\ nodes\end{tabular}          & \xmark        & \checkmark       & GC            & Single UAV         & Physical    \\ 
\cline{1-1} \cline{3-10}         

\cite{mozaffarid2d}  &          & \checkmark       & \checkmark                  & \begin{tabular}[c]{@{}c@{}}D2D \\ Density\end{tabular}               & \xmark                                               & \begin{tabular}[c]{@{}c@{}}Second part:\\ Number of\\  stops\end{tabular} & GC        & Single UAV        & Physical              \\
\cline{1-1} \cline{3-10}     

\cite{akram2016coverage} &      & \xmark            & \checkmark         & \xmark             & \xmark          & \xmark      & GC       & Single UAV        & Physical                                                                                  \\ 
\cline{1-1} \cline{3-10}                                    

\multirow{2}{*}{\cite{azarireliability}} &  & \multirow{2}{*}{\xmark}                                                       & \multirow{2}{*}{\checkmark} & \multirow{2}{*}{\begin{tabular}[c]{@{}c@{}}Nodes\\ density\end{tabular}}                                 & \multirow{2}{*}{\xmark}                                               & \multirow{2}{*}{\xmark}                                    & \multirow{2}{*}{GC}                                                         & \multirow{2}{*}{Single UAV}                                                                                & \multirow{2}{*}{Physical}                                                                 \\ \cline{2-2}
                                                                                                               &    Reliability                                                       &                                                                                              &                                            &                                                                                                          &                                                                                      &                                                                           &                                                                              &                                                                                                            &   
                                                                                       \\ \hline
        \cite{throughput6*}                                                                                     & \multirow{2}{*}{Throughput}                        & \checkmark        & \checkmark          & \xmark        & \xmark  & \checkmark           & GC                                                                          & Single UAV         & Physical    \\ \cline{1-1} \cline{3-10} 
                                        \cite{cyclical9}           & \multirow{2}{*}{}        & \multirow{2}{*}{\checkmark}                                                   & \multirow{2}{*}{\checkmark} & \multirow{2}{*}{\begin{tabular}[c]{@{}c@{}}DTs \\ density\end{tabular}}                                  & \multirow{2}{*}{\xmark}                                               & \multirow{2}{*}{\checkmark}                                & \multirow{2}{*}{GC}                                                         & \multirow{2}{*}{Single UAV}                                                                                & \multirow{2}{*}{Physical}                                                                 \\  \cline{1-1} 
\multirow{2}{*}{\cite{Performance7*}}                                                                                                    &                                                           &                                                                                              &                                            &                                                                                                          &                                                                                      &                                                                           &                                                                              &                                                                                                            &                                                                                           \\ \cline{2-10} 
                                                                                                                          & Delay                     & \checkmark                                                                    & \checkmark                  & \begin{tabular}[c]{@{}c@{}}UAVs \\ Density\end{tabular}                                                  & \checkmark                                                            & \xmark                                                     & AC                                                                          & UAVs Network                                                                                               & \begin{tabular}[c]{@{}c@{}}Physical/\\ MAC\end{tabular}                                   \\ \hline
 \textbf{Our work}              & \begin{tabular}[c]{@{}c@{}} Throughput \\ and delay \end{tabular}   & \checkmark              & \checkmark      & \begin{tabular}[c]{@{}c@{}}IoT \\ devices \\ Density\end{tabular}                              & \checkmark         & \checkmark      & AC  / GC         & UAVs Network     & \begin{tabular}[c]{@{}c@{}}Physical/\\ MAC/ \\ Network\end{tabular}                         
 \begin{comment}
\end{comment}
\\ \hline 
\end{tabular}
\label{TabCOM}
\end{center}
\end{table*}

 All the above-mentioned works focus exclusively on physical layer parameters and considered the Air-to-Ground channel (GC) to analyze the system performance in terms of coverage, reliability, and throughput. However, few researchers considered MAC layer parameters to analyze UAV network performance. For example, the work in \cite{Performance7*} proposes a new MAC protocol with directional and omnidirectional antennas to improve the performance of UAVs communication in terms of throughput as well as end-to-end delay (using Opnet).
Table \ref{TabCOM} classifies existing work related to UAV performance analysis, according to the considered parameters and corresponding OSI layer. 

As far as we are aware, none of the existing works on UAV-mesh Network has analyzed the QoS requirement, in terms of e2e throughput, reliability and e2e delay, by jointly incorporating the UAV design parameters (Altitude, density, Aperture angle) and the OSI model layers as well. Besides, to date, the impact of the UAV design parameters on the traffic stability is still not properly/sufficiently investigated. 
 Instead, there has been significant work on analyzing the performance metrics of ad hoc network taking into account OSI model parameters. First, the authors in \cite{2000} initiated a study on wireless ad hoc network capacity where the nodes are assumed to be fixed and distributed randomly following independent and identically distributed (i.i.d). They have shown that as the number of node increase the throughput for each node for a destination diminishes. Therefore, after this pioneering work, several researchers have continued studying fixed ad hoc network capacity taking into consideration different network parameters. Afterward, considerable works were made to study the mobile ad hoc network. Grossglauser and Tse \cite{1002} inset mobility variant into the network model. They have shown that regardless of network scalability, the throughput per source-destination can remain constant by using a two-hops route and mobility. The delay is also an important performance metric in various networking applications along with throughput. Indeed, using relaying across node mobility allows achieving a higher throughput in conjunction with a large delay.  In this context, the researchers were motivated to study the throughput-delay trade-off in ad hoc wireless networks to satisfy QoS requirements \cite{1004} \cite{1005} \cite{1006}. The authors in \cite{1001} exploited the results in \cite{2000} and \cite{1002} for determining the optimal throughput-delay trade-off in a static and mobile ad hoc wireless network. However, when a high network load is observed, the stability region conceives to be the most relevant parameters besides throughput and delay. 
The stability region of a system is defined as the ability to keep a quantity of interest in a bounded region \cite{1019}, which means that all queues-flows of the network shall remain finite in order to make the network operate in a stable region: unbounded delay and stable throughput.
Several studies widely adopted wireless network stability \cite{1019} \cite{1020} \cite{1021}. Besides, most of the existing works do not take into consideration the impact of traffic flows across intermediate forwarding nodes between source and destination, on network stability. Indeed, high traffic load entering an intermediate node may trigger congestion thereby collapsing the overall end-to-end network performance. To overcome this issue, the authors of \cite{1023} have studied the behavior of end-to-end throughput and forwarding queues stability on multi-hop wireless networks. Routing, random access in MAC layer and topology are considered in their proposed model. They have shown that when forwarding queues are stable, the end-to-end throughput of a given route is not affected by the intermediate nodes load. In \cite{1024}, the authors recovered the same model in \cite{1023} and investigated the interaction among PHY, MAC, and Network layers. 

Elkhoury el al. \cite{1025} have studied the end-to-end performance of a multi-hop wireless network for real-time application, using a cross-layer scheme including PHY, MAC, and Network layers.
\subsection{Our Main Contributions}
 The main contribution of this paper is to build a theoretic framework for performance evaluation of Flying Mesh Drone Network (FMDN), considering both GC links and AC links. %complete framework model including twofold links $AA$ and $AG$, to analyze Flying Mesh UAV Network under general realistic considerations. 
 Specifically, we derive a probabilistic model jointly considering the air channel features and cross-layers modeling. Our contribution is fourfold:
\begin{itemize}
\item \textbf{End-to-End performance metrics: } Our model allows to predict the average end-to-end throughput and average end-to-end delay. These two key quality of service metrics provide interesting insights on how to plan and properly size the flying backbone given the ground IoT node density, their packet generation distribution, UAV trajectory, etc.

\item \textbf{Cross-layer model: } Our cross-layer model assumes a synergy between APP, NET, MAC and PHY layers allowing the protocol stack to share specific information. Such a paradigm enables a context-aware fine-tuning of internal parameters, allowing thereby to meet optimal performance;

\item \textbf{Efficient trade-offs: } Analyzing the impact of the key parameters we come out several trade-offs could be defined. For instance, one could easily optimally control the drone altitude while granting a target min-throughput/max-delay. This is particularly useful to support delay sensitive bi-directional applications with low bit-rate and stochastic duty cycles;

\item \textbf{Support of LoRaWAN/Sigfox/NB-IoT \cite{arabi} standards: } LoRaWan, SigFox and NB-IoT are mainly based on the aloha random access to serve class A-like applications. Since our model consider IEEE 802.11 DCF for ground-UAV link. Then, it can be straightforwardly extended to support the three aforementioned IoT standards. The changes to be done are including the spreading factor at PHY layer, and replacing the IEEE 802.11 link between ground IoT devices and UAV, by a time-frequency aloha link at MAC level.

%\item \textbf{Average End-to-End Delay}: 
%is considered the second important performance metric that evaluates the network congestion condition. In this paper, 
%we derive the expression of the e2e delay of  data transmitted over the drone network to reach its destination for two link: uplink and Downlink. This metric will be analyzed for the purpose to establish numerous insights on parameters setting to meet  UAV  application needs.
%, and the time of its return if it required. 
%\textcolor{red}{This metric allows operators to previous knowledge of the range of network parameters that ensure a delay limit required by an application}. %Throughput Analysis of Network Coding in Multi-Hop Wireless Mesh Networks Using Queueing Theory%to study the throughput of inter- flow network coding in multi-hop wireless mesh networks.
\end{itemize}
 %that the system can carry in stable conditions.
 \subsection{Paper organization}
The remainder of this paper is organized as follows: Section II describes the problem formulation which includes three main parts: network topology, channel model, and network model layers. Then, the traffic intensity, end-to-end throughput and end-to-end delay expressions are derived in section III, the simulation results are presented in section IV. Section V presents  discussion and insights. Finally, section VI concludes this paper.

\section{Problem Formulation}
The system model, including the network topology, channel model and PHY/MAC/NET cross-layer models, is investigated in this section.
  \begin{figure*}[]
  \begin{center}
\includegraphics[trim = 0cm 0cm 0cm 0cm, scale=0.5]{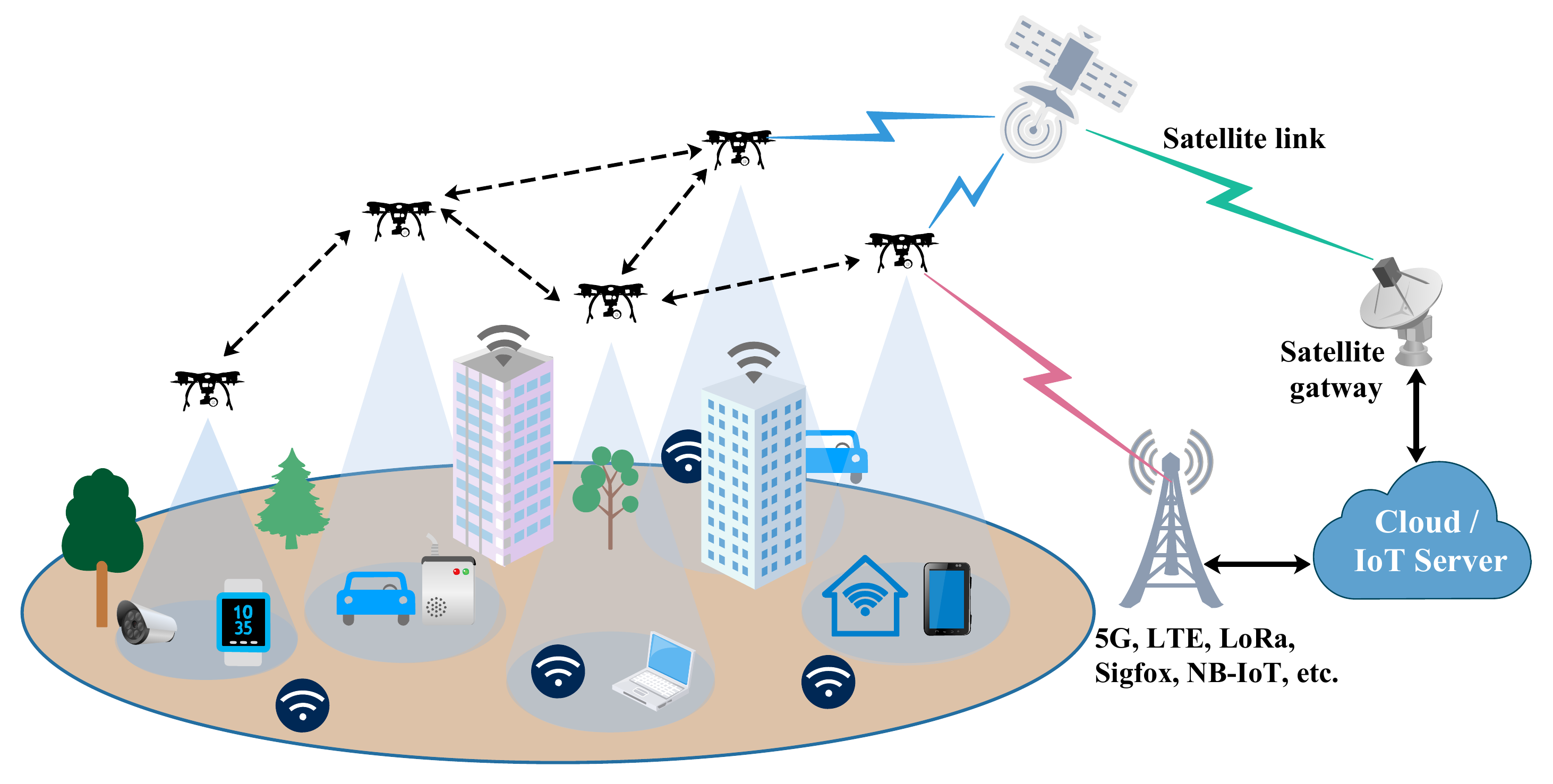}
  \end{center}
  \caption{Network architecture and flying backbone for ground data collection.}
\label{A} 
\end{figure*}

\subsection{Network Architecture}
\begin{figure}[h!] 
\begin{center}
\includegraphics[trim = 1cm 0cm 0cm 0cm, scale=0.55]{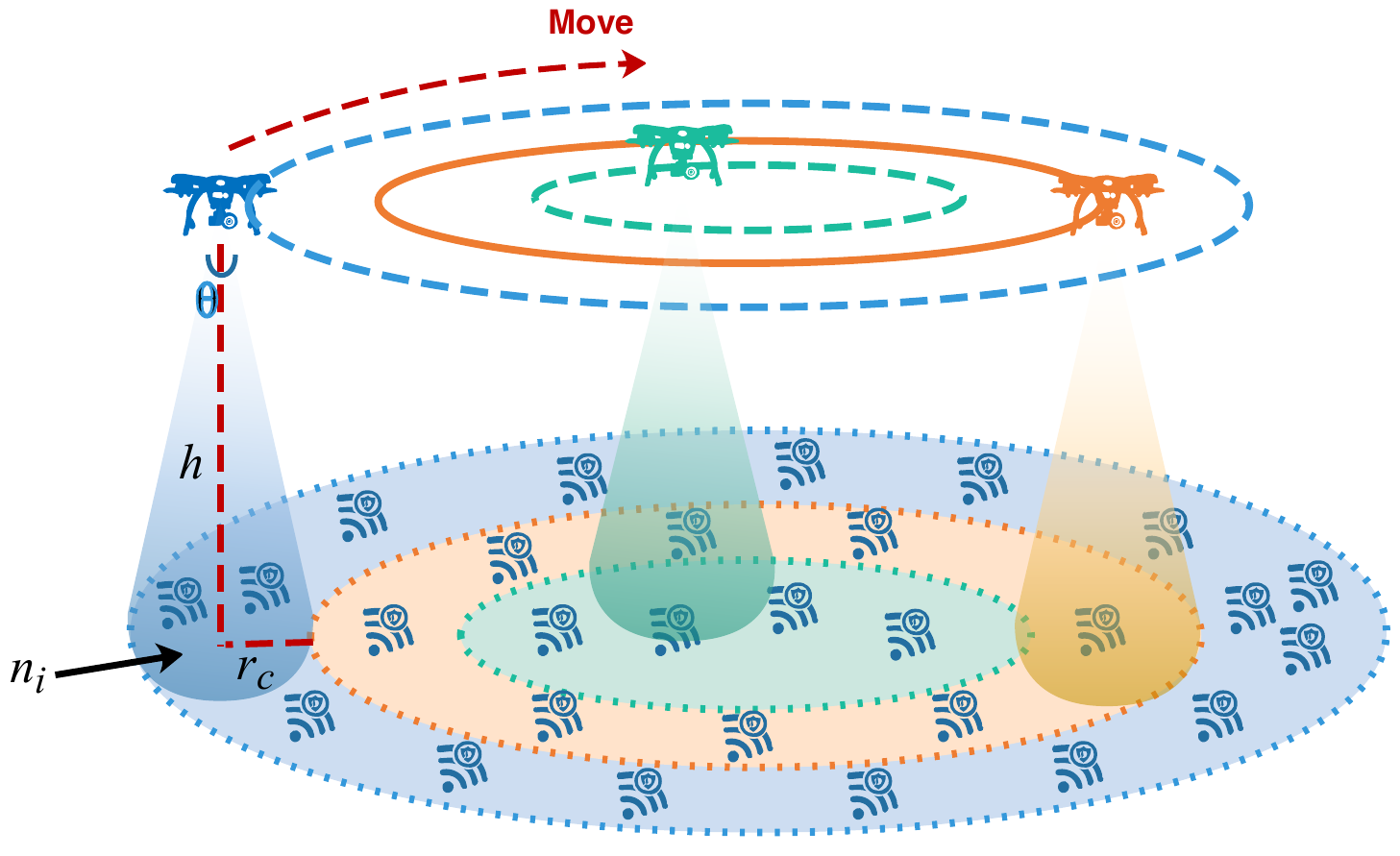}
\end{center}
\caption{Covered IoT at different positions.} 
\label{N}
\end{figure}

We consider a remote geographic area, where a set of $\mathcal{N}= \left\{1, 2, \ldots, N \right\}$ IoT devices (e.g., temperature/pressure sensors, cameras, etc.) are randomly scattered on the ground. A set of $\mathcal{M}=\left\{1, 2, 3, \ldots,M\right\}$ low-altitude UAVs are deployed to operate as aerial base stations to collect data from the ground IoT devices as depicted in figure {\ref{A}},  where each UAV follows  a  pre-programmed trajectory. This scheme emulates the UAV's real movements before flying, which is used in practice for several UAV applications, to overcome the problem of a random path while avoiding collision between a fleet-UAVs cooperating in such a mission completion. Without loss of generality, we assume that each UAVs follows a circular trajectory, we point out that the circular path is mostly studied in the literature due to an efficient tradeoff between rate maximization and energy minimization by regularizing the circle radius {\cite{CircularPath}.
 Trajectory radius are strategically calculated to meet full coverage of the target area with no coverage-hole both in the center and in the borders of the area as depicted in figure {\ref{N}}}. For the sake of simplicity, we assume that UAVs are numbered increasingly, where UAV $1$ follows the trajectory with the smallest rotational radius.
Depending on the running application, the data provided by IoT devices comes in different forms include log files, audio streams, video feeds, surveillance data, etc. Collected data has to reach the application system deployed in the cloud. Thus, due to the lack of a direct link between every UAV and cloud system, a UAV has to forward the data packets in a multi-hop fashion until they get to the base station/satellite. Without loss of generality, we assume a static routing, where UAV $i$ transmits its packets to UAV ($i+1$) to reach the gateway, and vice-versa. Note that deploying a multi-hop scheme instead of connecting directly each UAVs to BS exhibits nice features in terms of QoS, lower transmission cost, better energy efficiency, longer ground devices’ lifetime, higher spectrum efficiency/utilization, and better self-organizing capability.

Whereas the gateway UAV is responsible for aggregating data from all ground IoT devices, eventually making it the bottleneck of the network. Here, we assume that the UAV $GW$ is connected to the end-system/backend via a separate reliable link (e.g 5G) to transfer more data over the air at a faster speed, reduced congestion, and lower latency \cite{5Guav}. That said, massive IoT deployment might cause network failures due to Denial of Service (DoS) attack. However fortunately, most of IoT services have lenient QoS requirements (e.g., few bits per second). Combining wireless technologies (e.g., IEEE 802.11, 802.15, Zigbee, etc.) seems to solve such bottleneck issues by bringing the network closer to IoT devices using widespread UAV deployment.\\

%Thus, UAV-fleet is meant to provide wireless connectivity between a device and wireless end system (In our case it represent Cloud System C), owing to the lack of reliable direct communication link between these two components. Drone-network or FANET operates as a multi-hop wireless network, each UAV not only acquire data from ground devices but also act as a router to forward a packet to its neighbors, in order to reach the end system (C). 
%Here, the Cloud system (C) receives data from a gateway which act as a relay between UAV-Network and $c$, this gateway can send their data via satellite, cellular, LoRa.. etc to the terrestrial base station. In the considered architecture, we assume the case of no packets losses between the cloud system and the gateway, otherwise,  the packets that leave GW are undoubtedly received by $c$.
We consider both the uplink and the downlink streams. The uplink is used to forward collected data from covered ground IoT devices to the gateway. Whereas, in the downlink, IoT control instructions ( e.g. reboot, turn-off and self-test, or feedback acknowledgments, parameters estimation, etc.) are transmitted from the cloud to the targeted IoT over the same multi-hop UAVs network.  
 %\textcolor{red}{Changer stops par time slots. Ne plus parler de hover, mais plutot de flying, where the position of the UAV changes from a time slot to another} 
  At each time slot, the  UAV $i$ covers $n_{i}$ ground IoT devices as shown in figure$~\ref{N}$, which are distributed according to a homogeneous Poisson point process (PPP) with density $\lambda$, measured in IoT devices/$m^{2}$.
 %\textcolor{red}{the devices always have something to transmit (i.e., the saturation condition).} 
 It is worth noting UAV $GW$ performs just as a gateway, it does not cover ground devices. Then, at each stationary position, UAV $i$ can serve $n_{i}$ IoT devices. Namely
\begin{equation}
n_{i}= \lambda \, \pi \, \left(r_{i}^{c}\right)^{2}= \lambda \pi \tan^{2} \left(\frac{\theta_{i}}{2} \right ) h_{i}^{2},
\end{equation}
where $r_{i}^{c} $ is the radius of the area covered by UAV $i$, $\theta_{i}$ is the antenna beamwidth used to communicate with ground devices and $h_{i}$ stands for the altitude of $UAV_{i}$. Subsequently, and during every round-trip, each UAV $i$ covers $N_{i}=\frac{4 \pi \,r_{i} }{r_{i}^{c} } n_{i}$; where $r_{i}$ denotes the rotation radius of UAV $i$ ( $r_{i}=\frac{V_{i}}{\omega_{i}}$). $V_{i}$ and $\omega_{i}$ are the linear velocity and the angular velocity of {$UAV_{i}$} respectively. Furthermore, exposing IoT devices and UAVs to the public (legitimate and illegitimate users) might introduce some security difficulties and privacy/data protection challenges. The security aspect goes beyond the scope of this work. However, it is worth noting that several recent works deal with these issues from different perspectives. Indeed, considerable research efforts have been spent on developing security concepts that must be implemented by UAVs and IoT devices. Namely, one must consider authentication, access control, authorization, etc. \cite{uavsecurity}. Yet, other studies focus on the secrecy rate of UAV-IoT commination systems under spatially distributed eavesdroppers \cite{securityIoT}.\\
\vspace{-0,5cm}
\subsection{Channel Model}
In this paper, IEEE 802.11s mesh networking standard is adopted for drone interconnection. Meshed communication provides the best benefits in terms of flexibility, reliability, and performance in comparison with other architectures \cite{FANET}, \cite{StandarS} and \cite{MeshCom}. Referring to IEEE 802.11s architecture, each UAV acts as a mesh access point, providing a local relaying service to various IoT devices that stand for basic service set. In such an architecture, two separate channels must be considered: 1) the Air-to-Ground channel (GC), and 2) Air-to-Air channel (AC). We assume that GC and AC links use different bands. In the following, we mainly focus on the characterization of reliable channel models of UAV communication links.\\

\noindent\textbf{Coverage Probability over Air-to-Ground Channel: } 
In line with literature {\cite{los}},  we assume that LoS link is predominant in Air-to-Ground channel. Under this plausible assumption, the probability that an IoT device experiencing an LoS link with a drone is expressed as follows {\cite{1018}}:
\begin{equation}
P_{LoS}={\beta_{1}\bigg(\frac{5 \pi}{12} - \theta_{e}\bigg)^{\beta_{2}} },
\end{equation}
where $\beta_{1}$ and $\beta_{2}$ represent the environment parameters and $\theta_{e}$ is the elevation angle between drone and ground device. Besides, the coverage  probability is considered as an important metric in radio channel characterization, which is influenced also by LoS probability. For the sake of simplicity, we assume that $P_{cov}^{Down}\equiv P_{cov}^{Up} \equiv  P_{cov}$. Then, the coverage probability of a tagged UAV writes \cite{1018}:
\begin{equation}
\begin{split}
P_{cov}(h,\lambda)= & 2\pi \lambda h^{2} \int_{0}^{\frac{\theta}{2}} Q\bigg(\frac{\mu_{LoS}- \psi(\theta_{c}) }{\sigma_{LoS}(\theta_{c})}\bigg )  
 \underbrace{\beta_{1}\bigg(\frac{5 \pi}{12} - \theta_{c}\bigg)^{\beta_{2}} }_\text{$P_{LoS}$}      \\ 
 & \times 
\frac{\sin(\theta_{c})}{\cos^{3}(\theta_{c})} e^{- \lambda \pi h^{2}  \tan^{2}(\theta_{c})} \,\,\,\, d\theta_{c},
\end{split}
\end{equation}
where $\mu_{LoS}$ and $\sigma^{2}_{LoS}$ are respectively the mean and variance of the shadow fading for LoS link. Let $\sigma_{LoS}$= $a_{LoS}\, e^{b_{LoS}\, \theta_{c}}$, with  $a_{LoS}$ and $b_{LoS}$ being frequency and environment dependent parameters, $\theta_{c}$ is the random variable angle that connect any IoT in $N$ to the closest UAV, and $\psi(\theta_{c})$ can be given by:
\begin{equation}
\psi(\theta_{c}) = 10\, \log_{10}\bigg(\frac{P_{t}}{N_{o} . \,L_{f} . \,SNR_{th}}\bigg), 
\end{equation}
where $P_{t}$ is the transmit power (ground IoT or UAV) and $N_{0}$ is the noise power and  $L_{f}=\bigg(\frac{4 \pi f d}{c}\bigg)^{2}$with $f$ being the used frequency, $c$ the speed of light and $d$ the distance between transmitter (IoT) and receiver (UAV) and $SNR_{th}$ is the threshold SNR.\\

\noindent\textbf{Contact Probability over Air-to-Air Channel:} 
Air-to-Air Channel is the limelight for UAV-fleet network deployment when a communication between drones is required. The literature lacks pioneering research dealing with air-to-air channel \cite{U2U} compared to air-to-ground. It is mainly affected by both LoS and NLoS link, shadowing, interference, ground reflection, etc. \cite{1027}. The authors of \cite{LayerA2A} consider two  A2A channel communication forms: 1) when drone transmitter and drone receiver are situated in the same layer (same altitude), and 2) when the drone receiver and transmitter are located in different layers (different altitude). Here, we assume that UAV $i$ is in contact with UAV $j$ ($j=i\pm 1$ for one-tier neighbors) when the separation distance $d_{i,j}(t)$ between the two drones is lower than the transmission range $d_{tx}$ either along the drone's flying time or only for a series of sporadic time slots (otherwise). Then, the contact probability $ \xi_{i,j}$ between UAV $i$ and $j$ is can be given by:
%\begin{equation}
 %\xi_{i,j}= \left \lbrace 
% \begin{align}
% &1 \qquad\qquad\qquad \text{if} \qquad d_{i,j}(t)\le d_{tx},  \quad \forall t\\
% &0 \qquad\qquad\qquad \text{if} \qquad d_{i,j}(t) > d_{tx},  \quad \forall t \\
 %&\frac{1}{\pi}\arccos\left(\frac{r_{i}^{2}+r_{j}^{2}-d_{tx}^{2}}{2\, r_{i}\, r_{j}}\right)   \qquad  \text{Otherwise} 
%\end{align}
%\right.
%\end{equation}

 \begin{eqnarray}
 \xi_{i,j}=\begin{cases}
               1 \qquad \qquad  \qquad \text{if}  \qquad d_{i,j}(t)\le d_{tx}, \quad \forall \, \, t \\
             0 \qquad \qquad \qquad \text{if}  \qquad d_{i,j}(t) > d_{tx},  \quad \forall \, \, t\\
             \frac{1}{\pi}\arccos\left(\frac{r_{i}^{2}+r_{j}^{2}-d_{tx}^{2}}{2\, r_{i}\, r_{j}}\right)   \qquad  \text{Otherwise} 
            \end{cases}
 \end{eqnarray}

For the sake of simplicity, we assume that the contact distance derived, for circular trajectories, is the sufficient isolation distance allowing to mitigate co-channel interference generated by a fleet of UAVs operating within a given area using the same frequency at the same time, see {\cite{Co-channel}} for detailed analysis on separation distance between drones.
\subsection{NET/MAC/PHY cross-layer model}
The dramatic development of drone communications sustains the need for prior efficient network design aiming to covey data with a satisfactory QoS. Mathematical modeling is adopted as a tool to analyze the network dynamics in terms of e2e throughput and e2e delay. UAV Network is seen as shared resources among multiple users. This is why we model such a complex system as a network of interconnected queues, where communication protocols present a meaningful function to ensure seamless data transmission. Throughout this paper, we simultaneously consider the network layer, MAC layer and Physical layer for each UAV.\\
\begin{figure}[!h] \centering
\includegraphics[trim = 0cm 0cm 0cm 1cm, scale=0.39]{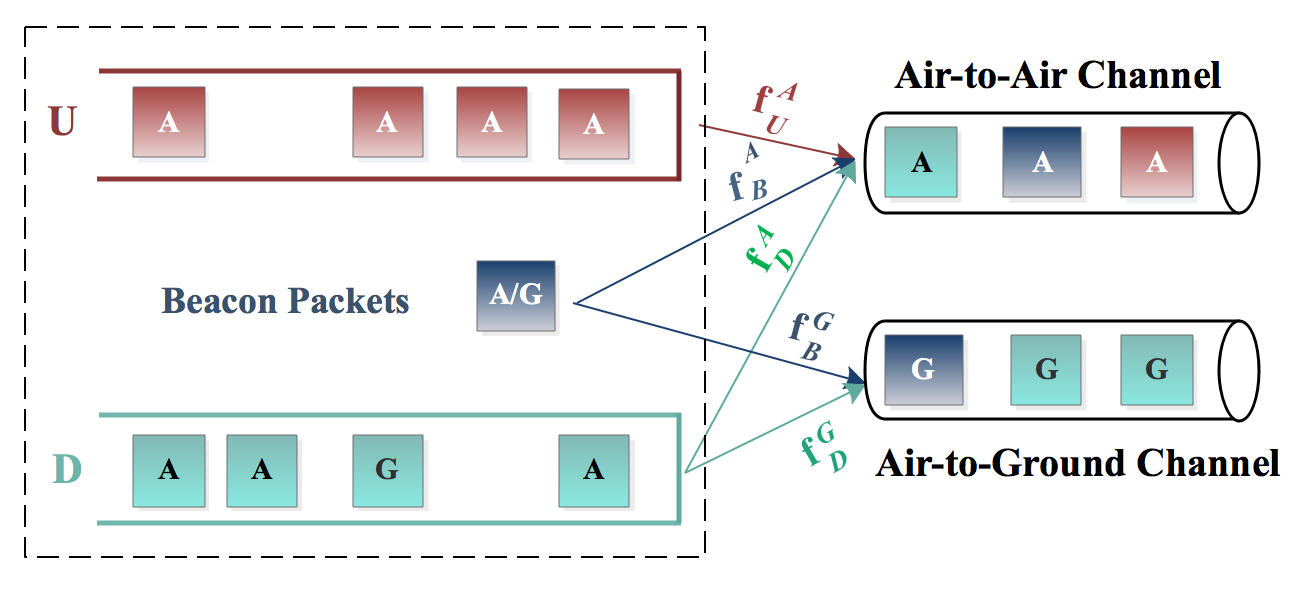}
\caption{NET/MAC/PHY cross-layers synergy.} 
\label{Queues}
\end{figure}

\noindent\textbf{Network layer: }
The network layer is the top layer of our cross-layered architecture. It defines sources and destination addresses of data packets and is responsible for routing across the UAV network. As illustrated in figure$~\ref{Queues}$, each drone $i$ handles two queues: 1) the uplink queue $U_{i}$ carries packets originating from IoT devices and then forwards them through the UAVs until reaching the cloud $c$; and 2) the downlink queue $D_{i}$ carries data coming from $c$ and to be delivered to their final destination (IoT devices). The data in downlink queue mainly includes the acknowledgment of uplink packets, control and command messages (e.g. Reboot, turn-off, adjust transmit power of IoT) \cite{2001}. Of course each UAV must periodically advertise its presence by transmitting beacon packets. Downlink and uplink queues are assumed to have infinite storage capacity, which allows to avoid the packet losses due to overflow. A First In First Out scheduling is considered. Moreover, the NET layer employs a Weighted Fair Queuing WFQ to manage the data to be transmitted over each single cycle. This mechanism allows some flexibility and enables QoS and packet prioritization. Moreover, UAV $i$ transmits over the air channel with probability $f^{A}_{U_{i}}$ when the uplink queue $U_{i}$ is not empty. It follows that the UAV $i$ transmits over the air channel and/or ground channel with probability $f^{A}_{D_{i}}$ and $f^{G}_{D_{i}}$ respectively. When UAV $i$ needs to send a beacon packet, it forwards it over air channel and/or ground channel with probability $f^{A}_{B_{i}}$ and $f^{G}_{B_{i}}$, with $f^{A}_{B_{i}}= 1-f^{A}_{U_{i}}-f^{A}_{D_{i}}$ and  $f^{G}_{B_{i}}+f^{G}_{D_{i}}=1$.\\
\begin{figure*}[!htb]
\centering
    \includegraphics[scale=0.55]{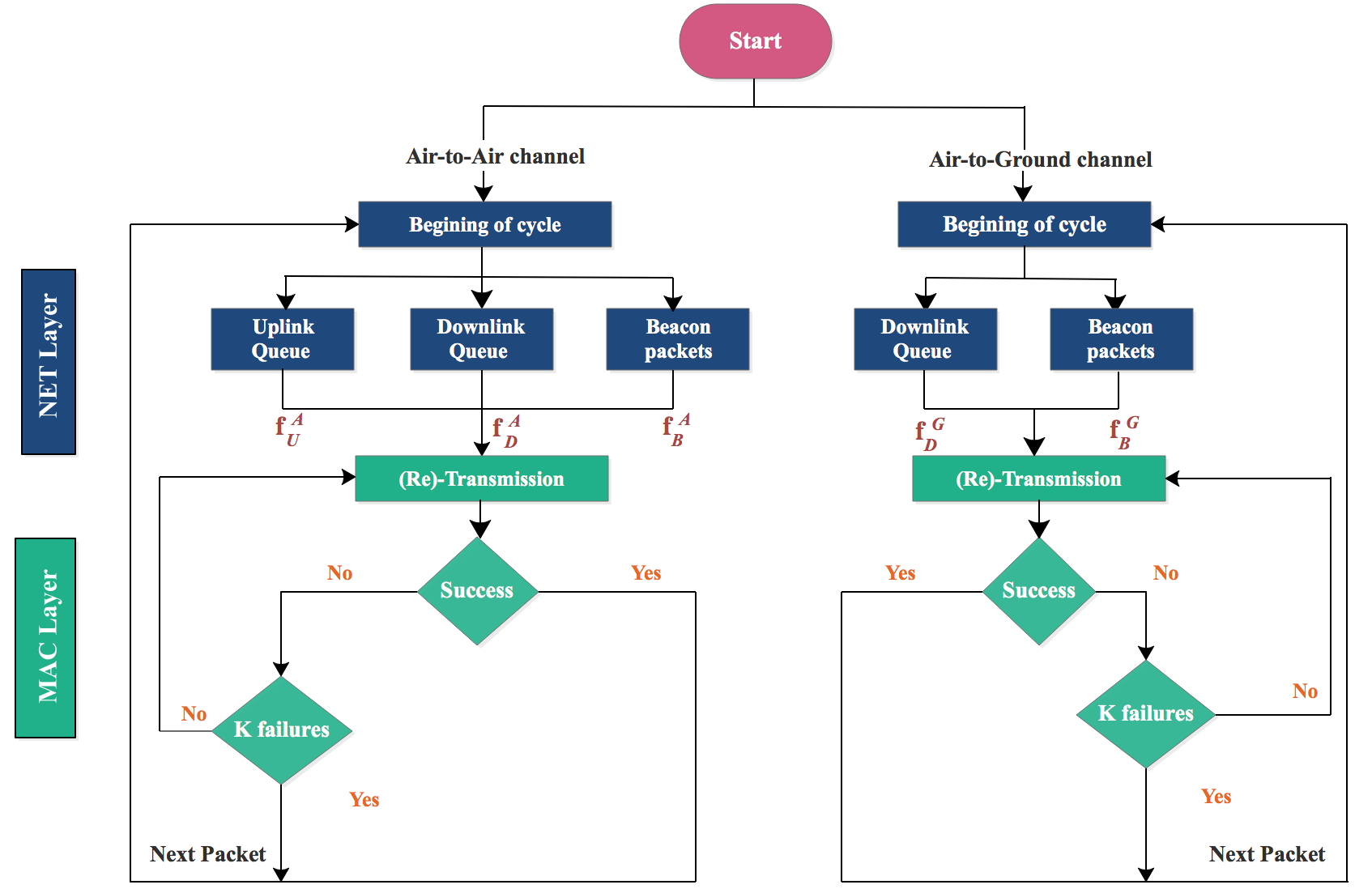}
\caption{Cross-layer flow chart and transmission cycle.}
\label{org} 
\end{figure*}

\noindent\textbf{MAC layer: } MAC layer defines the rules of how the communication medium is to be shared among active ground IoT devices. Here, we consider Distributed Coordinated Function (DCF) access method for both Air-to-Ground channel and Air-to-Air channels. However, when more than one IoT device attempt to access the same channel; a collision may occur. For this reason, the Carrier Sense Multiple Access/Collision Avoidance (CSMA / CA) used to solve collisions. At the beginning of every time slot, each node (UAV or IoT) attempts the ground channel with a probability $p_{i}^{G}$, whilst the air channel is attempted with probability $p_{i}^{A}$. The attempt probability depends on the number of neighbors, on the back-off mechanism and other parameters, it is given by \cite{2002}:
\begin{equation}
p^{A/G}_{i}= \frac{2(1-2\, p_{c})}{(1-2\, p_{c})(W+1)+p_{c}\, W (1-(2\, p_{c})^b)},
\end{equation}
where $p_{c}$ is the conditional collision probability, $W$ is the contention window and $b$ is the maximum back-off stage. UAV $j$ successfully receives a packet from UAV $i$, if drones $i$ and $j$ are getting in mutual contacts, and no concurrent communication is being from UAV $j$'s neighbors ($Ng(i)$). The success probability of from UAV $i$ to UAV $j$ is expressed by:
 \begin{equation}
P_{i,j} =  p_{i}^{A} \, \xi_{i,j} (1-p^{A}_{j})\prod_{k \in Ng(j) \backslash i} (1 - p_{k}^{A}\xi_{k,j}),
\end{equation}
Therefore, the probability that an IoT or UAV successfully transmits on Air-to-Ground channel is written: 
\begin{equation}
P^{s}_{i}= p_{i}^{G}\,P_{cov} \prod_{k \in \mathcal{N}_{i}\cup \, i}(1-p_{k}^{G}P_{cov_{k}})
\end{equation}
In case a collision is experienced, the packet is allowed to be re-transmitted for a limited number of consecutive attempts. Yet, after $K$ transmission failures, the packet is simply dropped. Then, the number of transmission attempts per packet on the Air-to-Air channel ($L^{A}_{i,j}$) and Air-to-Ground channel ($L^{G}_{i}$) write:
\begin{equation}
    L^{A}_{i,j}= \frac{1-(1-P_{i,j})^K}{P_{i,j}}
    \quad  \text{and}  \quad 
       L^{G}_{i}=\frac{1-(1-P^{s}_{i})^K}{P^{s}_{i}}.
\end{equation}

\noindent\textbf{Physical layer: } As aforementioned, we consider two separate channels air-to-air (AC) and air-to-ground GC over two separate frequency bandwidths. Also, we consider that each drone is equipped with two transceivers and two antennas: the UAV collects data from the ground using a directional antenna pointing downward towards the ground, while an omnidirectional antenna  allows to communicate with its neighboring drones. For sack of simplicity, the transmit power is assumed to be the same for all UAVs.\\

\noindent\textbf{Cross-layer design: } The organizational chart in Figure$~ \ref{org}$ is drawn to better understand the interaction between NET, MAC and PHY layers. It is worth noting that several time slots constitute a transmission cycle that either ends up with a transmission success or a failure/drop.

\section{Queuing Analysis and Performance Metrics}
We are constructing in this section the queuing networks associated to considered flying mesh drone network, in which multi-class of packets are supported. The queuing network model allows to predict the end-to-end performance metrics in terms of throughput and delay of every ongoing stream in the FMDN. We start by deriving the traffic intensity and then derive e2e throughput and e2e delay at the steady state. 
 \subsection{Traffic Intensity}
At each time slot, the uplink and downlink queues of each UAV experience some instantaneous loads that converges to the average value at the steady state. In order to analyze the queueing network stability, individual loads need to be derived for every single stream. We denote by $\pi^{U}_{i}$, respectively $\pi^{D}_{i}$, the probability that uplink queue of UAV $i$, respectively the downlink queue of UAV $i$, is non-empty. Stability of an infinite capacity queue is granted when the departure rate $d_{i}^{U/D}(t)$ is equal to the arrival rate $a_{i}^{U/D}(t)$. Namely, the stability condition writes \begin{equation}
\lim_{t \to +\infty}d_{i}^{U/D}(t)=\lim_{t \to +\infty} a_{i}^{U/D}(t).
\end{equation}
For high flexibility and accuracy, we are going to derive the departure/arrival rates expression for every active stream crossing a tagged UAV, and use the same cycle approach developed in \cite{1024} and \cite{1022}. Table~\ref{tab1} summarizes the main notation used in this section.\\
\begin{table}[h!]
\caption{Main notations and symbols.}
\label{tab1}
\centering
\begin{tabular}{|p{1cm}|p{7cm}|}
\hline
\textbf{Symbol} & \textbf{Meaning} \\
\hline
$n_{i}$  &  Instantaneous number of IoT devices covered by  UAV $i$ \\
\hline
$h_{i}$ & Altitude of UAV $i$ \\
\hline
$r_{i}^{c}$ & Coverage radius at each stop \\
\hline
$\lambda$ & Density of ground IoT devices \\
\hline
$\theta_{i}$ & Aperture angle of the antenna equipping UAV $i$ \\
\hline
$P_{cov}$ & Coverage probability\\
\hline
$\xi_{i,j}$  &  Contact Probability between UAV $i$ and UAV $j$\\ 
\hline
$p^{A/G}$   & Attempt probability on Air/Ground channel                                 \\ 
\hline
$P_{i,j}$ & Success probability from UAV $i$ to UAV $j$  \\
\hline
$P^{s}_{i}$ &  Success probability from a ground IoT device to UAV $i$\\
\hline
$K$  & Maximum attempts permitted per packet\\
\hline
$f^{A/G}_{U_{i}}$ & Probability to pick a packet from the uplink queue $U_{i}$ of UAV $i$, to be transmitted over Air/Ground channel\\ 
\hline
$f^{A/G}_{D_{i}}$ & Probability to pick a packet from the downlink queue $D_{i}$ of UAV $i$, to be transmitted over Air/Ground channel\\
\hline
$\pi^{U/D}_{i,s,d}$ & Probability that queue \textit{U} or queue \textit{D} has a packet, from source $s$ to destination $d$, at the head of line \\
\hline
$L^{A/G}_{i,j}$ & Number of attempts per packet over A/G channel from $i$ to $j$ \\ 
\hline
$\overline{R^{A/G}_{i}}$& Mean residual time over air/ground channel \\ 
\hline
$\tau^{A/G}_{D/B/U}$ & Service time of either beacon, downlink or uplink packet over GC or AC\\ 
\hline
$\tau^{A}_{B,U}$ & Service time of beacon and uplink packets, served before a packet at the head of downlink queue over air channel \\ 
\hline
$\tau^{A}_{B,D}$ & The mean service time of beacon and downlink packets  over Air-to-Air channel   \\ 
\hline
\end{tabular}
\end{table}

\subsubsection{\textbf{Uplink Queue Model}}

As mentioned earlier, the uplink queue carries packets from ground IoT devices to the cloud system through UAV relays. The inter-arrivals times of packets have a general distribution. The service time is the time elapsed between the instant a given packet reaches MAC layer until it is transmitted with success or it is dropped, it has a also general distribution that we will derive later. Since MAC layer can only handle a single packet at once, then it is natural to consider a single server modeling. Thus, uplink queue $U_{i}$ is captured as a G/G/1 queuing system. \\

\noindent \textbf{Departure rate: } This performance metric measures the rate at which packets are removed from queue $U_{i}$ following a successful transmission or a drop, per time slot. Then, 
\begin{lemma}
The average departure rate of the uplink queue $U_{i}$, on the route from IoT devices served by UAV $s$ to the IoT server $c$, is given by:
 \begin{equation}
	d^{U}_{i,s,c}= \begin{cases}
   \frac{ \pi^{U}_{i,s,c} \,\,  f^{A}_{U_{i}}}{\bar L_{i}^{A}} \, p_{i}^{A}\,   \xi_{i,i+1}    \qquad  \forall \, \, \, s\in\mathcal{M} \backslash  GW\\
 \pi^{U}_{i,s,c} \hfill i=GW
 \end{cases}
  \end{equation}
where the average cycle length $\bar L^{A}_{i}$, in slots, over air-to-air channel is given by:
\begin{multline}
\bar L_{i}^{A}=\sum_{s=1}^{i} \pi_{i,s,c}^{U}\,\,   f^{A}_{U_{i}} \, L^{A}_{i,i+1} +\sum_{d=1}^{i} \pi_{i,c,d}^{D}\,\,  f^{A}_{D_{i}} \, L^{A}_{i,i-1} + \\
\bigg(\underbrace{1-\sum_{s=1}^{i}\pi_{i,s,c}^{U}\,\,   f^{A}_{U_{i}}- \sum_{d=1}^{i}  \pi_{i,c,d}^{D}\,\,  f^{A}_{D_{i}}}_{\textnormal{\small{Probability to get a beacon packet}}}\bigg).
\end{multline}
\label{Lemma1}
\end{lemma}
\begin{proof}
We could prove this lemma using the cycle approach introduced in \cite{1025, 1025}. Here, we sketch a simpler alternative proof based on events decomposition. Let introduce the events A,B and C defined as follows:\\
\begin{itemize}
    \item Event A: \textit{A packet from source $s$ is at the head of line of the Uplink queue $U_{i}$};
    \item Event B: \textit{The Uplink queue $U_{i}$ is chosen to transmit in the ongoing cycle};
    \item Event C: \textit{UAV $i$ and UAV $i+1$ are currently in their mutual transmission ranges.}
\end{itemize}
We can check easily that:\\

\hspace{0.3cm} $P(A)= \pi^{U}_{i,s,c}$,\,\, $P(B)= f^{A}_{U_{i}}$, \,\,and $P(C)= \xi_{i,i+1}$.\\

\noindent It is worth noting that $P(A \cap B \cap C )$ is exactly the probability at which packets of active stream $(s,\, c$ depart from queue $U_{i}$. Then, the departure rate from $s$ towards the IoT server $c$ can be written as:
\begin{equation}
d^{U}_{i,s,c}=\frac{P(A \cap B \cap C )}{\tau^{A}_{i}} \hspace{0.3cm} with \hspace{0.3cm}  \tau^{A}_{i}=\frac{\bar L_{i}^{A}}{p_{i}^{A}}
 \end{equation}
 where $\tau^{A}_{i}$ is the average service time of a packet forwarded over the air channel.\\
\end{proof}
The total departure rate from UAV $i$ towards the cloud is then expressed as:
 \begin{eqnarray}
  d^{U}_{i} &=& \sum_{s \in \mathcal{M}} d^{U}_{i,s,c}\nonumber \\ &=&\begin{cases}
               \sum\limits_{s\in  \mathcal{M}}\frac{ \pi^{U}_{i,s,c} \,\,  f^{A}_{U_{i}}}{\bar L_{i}^{A}} \, p_{i}^{A}\,   \xi_{i,i+1}    \quad   \forall \, \, i \in\mathcal{M} \backslash  GW  \vspace{0.3cm}\\
             \sum\limits_{s \in \mathcal{M}} \pi^{U}_{i,s,c}  \hfill i=GW 
            \end{cases}
 \end{eqnarray}

\noindent\textbf{Arrival rate: } The arrival rate is defined as the rate at which packets arrives at queue $U_{i}$ per time slot. We distinguish two independent arriving flows: 1) the first flow is composed of packets coming from ground IoT devices, served by their own UAV $i$ and transmitted via the Air-to-Ground channel; and 2) the second flow composed of packets generated by another UAV $s$. Here, the UAV $i$ plays the role of a cooperative relay to forward data packets to the cloud. Let $S$ be the normalized throughput of the system composed of a single UAV and its covered ground devices at each stationary stop. $S$ is expressed as follows \cite{2002}:
\begin{equation}
    S=\frac{P^{s}E[Payload]}{(1-P_{tr})\sigma+P^{s}T_{s}+(P_{tr}-P_{s})T_{c}},
\end{equation}
were $E[Payload]$ denotes the average payload length, $P_{tr}$ is the channel busy probability. $\sigma$, $T_{s}$ and $T_{c}$ are the idle time, successful time and collision duration respectively. We have the following:
\begin{lemma}
The long term average arrival rate into queue $U_{i}$ from a source $s$ to the IoT server $c$ writes:
\begin{equation}
  \hspace{-0.2cm}a^{U}_{i,s,c}=\begin{cases}
                \frac{n_{i}\,S}{n_{i}+1}P_{cov} \hfill \text{if} \quad i=s \\
              \frac{n_{s}\,S}{n_{s}+1} P_{cov} \pi^{U}_{s,s,c} f^{A}_{U_{s}} \underbrace{\prod_{j=s}^{i-1}\left(1- (1 - P_{j,j+1})^K\right)}_{\substack{\textnormal{\small{successful transmissions over}} \\ \textnormal{\small{successive hops until UAV $i$}}}}\\
              \hfill \forall s=1, 2, \cdots, i-1
            \end{cases}
            \hspace{-0.5cm}
\end{equation}
\label{Lemma2}
\end{lemma}
\begin{proof}
Here, we sketch again a proof based on events decomposition. Let introduce the events A, B, C and D defined as follows:\\
\begin{itemize}
    \item Event A: \textit{The traffic arriving to UAV $i$ is generated by its covered ground IoT devices};
    \item Event B: \textit{The traffic generated by IoT devices covered by UAV $s$ has been actually departed from Uplink queue $U_{i}$};
    \item Event C: \textit{The transmission over successive hops from UAV $s$ until intermediate UAV $i$ are all successful}.
\end{itemize}
We can easily check that:\\

\hspace{0.3cm} $P(A)=  \frac{n_{i}\,S}{n_{i}+1}P_{cov} $,\,\, $P(B)= \pi^{U}_{s,s,c}\,f^{A}_{U_{s}}$,\,\, and $P(C)=\prod_{j=s}^{i-1}\left(1- (1 - P_{j,j+1})^K\right)$\\

\noindent Consequently, $P(A \cap B \cap C )$ is the packets  rate entering into uplink queue for a stream from $s$ to intermediate UAV  $i$. Then , the arrival rate can given by :
\begin{equation}
a^{U}_{i,s,c}=P(A \cap B \cap C),
 \end{equation}
 which completes the proof.\\
\end{proof}
\noindent\textbf{Steady state and rate balance equation: } When the setady state is reached, the long term arrival rate equals the long term departure rate. Yet, this provides the rate balance equation, i.e., $a^{U}_{i}=d^{U}_{i}$ $\forall i \in \mathcal{M}$. Since, we formally derived the two latter metrics, we can then find out the expression of average load $\pi^{U}_{i}$ at each UAV. We show in Appendix \ref{appendix:RBE}, that the rate balance equation yields a linear system. Namely, the uplink queueing system load $\boldsymbol{\Pi}^{U}=(\pi_{1}^{U}, \pi_{2}^{U}, \cdots, \pi_{M}^{U})$ is given by:
\begin{equation}
  \boldsymbol{\Pi}^{U}=\textbf{X}^{-1}\cdot \textbf{Y},
\end{equation}
where $\textbf{X}$ is a $M\times M$ matrix and $\textbf{Y}$ is a column vector with dimensionality $M\times 1$. The computation derivation is given is Appendix \ref{appendix:RBE}.\\

\subsubsection{\textbf{Downlink Queue Model}}
The downlink queue carries data packets originated from the cloud towards ground IoT devices as final destinations. It is worth noting that both the inter-arrival times of packets and the service time follow general distributions. Interestingly, the downlink case exhibits two servers as each UAV is multi-homed and might transmit two different packets at the same time, i.e., both over air channel and ground channel. Thus, the $D_{i}$ can be captured using a G/G/2 model.\\

\noindent \textbf{Departure rate:} Packet depart from downlink queue $D_{i}$ to either another UAV via the air channel, or to reach the final destination via the ground channel. We recall that the set of covered IoT devices by UAV $i$ is denoted $\mathcal{N}_{i}$. After some algebra, we have the following result:
\begin{lemma}
The long term average departure rate from downlink queue of UAV $i$ is given by:
 \begin{equation}
d^{D}_{i,c,d} = \begin{cases}
\frac{ \pi^{D}_{i,c,d} \, f^{A}_{D_{i}}}{\bar L_{i}^{A}} \, p_{i}^{A} \,   \xi_{i,i-1}  \qquad  \qquad   \text{if}  \quad d\in \mathcal{M}\\
\frac{ \pi^{D}_{i,c,d} \, f^{G}_{D_{i}}}{\bar L_{i}^{G}} \, p_{i}^{G} \,   P_{cov}   \hfill \text{if}  \quad d\in \mathcal{N}_{i}
 \end{cases}
\end{equation}
\label{Lemma3}
where $\bar L_{i}^{G}$ is the average length in slots of a cycle on Air-to-Ground channel. It can is expressed as: 
\begin{equation}
\bar L_{i}^{G}=\sum_{d\in \mathcal{N}} \pi_{i,c,d}^{D}\,\,  f^{G}_{D_{i}} \, L^{G}_{i,j} + 
\sum_{d\in \mathcal{N}}(1-  \pi_{i,c,d}^{D}\,\, f^{G}_{D_{i}}).
\end{equation}
\end{lemma}
\begin{proof}
Consider the events A, B, C, D and E defined as:\\
\begin{itemize}
    \item Event A: \textit{A packet from source $s$ is currently at the head of line of the downlink queue $D_{i}$};
    \item Event B: \textit{The Downlink queue of UAV $i$ is chosen to transmit in the ongoing cycle over air channel};
    \item Event C: \textit{UAV $i$ and UAV $(i-1)$ are currently in their mutual transmission ranges};
    \item Event D: \textit{The Downlink queue of UAV $s$ is chosen to transmit in the ongoing cycle over ground channel};
    \item Event E: \textit{The ground IoT devices are covered by UAV $i$}
\end{itemize}
We have the following:\\

\hspace{0.3cm} $P(A)= \pi^{D}_{i,c,d}$,\,\, $P(B)= f^{A}_{D_{i}}$, \,\, $P(C)=\xi_{i,i-1} $, \,\,  $P(D)= f^{G}_{D_{i}}$ , \,\, and $P(E)= P_{cov}$ \\

\noindent Therefore, $P(A \cap B \cap C )$ is the packets rate departing from downlink queue $D_{i}$ over air channel. Similarly, $P(A \cap D \cap E )$ is the packets rate departing from queue $D_{i}$ over channel channel. Then , the departure rate can given by :
\begin{equation}
d^{U}_{i,s,c}= \frac{P(A \cap B \cap C )}{\tau^{A}_{i}} + \frac{P(A \cap D \cap E )}{\tau^{G}_{i}}
 \end{equation}
 with $\tau^{G}_{i}= \frac{\bar L_{i}^{G}}{p_{i}^{G}}$ being the average service time of a packet sent over the ground channel.
 \end{proof}
 The total downlink departure rate from UAV $i$ writes:
\begin{equation}
\hspace{-0.1cm} d^{D}_{i}=\begin{cases}
               \sum\limits_{\substack{{d=1}\\{d\neq i}}}^{i-1}  \, \frac{ \pi^{D}_{i,c,d} \, \, f^{A}_{D_{i}} }{\bar L_{i}^{A}} \, p^{A}_{i} \,   \xi_{i,i-1}  + \frac{ \pi^{D}_{i,c,i} \, f^{G}_{D_{i}}}{\bar L_{i}^{G}} \, p^{G}_{i} \,   P_{cov} \quad \\ \hfill \forall \,i\in \, \mathcal{M} \backslash GW \\
            \sum\limits_{\substack{{d=1}}}^{i-1}  \, \frac{ \pi^{D}_{i,c,d} \, f^{A}_{D_{i}}}{\bar L_{i}^{A}} \, p^{A}_{i} \, \xi_{i,i-1}  \hfill \text{if} \quad   i=GW
            \end{cases}\hspace{-0.2cm}
\end{equation}

\noindent \textbf{Arrival rate: } We consider the each single packet reaching the cloud/server, must be acknowledged. Here, we assume that source $s$ will be acknowledged by a tiny packet whose size is a fraction $\upsilon_{ack}$ of the uplink flow $d_{GW,s,c}^{U}$. Moreover, the server may periodically send some control/command packets to IoT devices or to UAVs theme-selves. We denote by $\Upsilon_{k}^{c}$ the probability that the cloud server sends a control packet to a given IoT device/UAV $k$. Let $\zeta_{_{GW}}$ be the packet success probability experienced by UAV $GW$ over legacy network (5G, LoRaWAN, Sigfox, NB-IoT, satellite, etc.) to reach the cloud. $\zeta_{_{GW}}$ is a sigmoidal function with the signal to interference plus noise ratio, depending on channel bandwidth, the transmit power, channel gain, interference level, packet length, etc. Yet
\begin{lemma}The long term arrival rate from the IoT server $c$ into UAV $d$ is given by:
\begin{equation}
  a^{D}_{i,c,d}= \begin{cases}
               \left(\upsilon_{ack}\, d_{GW,s,c}^{U} + \Upsilon_{GW}^{c}\right)\cdot\zeta_{_{GW}} \quad \text{if} \,\,   i=GW\\
            \left(\upsilon_{ack} \, d_{GW,s,c}^{U}\, +\sum\limits_{k\in \cup_{l=1}^{i} \mathcal{N}_{l}} \Upsilon_{k}^{c}\right)\cdot\zeta_{_{GW}} \cdot \\
     \prod\limits_{j=GW}^{i}\left(1- \left(1 - P_{j,j-1}\right)^K\right)  \hfill \forall \, i\in \, \mathcal{M} \cup \mathcal{N}
            \end{cases}
            \hspace{-0.2cm}
\end{equation}
\label{Lemma4}
\end{lemma}
\begin{proof}
The proof follows using the same approach based on events decomposition as presented in lemmas 1-3.\\
\end{proof}
%%%%%%%%%%%%%%%%%%%%%%%%%%%%%%%%%%%%%%%%%%%%%%%%%%%%%%%%%%%%
%%%%%%%%%%%%%%%%%%%%%%%%%%%%%%%%%%%%%%%%%%%%%%%%%%%%%%%%%%%%
%%%%%%%%%%%%%%%%%%%%%%%%%%%%%%%%%%%%%%%%%%%%%%%%%%%%%%%%%%%%
%%%%%%%%%%%%%%%%%%%%%%%%%%%%%%%%%%%%%%%%%%%%%%%%%%%%%%%%%%%%
\noindent It follows that the total arrival rate to downlink queue $D_{i}$ can be written as:
\begin{equation}
  a^{D}_{i}= \begin{cases}
               \left(\upsilon_{ack}  \sum\limits_{\substack{{s=1}}}^{i-1} \, d_{GW,s,c}^{U} + \Upsilon_{GW}^{c}\right)\cdot\zeta_{_{GW}} \quad \text{if} \,\,   i=GW\\
            \left(\upsilon_{ack}  \sum\limits_{\substack{{s=1}}}^{i-1} \, d_{GW,s,c}^{U}\, +\sum\limits_{k\in \cup_{l=1}^{i} \mathcal{N}_{l}} \Upsilon_{k}^{c} \right) \cdot\zeta_{_{GW}}\cdot \\
     \prod\limits_{j=GW}^{i}\left(1- \left(1 - P_{j,j-1}\right)^K\right)  \hfill \forall \, i\in \, \mathcal{M} \cup \mathcal{N}
            \end{cases}
            \hspace{-0.2cm}
\end{equation}
%The result can be obtained using a similar approach used in Lemma \ref{Lemma1}'s and \ref{Lemma2}'s  proof. \\
%%%%%%%%%%%%%%%%%%%%%%%%%%%%%%%%%%%%%%%%%%%%%%%%%%%%
%%%%%%%%%%%%%%%%%%%%%%%%%%%%%%%%%%%%%%%%%%%%%%%%%%%%
\noindent \textbf{Steady state and rate balance equation: } The rate balance equation $a^{D}_{i} = d^{D}_{i}$, $\forall i\in\mathcal{M}$ remains true when the steady state is achieved. Hence, the traffic intensity of downlink queues can be written in a matrix form as (see Appendix \ref{appendix:RBE}):
\begin{equation}
  \boldsymbol{\Pi}^{D}=\textbf{W}^{-1}\cdot \textbf{Z},
\end{equation}
where $\textbf{W}$ is a $M\times M$ matrix and $\textbf{Z}$ is a column vector with dimensionality $M\times 1$. The computation derivation is given in Appendix \ref{appendix:RBE}.

%%%%%%%%%%%%%%%%%%%%%%%%%%%%%%%%%%%%%%%%%%%%%%%%%%%%
%%%%%%%%%%%%%%%%%%%%%%%%%%%%%%%%%%%%%%%%%%%%%%%%%%%%
\subsection{End-to-End throughput}

Now, we are interested in deriving the normalized end-to-end throughput of an active streams. This latter measures the amount of data packets successfully sent, per time slot, from a given source $s$ to a tagged destination $d$. It is worth noting that the e2e throughput of a given stream is defined as the long term arrival rate at the final destination.\\

\noindent \textbf{Uplink End-to-End Throughput: }
The average e2e throughput of an active stream initiated by an IoT device, covered by UAV $s$, towards the IoT server within the cloud is denoted by $\Theta^{\uparrow}_{s,c}$. It equals the arrival rate at the gateway UAV $GW$ times the packet success rate over the IoT legacy network (4G, 5G, LoRaWAN,Sigfox, NB-IoT, etc.). Therefore, $\Theta^{\uparrow}_{s,c}$ writes:
\begin{eqnarray}
    \Theta^{\uparrow}_{s,c}&=& a^{U}_{GW,s,c}\cdot\zeta_{GW}\\&=&\frac{n_{s}\,S\,\zeta_{GW}}{n_{s}+1} P_{cov} \pi^{U}_{s,s,c} f^{A}_{U_{s}} \prod_{j=s}^{M}\left(1- \left(1 - P_{j,j+1}\right)^K\right).\nonumber
\end{eqnarray}
\noindent \textbf{Downlink End-to-End Throughput: } The average e2e throughput of an active stream initiated by the IoT server $c$ towards an IoT device served by UAV $d$, is denoted by $\Theta^{\downarrow}_{c,d}$. It equals the arrival rate at the serving UAV $d$ times the packet success probability between this latter and the ground IoT device. Namely,
\begin{equation}
  \Theta_{c,d}^{\downarrow} = 
        \frac{1}{n_{d}+1}a_{d,c,d}^{D} \, \, \left(1- \left(1-P^{s}_{d}\right)^K\right).
\end{equation}
%\begin{equation}
%  \Theta_{c,d}^{\downarrow} = \begin{cases}
%        \frac{1}{n_{d}+1}a_{d,c,d}^{D}    \hfill  \forall \,\, d\in \mathcal{U}_{i}\\
%        \frac{1}{n_{d}+1}a_{d,c,d}^{D} \, \, (1- (1-P^{s}_{d})^K)
%        \quad\, \text{destination} \\ 
%        \hfill\text{is an IoT device covered by UAV $d$.}\\
%            \end{cases}
%\end{equation}

%%%%%%%%%%%%%%%%%%%%%%%%%%%%%%%%%%%%%%%%%%%%%%%%%%%%
%%%%%%%%%%%%%%%%%%%%%%%%%%%%%%%%%%%%%%%%%%%%%%%%%%%%
\subsection{End-to-End delay}

We turn now to analyze the e2e delay of an active stream $(s,\,\,d)$. This crucial performance metric is defined as the time required for a packet to moves from the source node $s$, across a multi-hop network, until it reach its final destination $d$.  The sojourn  delay  $D_{i}$ at UAV $i$ is an accumulative metric that writes as the sum of processing delay $D_{i}^{Proc}$, Waiting delay $D_{i}^{Wait}$, transmission delay $D_{i}^{Trans}$ and propagation delay $D_{i}^{Prop}$. Namely
\begin{equation}
  D_{i}=D_{i}^{Proc}+D_{i}^{Wait}+D_{i}^{Trans}+D_{i}^{Prop}.
\end{equation}
For the sack of simplicity, this article will only consider the queuing delay and transmission delay as the processing time and propagation delay are negligible. We denote the sojourn delay  of a packet at UAV $i$ in the uplink and downlink queues respectively  by  $D_{i}^{\uparrow}$ and $D_{i}^{\downarrow}$.\\

\subsubsection{\textbf{Uplink Average e2e Delay}}
the e2e delay is the sum of partial sojourn delays $D^{\uparrow}_{i}$ experienced at every crossed UAV $i$ crossed by active stream $(s,\,\,d)$ which can be  given by:
\begin{equation}
   D^{\uparrow}_{s,d}=\sum_{i \, \in \,(s,\,d)} D^{\uparrow}_{i}.
\end{equation}
Each packet in the uplink queue $U_{i}$, at NET layer, has to wait a certain mean amount of time called waiting time. Then, it is passed to the MAC layer, in order to process its transmission over Air-to-Air channel. This latter corresponds to the service time. Moreover, the waiting time over the air channel is divided into two terms: 1) The mean residual time $\overline{R}^{A}_{i}$ seen by a newly entered packet, plus 2) the Queuing time $Q^{A}_{U_{i}}$ spent in the queue $U_{i}$ which is the time required for all the packets entered earlier to get served. Let denote by $\tau^{A}_{B_{i}}$ the average service time of a packet from node $i$ to $j$ in  MAC  layer over Air-to-Air channel. We have the following result:
\begin{proposition}
The end-to-end uplink delay along the route from source $s$ to the IoT server $c$ (hosted in the cloud) writes:
\begin{eqnarray}
 D^{\uparrow}_{s,c}&=&\sum_{i=s}^{GW}  D^{\uparrow}_{i} \\ &=&\mathlarger{\mathlarger{\sum}}_{i=s}^{GW} \left[\frac{ \overline{R}^{A}_{i} + \left(\frac{1-f^{A}_{U_{i}}}{f^{A}_{U_{i}}}\right) \, \tau^{A}_{(B,D)_{i}}}{1- a^{A}_{i}\, \,\left[\tau^{A}_{U_{i}}+ \left(\frac{1-f^{A}_{U_{i}}}{f^{A}_{U_{i}}}\right)\, \tau^{A}_{(B,D)_{i}}\right]} + \tau^{A}_{U_{i}}\right]\nonumber
 \end{eqnarray} 
\end{proposition}
\begin{proof}
A sketch of the proof is provided in Appendix \ref{appendix:Uplinke2eDelay}.
\end{proof}
%The  computation  derivation is provided in appendix \ref{appendix:UPLINK}.

\subsubsection{\textbf{Downlink Average e2e Delay}}
 is the sum of partial  sojourn delays encountered by a packet between the time of insertion into downlink queue of UAV $s$ and the time of delivery to IoT devices $d$ which can be expressed as:
\begin{equation}
   D_{s,d}^{\downarrow}=\sum_{i \, \in \,(s,\,d)} D^{\downarrow}_{i}.
\end{equation}
Each packet enters to the downlink queue has to spend a amount of time waiting another packet being served over the Air-to-Ground $ \overline{R}^{G}_{d} $\ Air-to-Air channel $\overline{R}^{A}_{i}$ which represent average residual time, and has to wait the time needed to serve the earliest packets (Queuing time), and the transmission time  in the MAC layer. Let $\tau ^{G}_{D}$ and $\tau ^{G}_{B}$ be the service time of a packet from downlink queue $D_{i}$ and the service time of a beacon packet, respectively. Notice that the UAV $d+1$ is the UAV located just before IoT serving UAV $d$ on the downlink side. Then, the delay for a packet to be transferred from source $s$ to destination $d$ is can be  given by the next lemma.
\begin{proposition}
The end-to-end downlink delay along the route from IoT server $c$ to IoT devices served by UAV $d$ is given by:
\begin{eqnarray}
D^{\downarrow}_{c,d} &=&
\sum_{i=c}^{d+1}  D_{i}^{\downarrow} \\
&=&
\mathlarger{\mathlarger{\sum}}_{i=c}^{d+1}\left[\frac{ \overline{R}^{A}_{i} + \left(\frac{1-f^{A}_{D_{i}}}{f^{A}_{D_{i}}}\right) \, \tau^{A}_{(B,U)_{i}}}{1- a^{A}_{i}\, \,\left[\tau^{A}_{D_{i}}+ \left(\frac{1-f^{A}_{D_{i}}}{f^{A}_{D_{i}}}\right)\, \tau^{A}_{(B,U)_{i}}\right]}+ \tau^{A}_{D_{i}}\right] \nonumber\\
&& +\frac{ \overline{R}^{G}_{d} + \left (\frac{1-f^{G}_{D_{d}}}{f^{G}_{D_{d}}}\right) \, \tau^{G}_{B_{d}}}{1- a^{G}_{d}\, \,\left[\tau^{G}_{D_{d}}+ 
\left (\frac{1-f^{G}_{D_{d}}}{f^{G}_{D_{d}}}\right) \, \tau^{G}_{B_{d}}\right]} +\tau ^{G}_{D_{d}}.\nonumber
\end{eqnarray}
\end{proposition}
\begin{proof}
A sketch of the proof is provided in Appendix \ref{appendix:Downlinke2eDelay}.
\end{proof}

\section{Simulation and numerical investigations}

%we choose to analyze UAVs at the height of 50m and 100m, which are the most practical cases in reality.
We consider a simple network with including a gateway and four homogeneous drones. Specifically, $h_{i} \equiv h$, $\theta_{i} \equiv \theta$, $P_{t_{i} }\equiv P_{t}$, $V_{i} \equiv V$, and $\omega_{i} \equiv \omega$ $\forall i=1, 2, 3, 4$. Next, we launch four bi-directional streams from/to ground IoT devices covered by the four UAVs, to/from the IoT server hosted in the cloud. (UAV $1 \leftrightarrow$ IoT server), (UAV $2 \leftrightarrow$ IoT server), (UAV $3 \leftrightarrow$ IoT server), and (UAV $4 \leftrightarrow$ IoT server). Table \ref{tab:setting} lists the main setting used in the simulation  \cite{mozaffarid2d} \cite{CircularPath} \cite{1016} \cite{Parameterssett} \cite{simulationv}.
\begin{table}[h!]
\begin{center}
\caption{Parameter settings.}
\label{tab:setting}
\begin{tabular}{|l|l||l|l|}
\hline
\textbf{Parameters} & \textbf{Values} & \textbf{Parameters} & \textbf{Values} \\
\hline
V          &   20 $m/s$    &       $d_{tx}$ &  15 $m$    \\
\hline
  $\omega$        &   8 $rad/s$     &         $P_{t}$   &    20 $dBm$    \\ 
\hline
 $N_{0}$        &   -150 dBm    &        $f$   &    2.4 GHz   \\
\hline
   $SNR_{th}$        &   5 $dBm$     &   $\mu_{LoS}$  &  1   \\ 
 \hline 
    $k_{1}$        &  11.95      &         $k_{2}$   & 0.136         \\
\hline
$\beta_{1}$        &  10.39      &         $\beta_{2}$   &    0.05   
\\ 
\hline
$Payload$        &     1184 bit   &       $\sigma$     &   50 $\mu$s    \\
\hline
$T_{c}$        &     1982 $\mu$s    &         $T_{s}$   &  1713 $\mu$s   \\ 
  \hline
$W$        &   8     &        $b$    &   4   \\
\hline
$Vack$        &   0.7     &         $K$   &   3    \\ 
\hline
\end{tabular}
\end{center}
\end{table}

\subsection{Normalized End-to-End throughput} 
%=============== Fig=================
%\begin{figure}
  %\begin{center}
  %\includegraphics[width=3.5in]{pdf/03.pdf}
 % \vspace{-15pt}
  %\caption{Simulated efficiency of reduced conduction angle half-wave rectifier versus $R_{DC} / R_s(f_0)$ for varying rectifier on-resistance.}\label{sim_opt_eff}
 % \end{center}
%\end{figure}

%%%%%%%%%%%%%%%%%%%%%%%%%%%%%%%%%%%%%%%%%%%%%%%%%%%%%%%%%%%%%%%%%%%%%%%%%%%%%%%%%%%%%%%%%%%%%%%%%%%%
\begin{figure}[H]
\begin{center}
\subfloat[Load of uplink queue Vs forwarding probability]
{\includegraphics[trim= 0cm 5cm 0cm 6cm,width=0.35\textwidth, height=5cm]{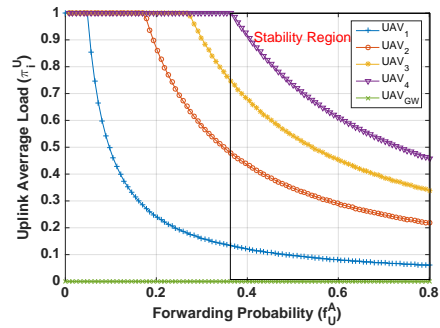}\label{aa}}\\
\subfloat[End-to End throughput of uplink queue Vs forwarding probability]
{\includegraphics[trim= 0cm 5cm 0cm 6cm, width=0.35\textwidth, height=5cm]{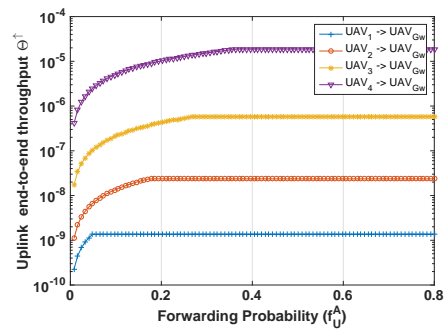}\label{bb}}
\caption[Optional caption for list of figures 5-8]{Uplink queue as a function of the Forwarding probability $f^{A}_{U}$}
\label{fig:subfigureExample2}
\end{center}
\end{figure}
\begin{figure*}
\centering
\subfloat[Load of downlink queue Vs forwarding probability$f^{A}_{D}$]
{
\includegraphics[trim= 0cm 5cm 0cm 6cm,width=0.3\textwidth, height=5cm]{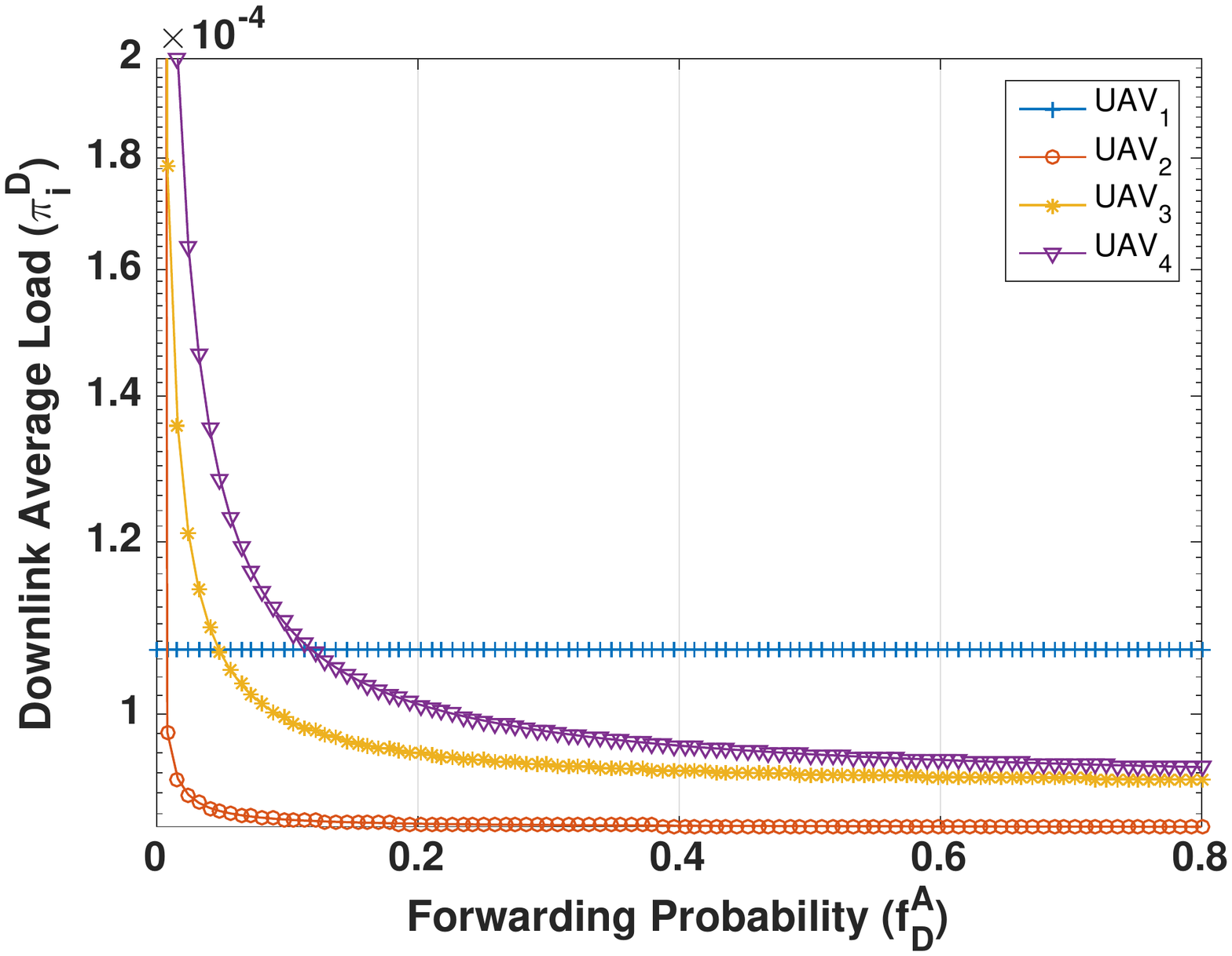}\label{cc}
}
\subfloat[Load of downlink queue Vs forwarding probability $f^{G}_{D}$]
{
\includegraphics[trim= 0cm 5cm 0cm 6cm,width=0.3\textwidth, height=5cm]{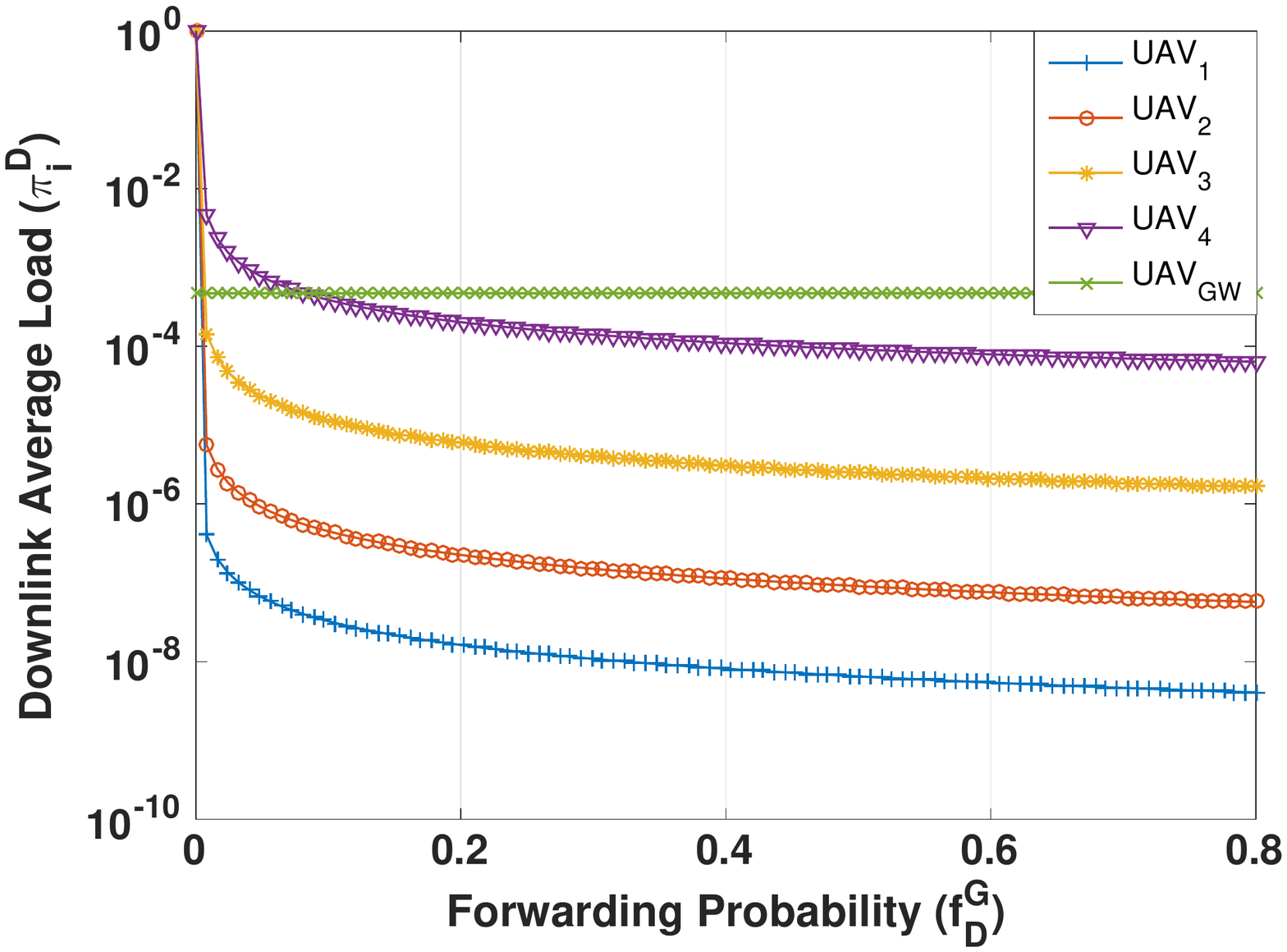}\label{dd}
}
\subfloat[End-to End throughput of downlink queue Vs forwarding probability     $f^{G}_{D}$]
{\includegraphics[trim= 0cm 5cm 0cm 6cm,width=0.3\textwidth, height=4.7cm]{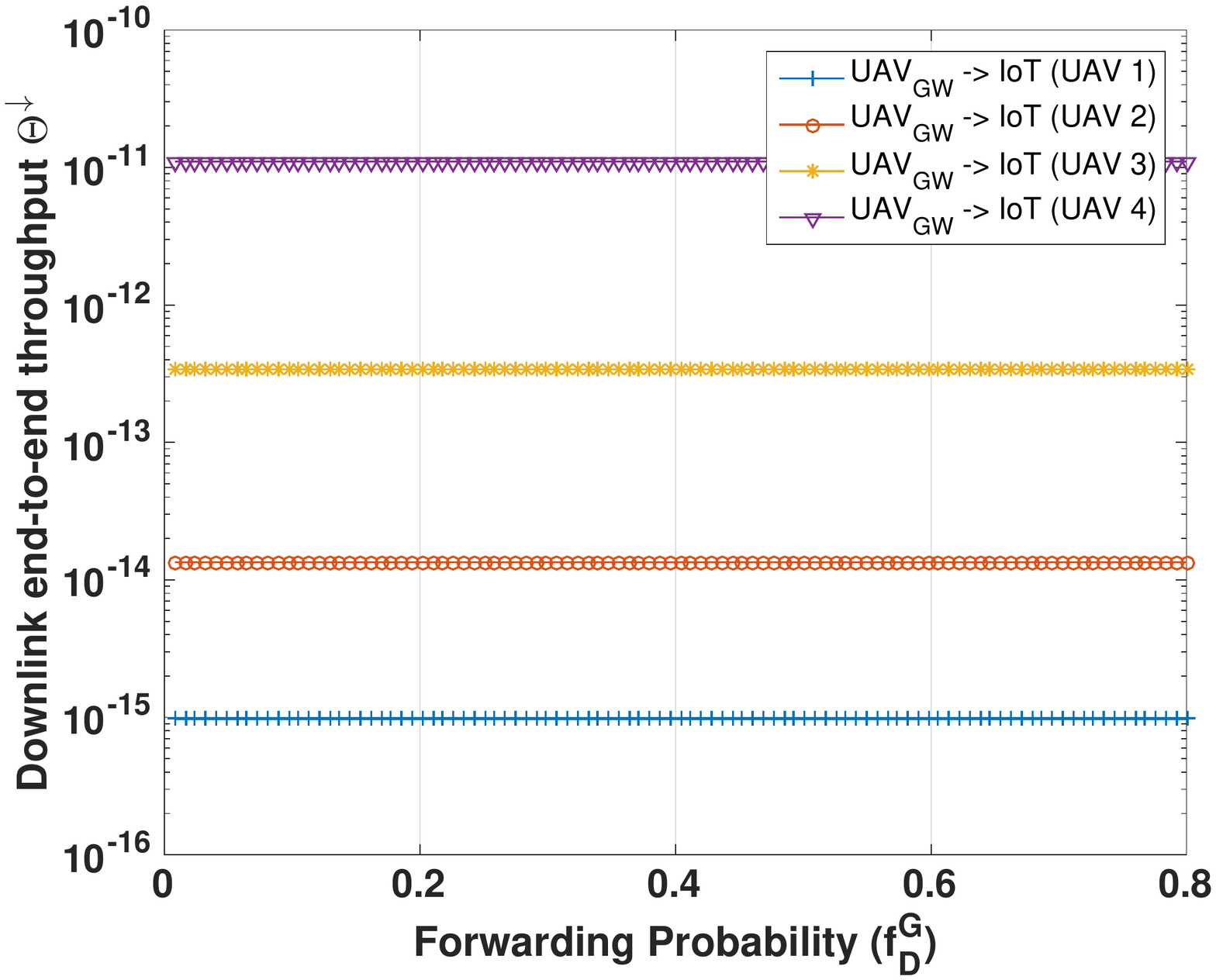}\label{ee}
}
\caption[Optional caption for list of figures 5-8]{Downlink queue as a function of the Forwarding probability}
\label{fig:subfigureExample3}
\end{figure*}
In figure.~$~ \ref{aa}$, we plot the load of uplink queue at each drone constituting the network against its forwarding probability $f^{A}_{U_{i}}$. We remark that drone far to the gateway is more relieved compared to the drone closer to the gateway, for example, the uplink queue of UAV $U$ is the busiest one because it accumulates the data gathered from its served IoT devices and the data acquired from the ground devices, that are covered by drone $1, 2, 3$, in order to forward them to Cloud system. Indeed, when $f^{A}_{U_{i}}$ takes a small value the queues remain unsaturated, and when it takes as a value higher than $0.4$, the queue becomes less relieved, in other words, through the greater likelihood of forwarding allow the stability of queue and then we can obtain the stability region.  This is because of the queue forward in a faster manner its packet toward the desired destination. We observe that the gateway is relatively empty because it doesn't cover any area, and it can transmit any time with success due to the proper channel between $GW$ and $c$.
\begin{figure}
 \centering
\subfloat[Load of downlink queue Vs Altitude]
{\includegraphics[trim= 0cm 5cm 0cm 5cm, width=0.25\textwidth, keepaspectratio]{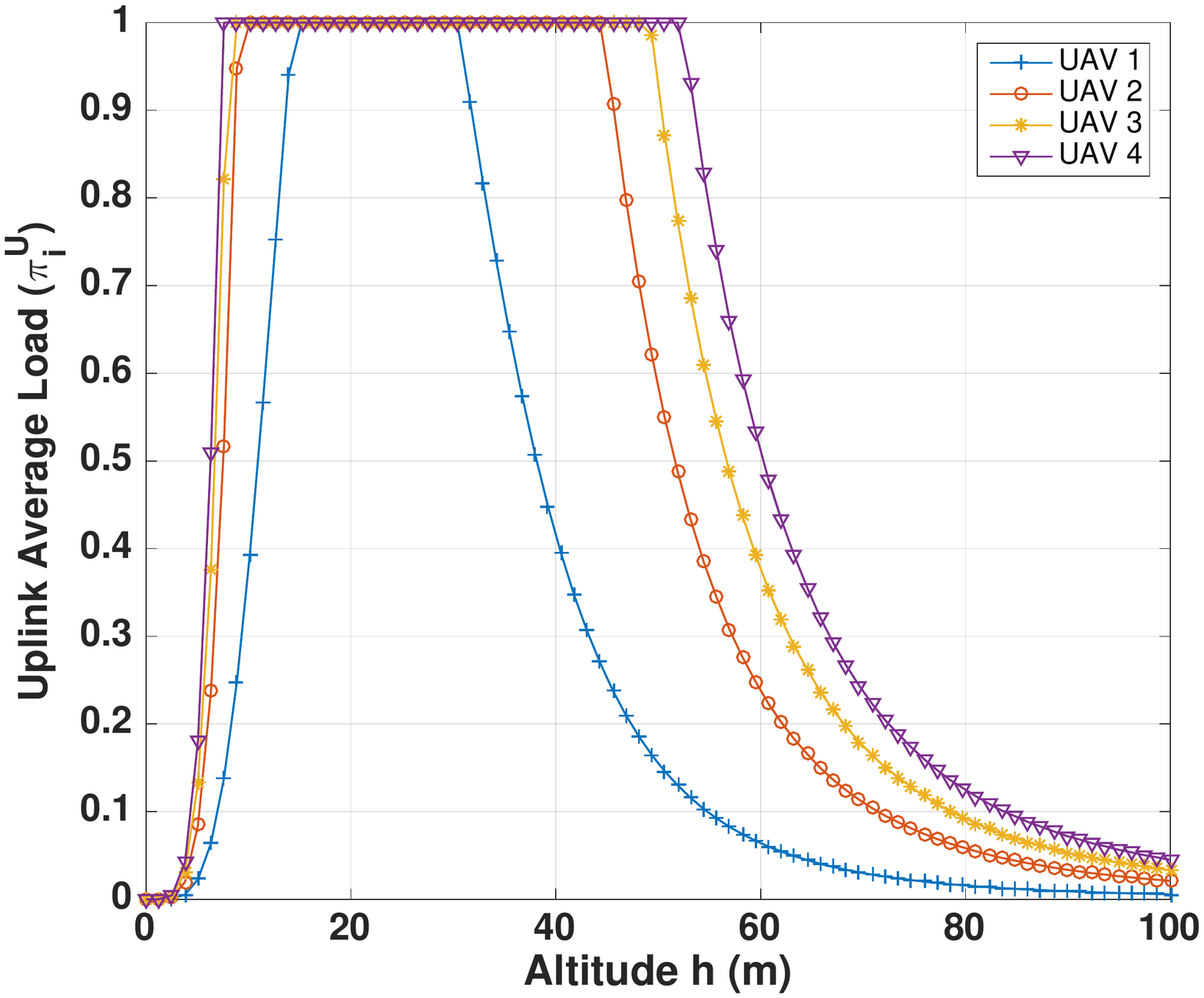}\label{ff}}
\subfloat[Load of downlink queue Vs Aperture Angle]{\includegraphics[trim= 0cm 5cm 0cm 7cm, width=0.25\textwidth, keepaspectratio]{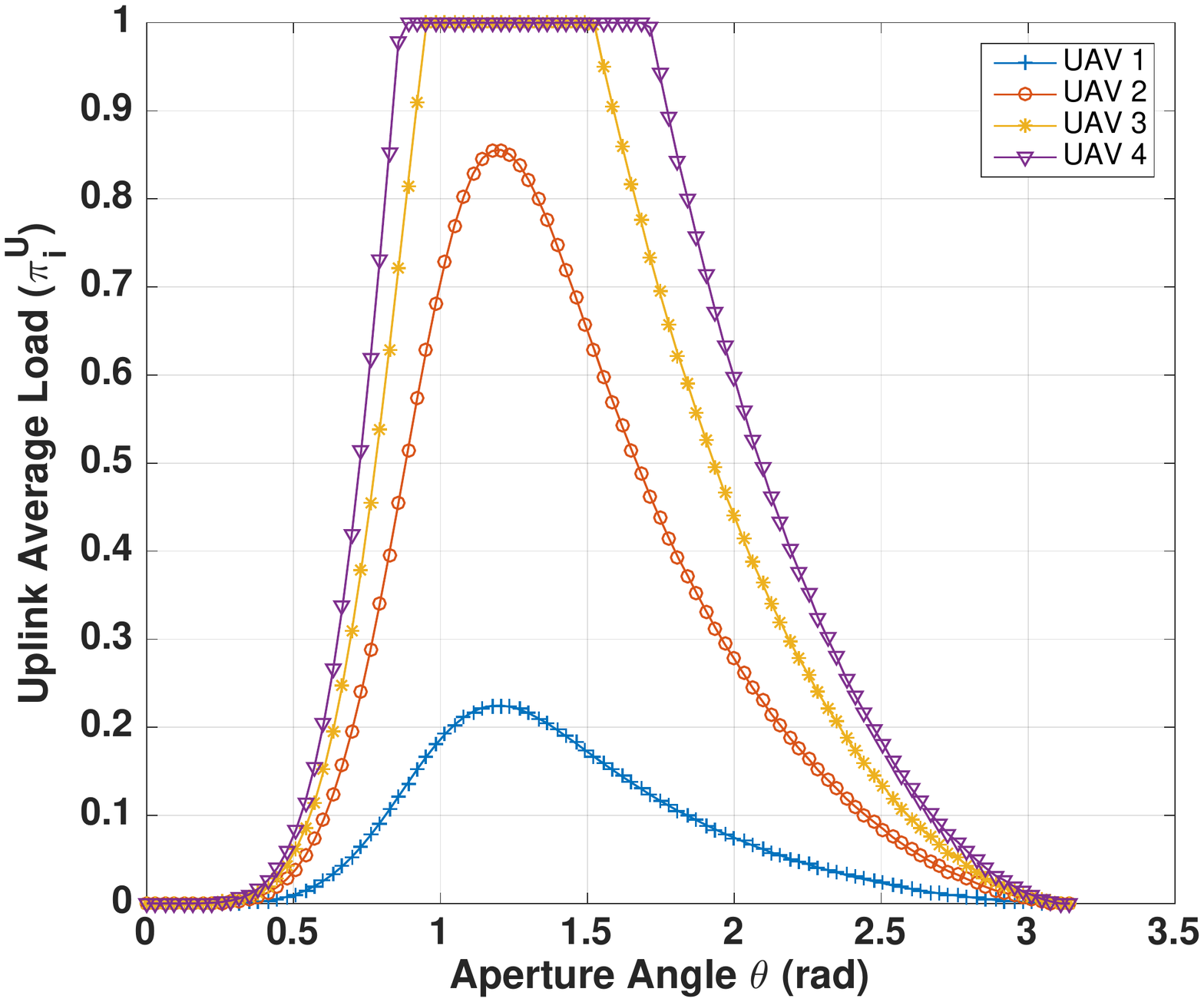}\label{gg}}\\
\subfloat[Load of uplink queue Vs Density]{\includegraphics[trim= 0cm 5cm 0cm 7cm, width=0.25\textwidth, keepaspectratio]{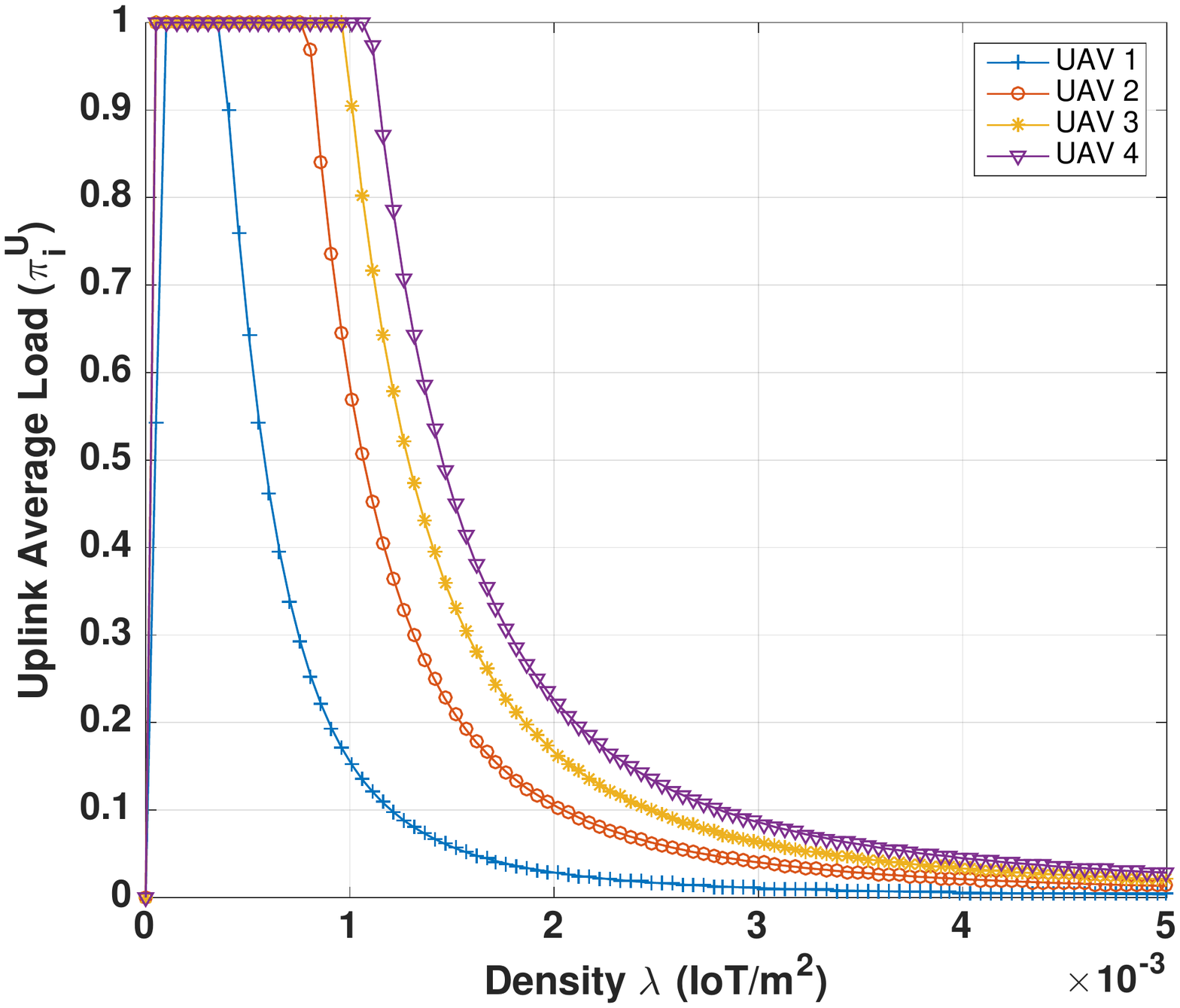}\label{hh}}
\subfloat[Load of uplink queue Vs Rotation Radius]{\includegraphics[trim= 0cm 5cm 0cm 7cm, width=0.25\textwidth, keepaspectratio]{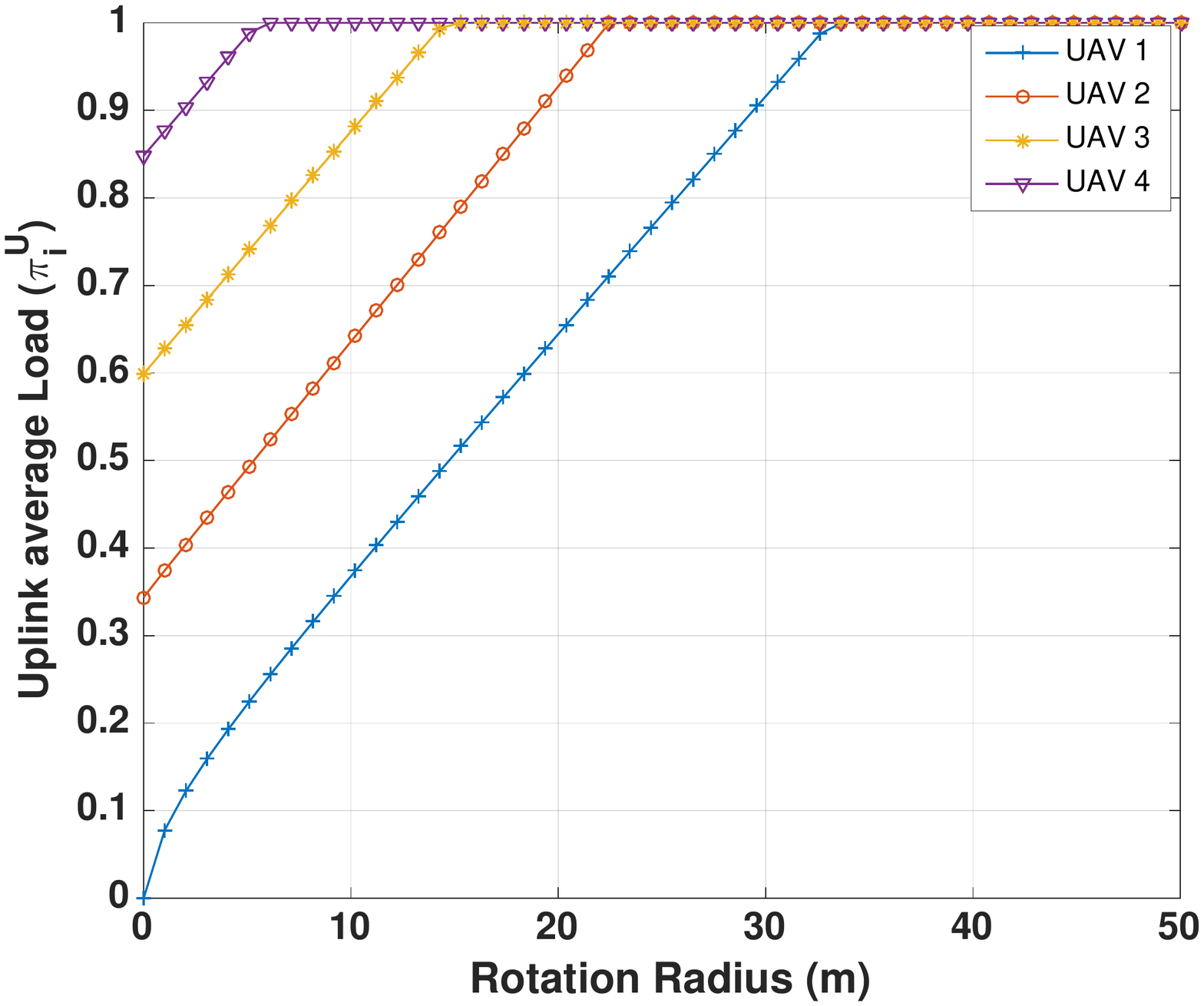}\label{ii}}\\
\begin{comment}
\subfloat[Load of uplink queue Vs Angular Velocity]{\includegraphics[trim= 0cm 5cm 0cm 7cm, width=0.25\textwidth, keepaspectratio]{Angularvelocity/LoadU.pdf}\label{jj}}
\end{comment}
\caption[Optional caption for list of figures 5-8]{Load of uplink Queue Vs UAV parameters: Altitude $h$, Aperture Angle $\theta$, Density $\lambda$ and Rotation Radius $r$ }
\label{kk}
\end{figure}
As shown in figure.~$~ \ref{bb}$, the throughput of uplink queue per connection versus its forwarding probability on Air-to-Air channel $f^{A}_{U}$ is plotted. The closest drone to the gateway containing the higher load because it carried packets of other previous drone neighbors, therefore it is considered as the busiest end-to-end throughput link, and moving away from $GW$, the average end-to-end throughput between a  UAV $i$   and $GW$ decrease. Indeed, the average throughput depends on forwarding probability, it increases with  $f^{A}_{U}$ until reaching its stability around $f^{A}_{U}= 0.4$

%%%%%%%%%%%%%%%%%%%%%%%%%%%%%%%%%%%%%%%%%%%%%%%%%%%%%%%%%%%%%%%%%%%%%%%%%%%%%%%%%%%%%%%%%%%%%%%%%%%

%that the system can carry in stable conditions.

The figure.~$~ \ref{cc}$, $~ \ref{dd}$ show the downlink queue load versus the forwarding probability either on Air-to-Air channel $f^{A}_{D}$ or Air-to-Ground channel $f^{G}_{D}$. The figures illustrate practically the same results. The load decreases when its forwarding probability $f^{A}_{D}$ and $f^{G}_{D}$ increase.  Is apparent that the downlink queues are almost stable, this is because it carried a small number of packets: A percentage of packets delivered in the Uplink, which are considered as an acknowledgment receipt, also the control messages which are sent periodically by the Cloud system. The downlink queue load of UAV $1$ remains constant regardless the value of $f^{A}_{D}$, this is due to its centralized position in the UAV network, which prevents it from sending any packet on the Air-to-Air channel. We observe also in figure.~$~ \ref{dd}$ that the downlink queue load of UAV $GW$ still steady for different values of $f^{G}_{D}$, which is fully justified by the fact that the gateway doesn't cover any ground devices, it acts just as a relay between the UAV network and the cloud system. Therefore, it doesn't send any packet on the Air-to-Ground channel.Next, we plot the average end-to-end throughput for each connection form the gateway to the IoT devices served by a UAV $i$ versus its forwarding probability $f^{G}_{D}$ on Air-to-Ground channel. %We 
We remark that this metric $\Theta^{\downarrow}_{s,d}$ still constant for any value of $f_{G}^{A}$, as shown in$~ \ref{ee}$.
\begin{figure}
\centering     %%% not \center
\subfloat[End-to End throughput of uplink queue Vs Altitude]{\includegraphics[trim= 0cm 5cm 0cm 7cm, width=0.25\textwidth]{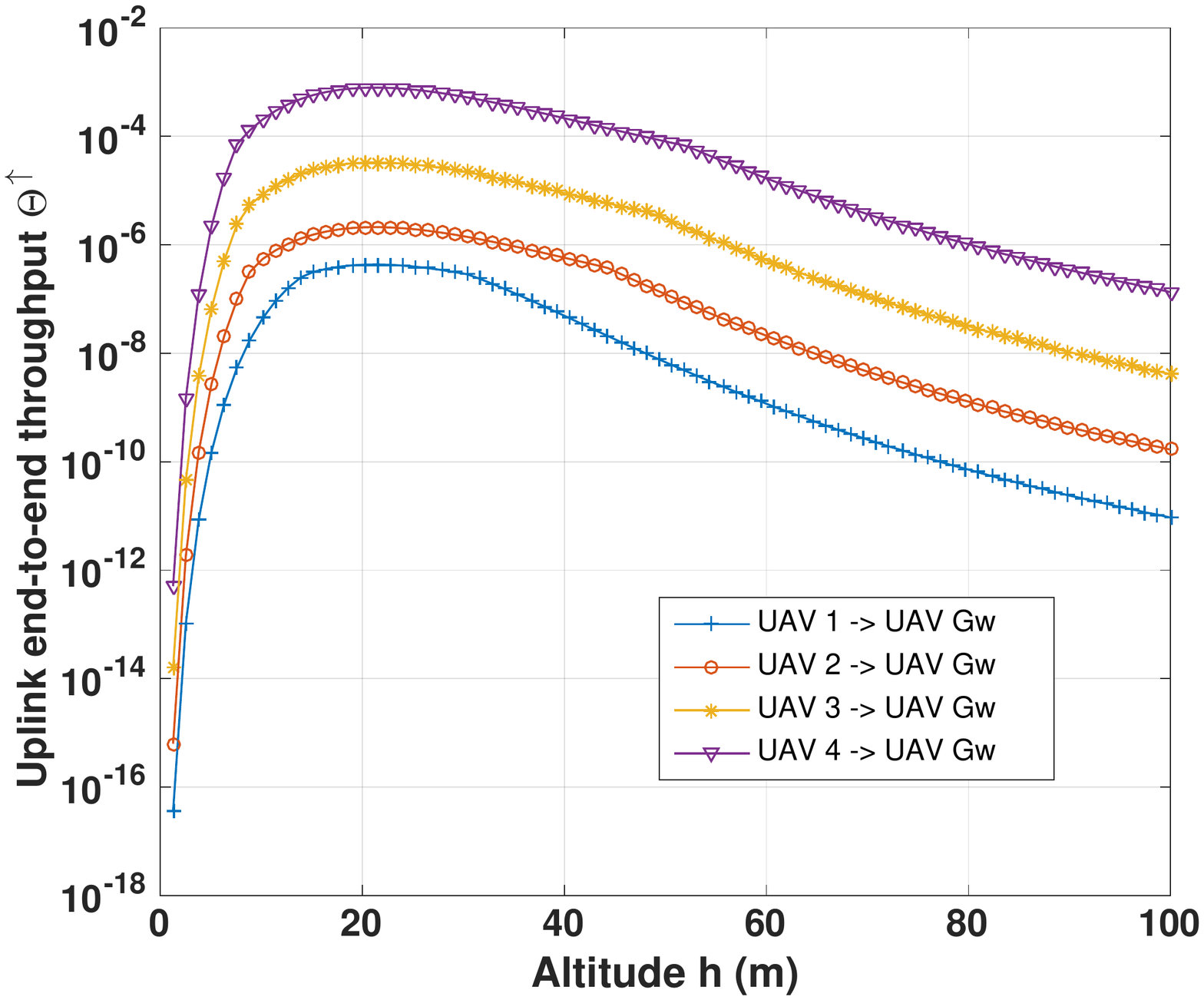} \label{ll}}
\subfloat[End-to End throughput of uplink queue Vs Aperture Angle]{\includegraphics[trim= 0cm 5cm 0cm 7cm, width=0.25\textwidth, keepaspectratio]{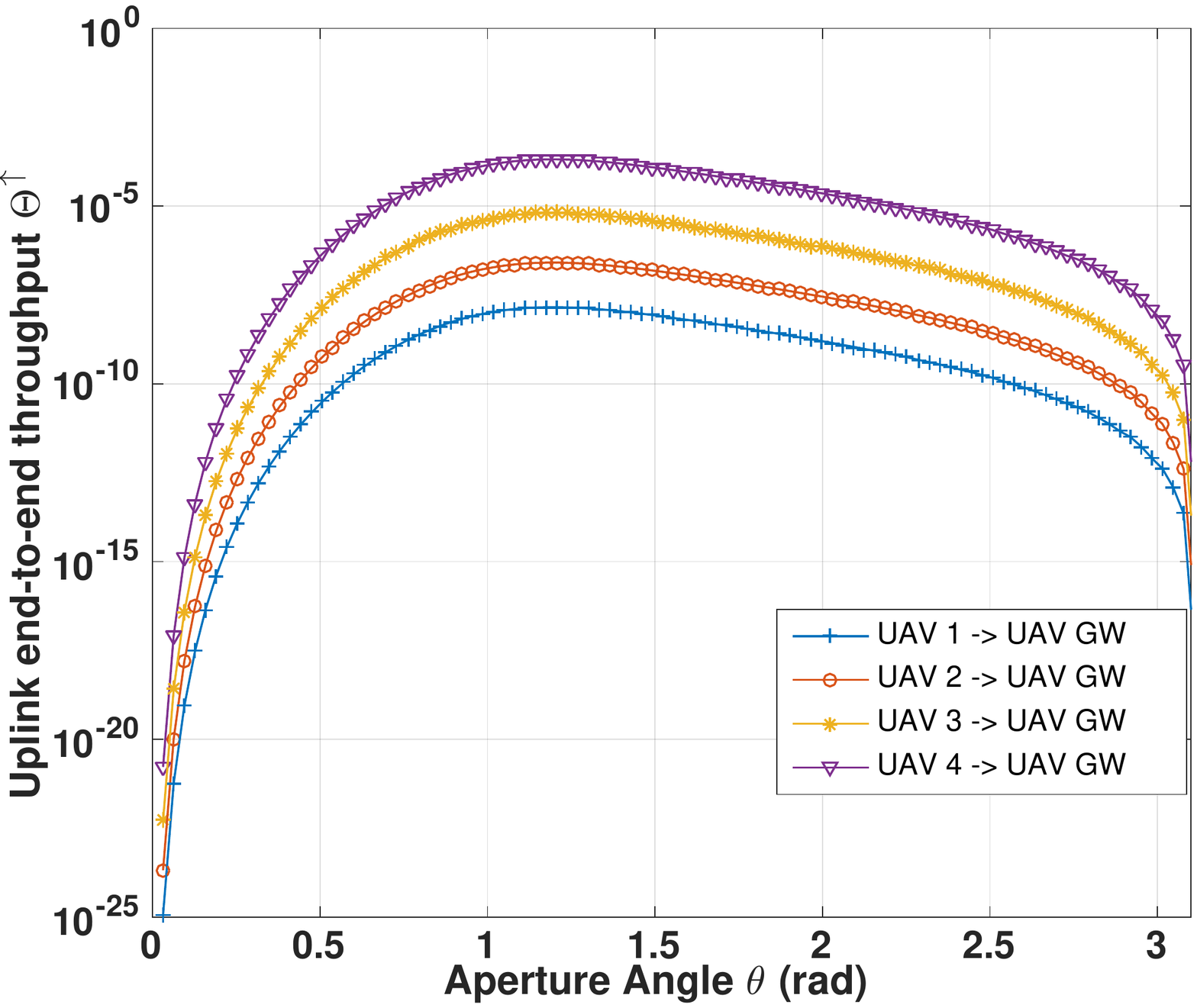}\label{mm}}\\
\subfloat[End-to End throughput of uplink queue Vs Density]{\includegraphics[trim= 0cm 5cm 0cm 7cm, width=0.25\textwidth, keepaspectratio]{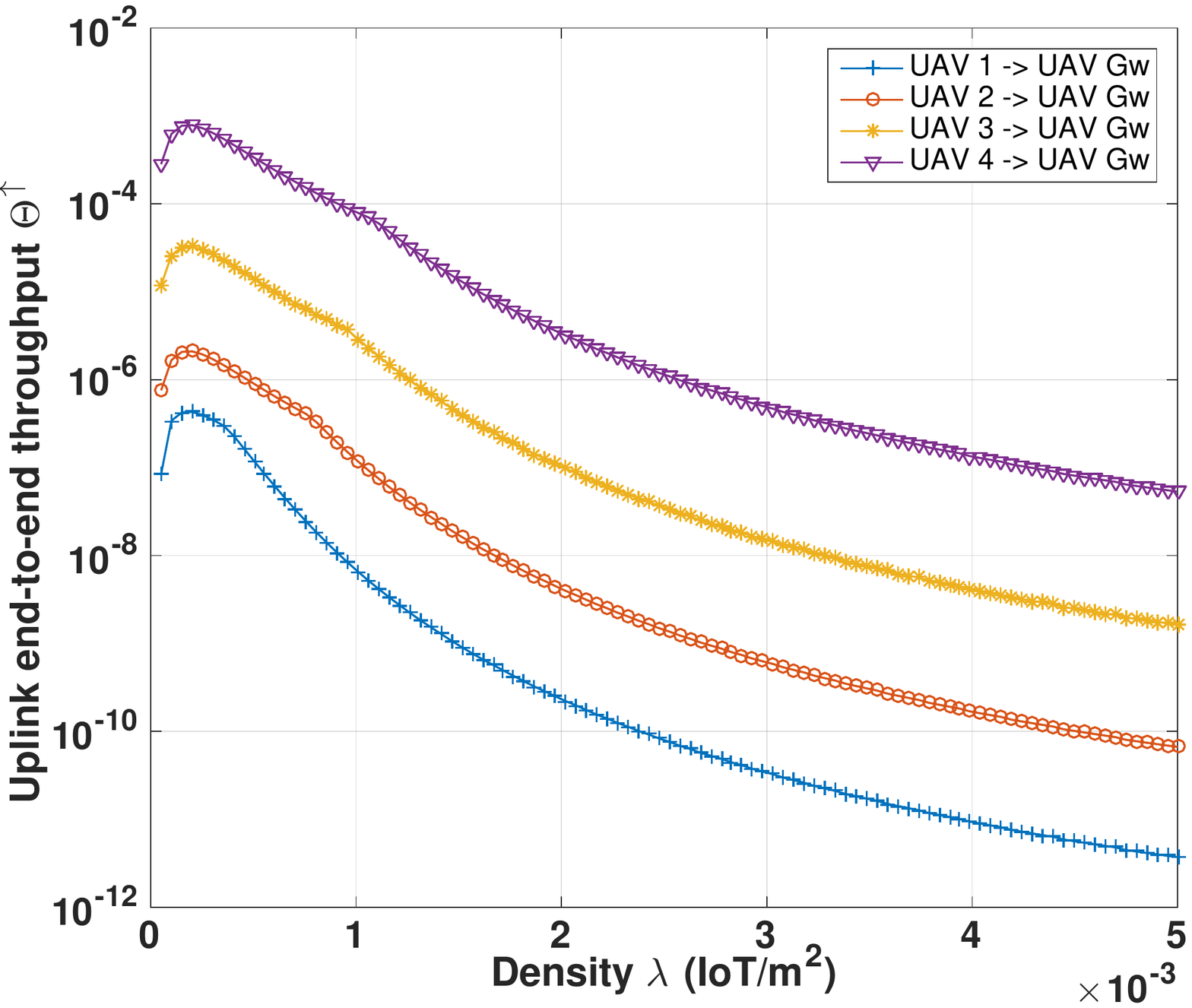}\label{nn}}
%\subfloat[Average End-to-End Throughput  Vs Rotation radius]{\includegraphics[trim= 0cm 5cm 0cm 5cm, width=0.33\textwidth, keepaspectratio]{Radius/ThpU.pdf}\label{pp}}
\subfloat[End-to End throughput of uplink queue Vs Angular Velocity]{\includegraphics[trim= 0cm 5cm 0cm 7cm, width=0.25\textwidth, keepaspectratio]{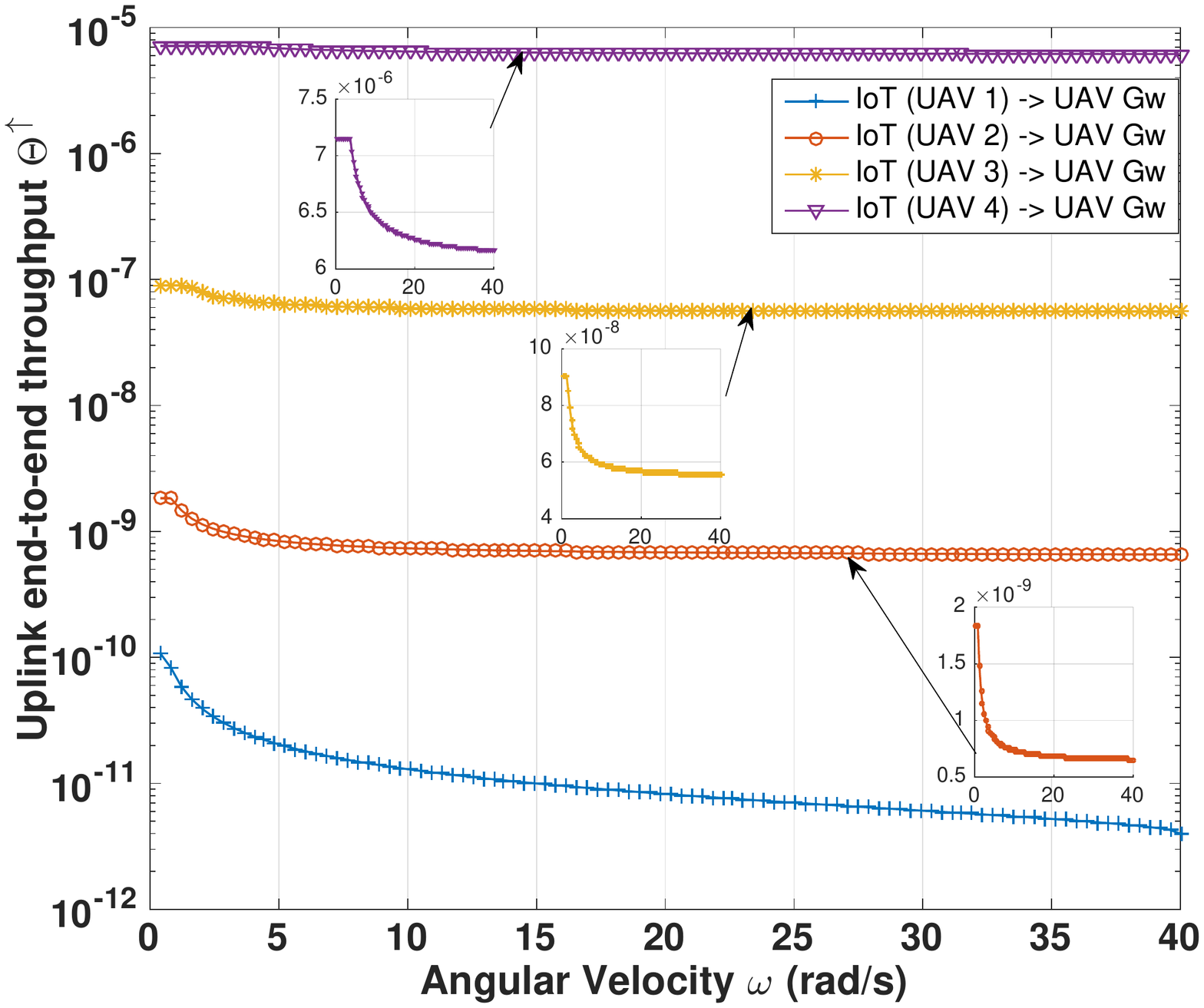}\label{oo}}
\caption{Average End-to-End throughput of uplink Queue Vs UAV parameters: Altitude $h$, Aperture Angle $\theta$, Density $\lambda$ and Angular Velocity $\omega$}
\label{qq}
\end{figure}
Therefore, the stability region is founded for any value of  $f^{G}_{D}$.
%The average end-to end throughput of each connection form gateway to the IoT served by an  UAV $i$  queue of each connection are increasing with $f^{A}_{D}$ until to be constant.

From the proposed analytic model, we plot the \textit{Uplink} queue load of several drones in the considered network as a function of five parameters: altitude $h$, Aperture Angle $\theta$, Density $\lambda$, Rotation radius $R$ and angular velocity $\omega$ as shown in figure.~$~ \ref{kk}$. We remark that this metric  $\pi_{i}^{U}$ either for $h$ or $\theta$ and $\lambda$  have the same behavior (see fig: $~ \ref{ff}$, $~ \ref{gg}$ and $~ \ref{hh}$ ). 
When $h$ and $\theta$ are low, the number of covered nodes is also low, resulting in an under-utilized queue. However, the load quickly increases with $h$, $\theta$ and $\lambda$ which might induce undesirable instability at queues. Beyond certain altitude, the mean number of covered nodes decrease due to poor coverage, and then the queues become less congested. In other words, traffic stability could be met by strategically locating UAVs and moving them closer or further away from the ground.

%The Uplink load of each drone increases as the altitude, apertures angle and density parameters increase. Figure $~ \ref{ff}$, $~ \ref{gg}$ and $~ \ref{hh}$ depicts the existence of a margin in which the load of uplink queue is maximized and then it decreases till getting annulled. This is because they control the coverage area. The low values allow more contending IoT devices to be covered,  which cause overloading of queues. Whereas, by increasing these parameters the queues become less congested. 

%when $5m<h_{m}<55m $ , $\frac{\pi}{3}<\theta_{m}<\frac{\pi}{2}$ and  $10^{-3}   IoT/m^{2}<\lambda_{m}<2.10^{-3} IoT/m^{2}$
The effect of rotation radius of uplink load queue is shown in $~ \ref{ii}$. When the drone flies with a low speed, the rotation radius increase dramatically $(R=\frac{V}{\omega})$, then the uplink queue load tends to be maximized ($\pi^{U}=1$). As the velocity of UAV increases, the $R$ decrease, the uplink load then encounters a sharp decrease. Otherwise, the queues are congested and unstable as $\omega$ takes a small value, $R$ then increases and the queues become more relieved when $\omega$ increase and $R$ decrease in turn. Explained by the fact that the higher velocity of UAV allows increasing the contact time between drones, this means that drone has enough time to forward their data and relieves it uplink queue. However, flying with high speed leads to the highest energy consumption of UAVs.

%The stability region is obtained when $R$ is too small. The load tends to be maximized ($\pi^{U}=1$) as $R$ tends to infinity, explained by the fact that the rotation manages the coverage ground area. If $R$ increases, the coverage area also widen and the number of IoT devices served by a UAV increase, what make unstable queues.   Hence, the rotation radius has to be small. 

%The average load versus angular velocity $\omega$ is depicted in figure.~$~ \ref{jj}$ ....................

%-----> \textit{Throughput of U queue Vs Altitude}\\
Next we depict the average end-to-end throughput per connection for different values of Altitude $h$, Aperture Angle $\theta$, Density $\lambda$ and Angular velocity $\omega$  (see figure.~$~ \ref{qq}$). The average end-to-end throughput per connection has likely the same behavior. It appears that this metric reaction is similar to the load as shown in figure.~$~ \ref{kk}$. As we
see in figure.~$~ \ref{ll}$,$~ \ref{mm}$ and$~ \ref{nn}$, for the low values of these three parameters, the throughput tends to increase until reach a peak that maximizes this metric and then decreases slightly. This corresponds to an increase of the numbers of IoT devices covered by each drone at these ranges. 
In figure.~$~ \ref{oo}$, the uplink end-to-end throughput achieved between a UAV $i$ and the gateway versus Angular velocity $\omega$ is plotted. It is observed that the throughput decreases dramatically till $\omega= 5 \, rad/s$ and it keep almost fixed while the angular velocity increases. This decline of uplink e2e throughput is justified by load relieving of the uplink queue, which in turn is due to an increase of the contact times between UAV.
\begin{figure}[ht]
\centering     %%% not \center
\subfloat[Load of downlink queue Vs Altitude]{\includegraphics[trim= 0cm 5cm 0cm 7cm, width=0.25\textwidth,keepaspectratio]{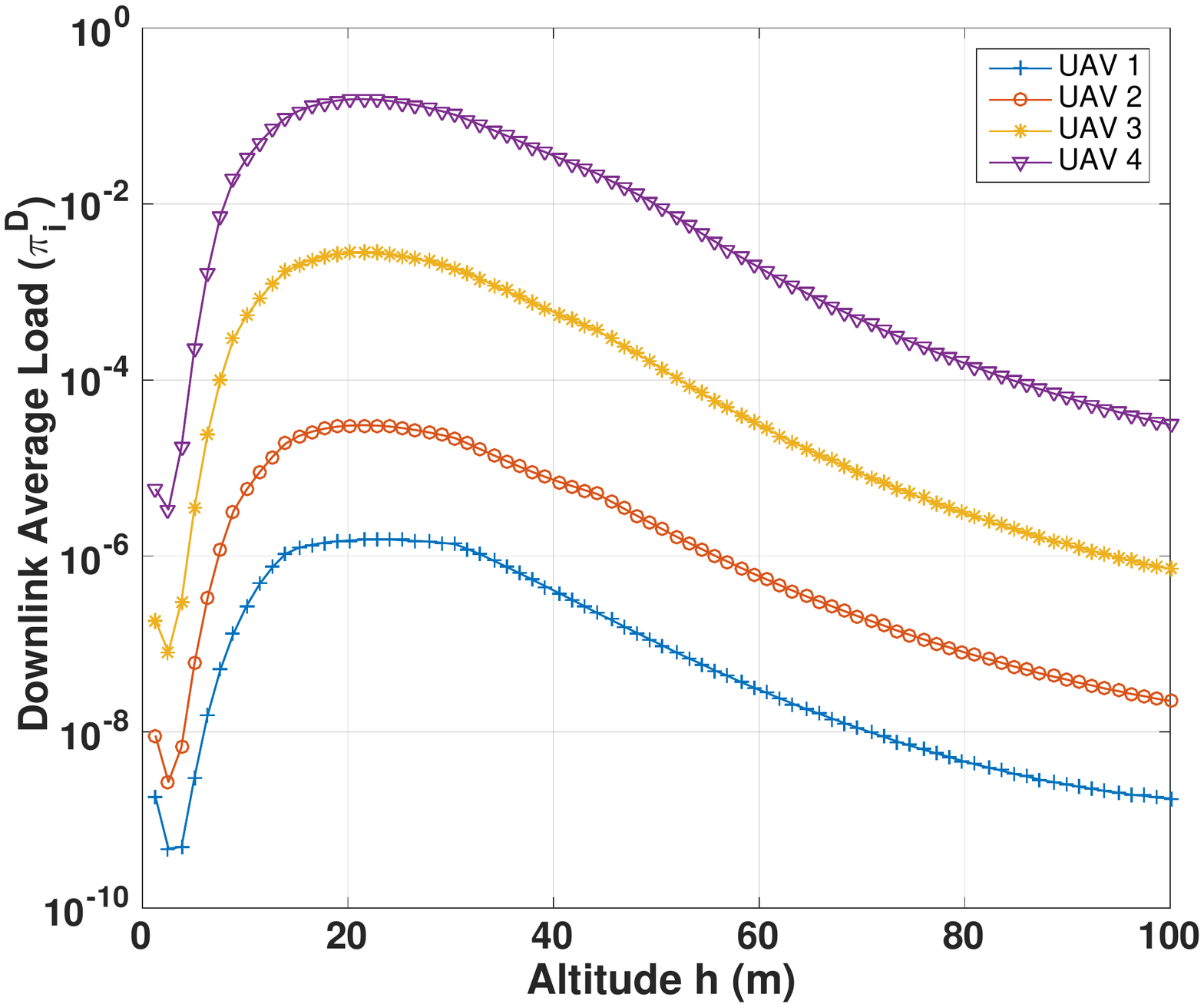}\label{rr}}
\subfloat[Load of downlink queue Vs Aperture Angle]{\includegraphics[trim= 0cm 5cm 0cm 7cm, width=0.25\textwidth, keepaspectratio]{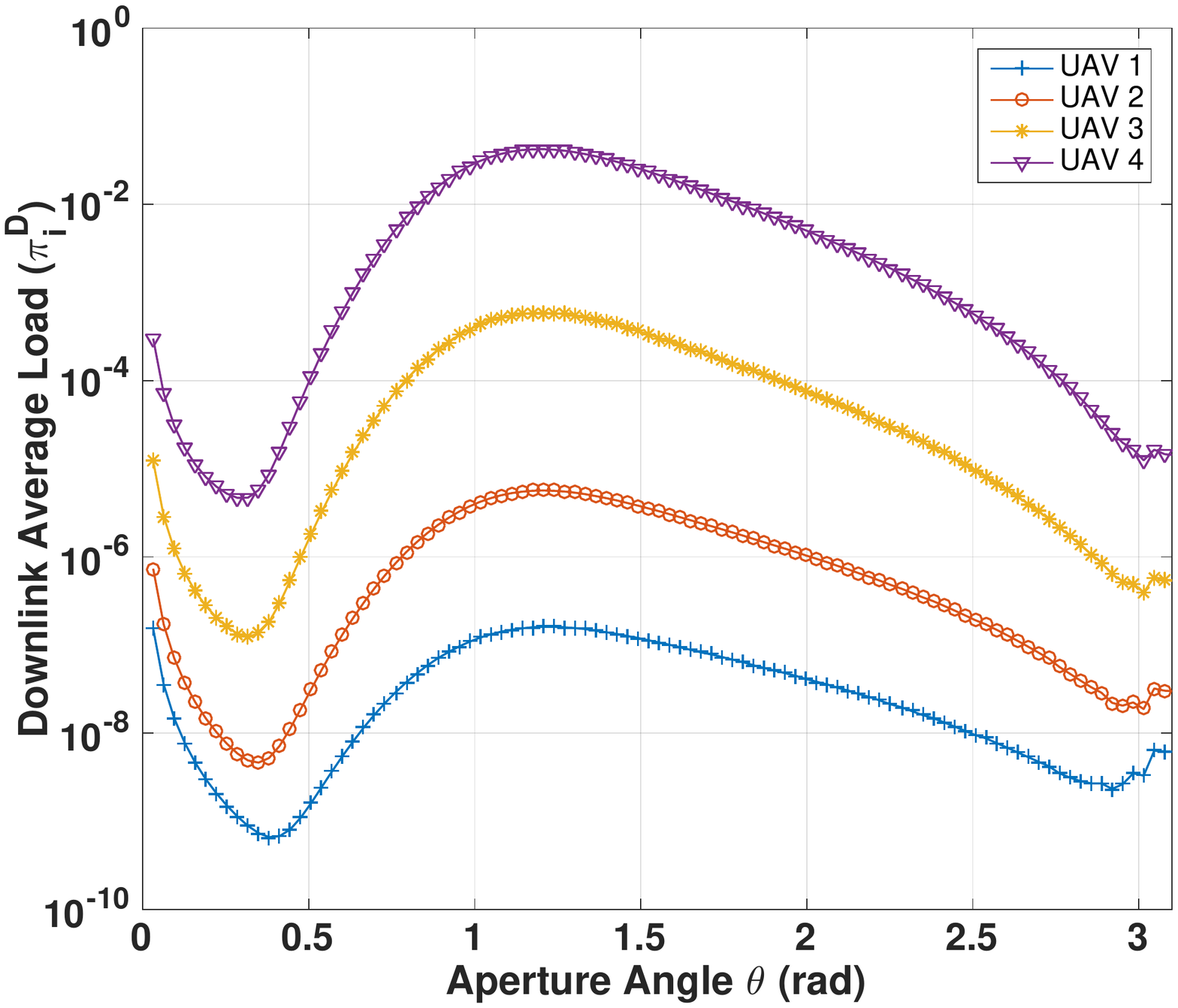}\label{ss}}\\
\subfloat[Load of downlink queue Vs Density]{\includegraphics[trim= 0cm 5cm 0cm 7cm, width=0.25\textwidth, keepaspectratio]{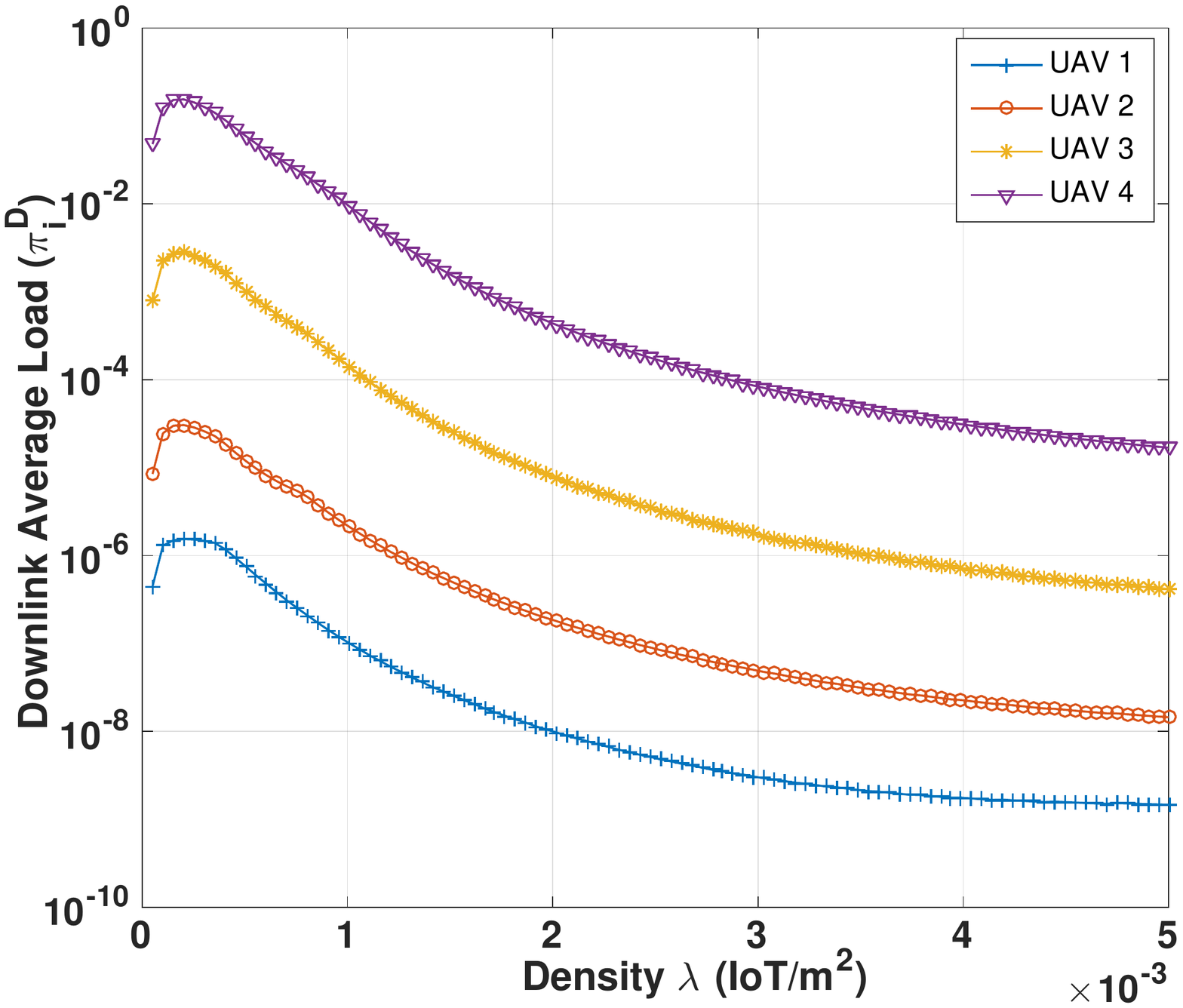}\label{tt}}
%\subfloat[Load of downlink queue Vs Rotation radius]{\includegraphics[trim= 0cm 5cm 0cm 5cm, width=0.3\textwidth, keepaspectratio]{Radius/LoadD.pdf}\label{uu}}
\subfloat[Load of downlink queue Vs Angular Velocity]{\includegraphics[trim= 0cm 5cm 0cm 5cm, width=0.25\textwidth, keepaspectratio]{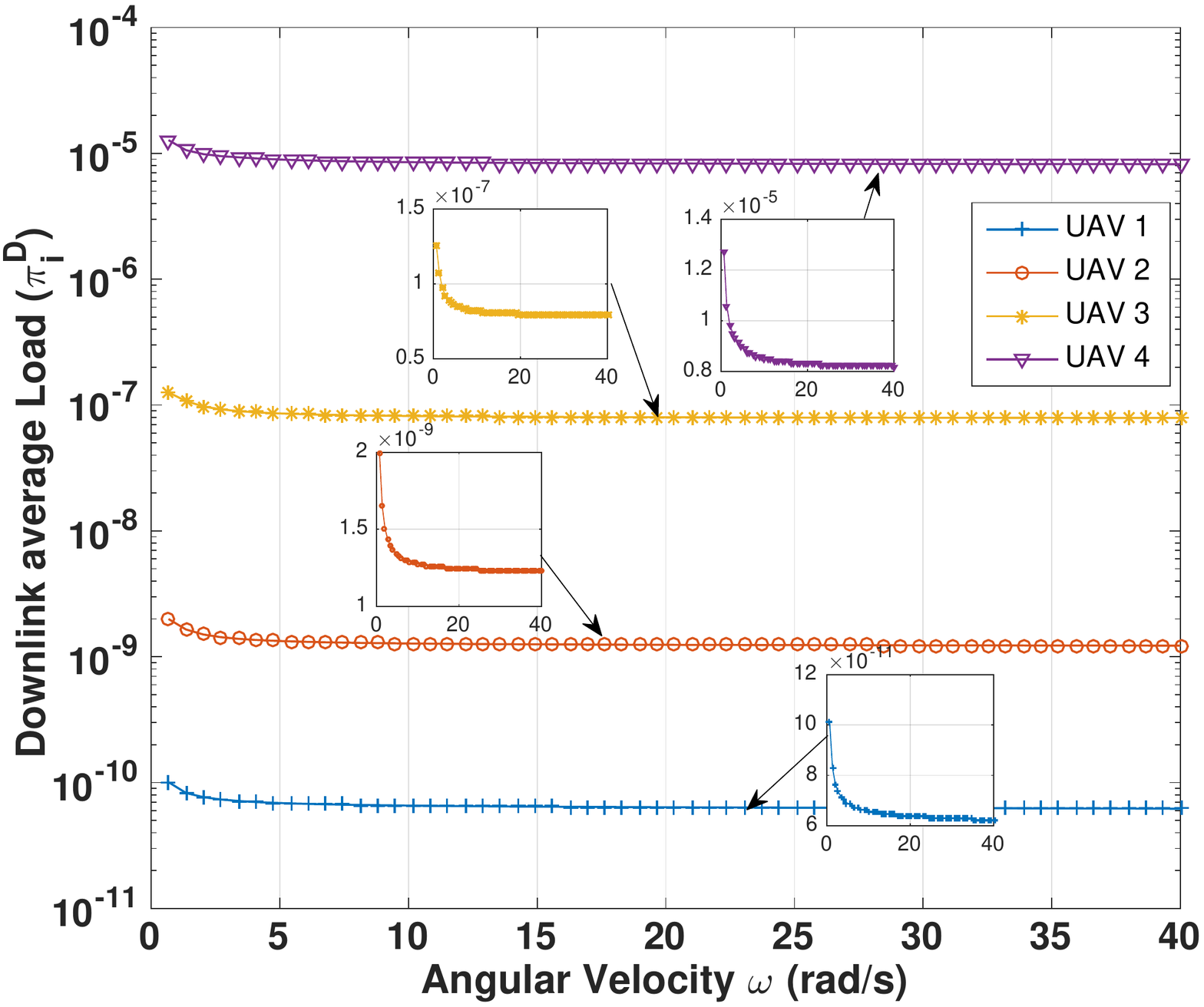}\label{vv}}
\caption{Load of downlink Queue Vs UAV parameters: Altitude $h$, Aperture Angle $\theta$, Density $\lambda$ 
and Angular Velocity $\omega$}
\label{ww}
\end{figure}
We turn now to plot  the downlink queue load as a function of altitude $h$, Aperture Angle $\theta$, Density $\lambda$ and Angular velocity $\omega$  (see figure.~$~\ref{ww}$). It is shown that the downlink load is below $1$, we can say that this queue remains stable regardless of the value of these parameters. This is expected since the fact that it carries a percentage of uplink load queue and a control messages which can be considered as small amount data compared to uplink queue. 
\begin{figure}
\centering     %%% not \center
\subfloat[End-to End throughput of Dowlink queue Vs Altitude]{\includegraphics[trim= 0cm 5cm 0cm 7cm, width=0.25\textwidth, keepaspectratio]{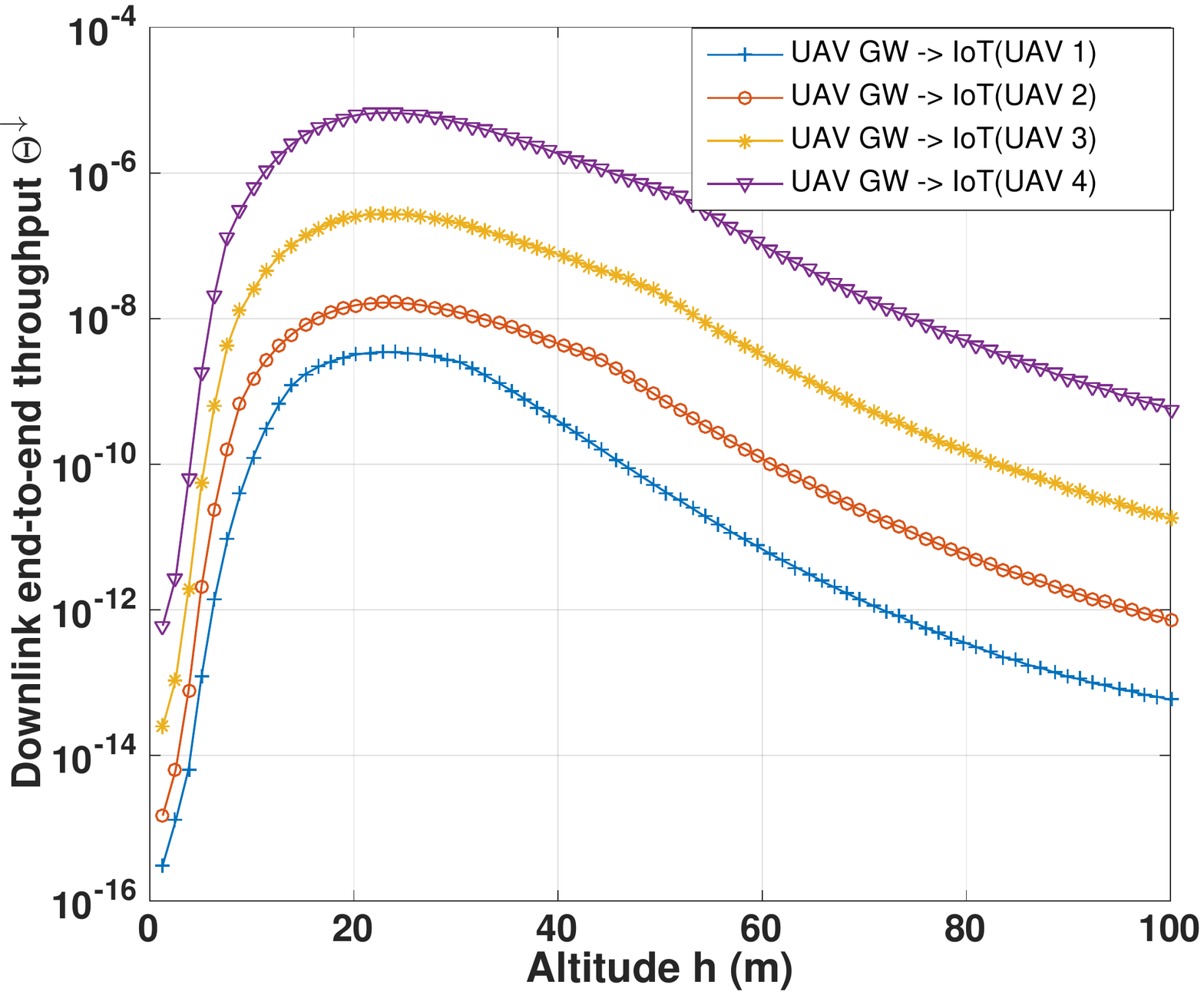}\label{xx}}
\subfloat[End-to End throughput of Dowlink queue Vs Aperture Angle]{\includegraphics[trim= 0cm 5cm 0cm 7cm, width=0.25\textwidth, keepaspectratio]{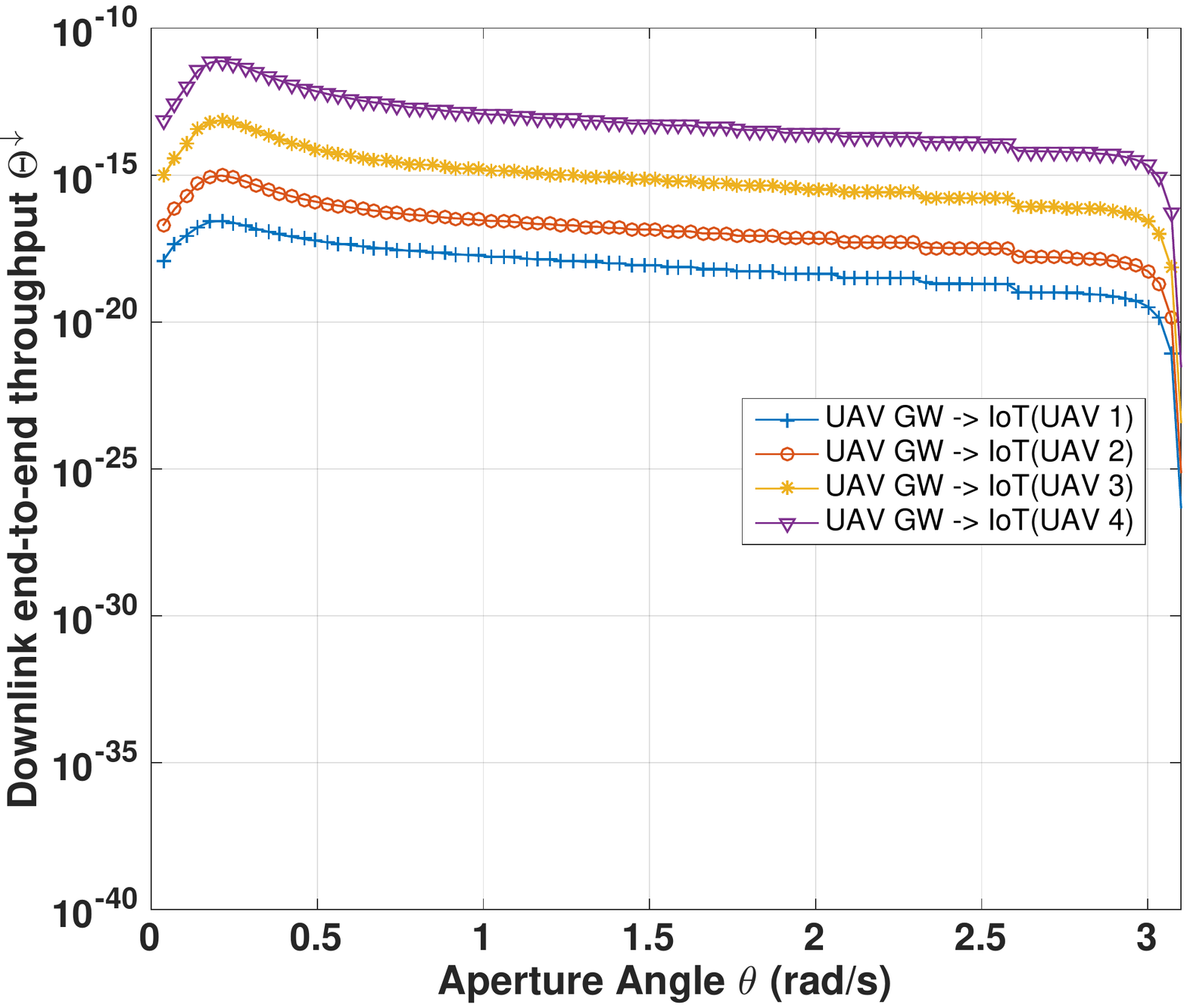}\label{yy}}\\

\subfloat[End-to End throughput of Dowlink queue Vs  Density]{\includegraphics[trim= 0cm 5cm 0cm 7cm, width=0.25\textwidth, keepaspectratio]{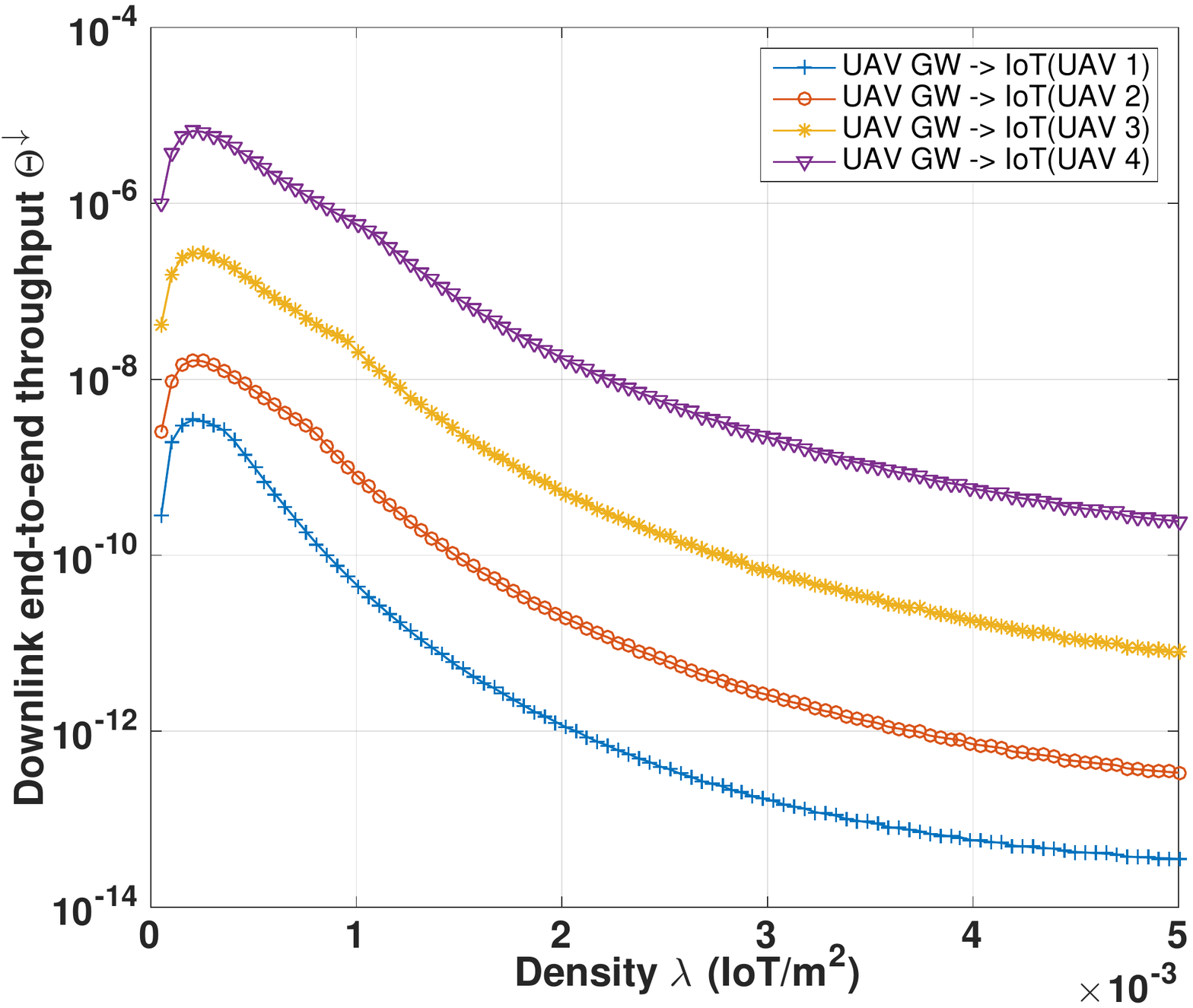}\label{zz}}
%\subfloat[$Thp^{D} $Vs Rotation radius]{\includegraphics[trim= 0cm 5cm 0cm 7cm, width=0.3\textwidth, keepaspectratio]{Radius/ThpD.pdf}\label{z1}}
%\subfloat[End-to End throughput of Dowlink queue Vs  Angular Velocity]{\includegraphics[trim= 0cm 5cm 0cm 7cm, width=0.25\textwidth, keepaspectratio]{Angularvelocity/ThpD.pdf}\label{z2}}
\caption{Average End-to-End throughput of downlink Queue Vs UAV parameters: Altitude $h$, Aperture Angle $\theta$ and Density $\lambda$ }
\label{z3}
\end{figure}

Note that similar behavior can also be observed for downlink normalized end-to-end throughput as $h, \theta, \lambda$ vary as shown in figure.\ref{z3}. The throughput per connection grows gradually to a maximum value till getting a maximum and decrease when the queue is more relieved as shown in figures. \ref{xx}, \ref{yy}, \ref{zz}. %In figure \ref{z2} the end-to-end throughput is high when the angular velocity is low and it decrease as the $\omega$ increases.
%From the figure.~$~ \ref{h}$ we can observe that the $D$ queue behavior is similar to the figure.~$~ \ref{f}$, except the intervals   that maximize the $D$ queue load, is small in comparison with that of $U$ queue. This will lead to widen the stability region. Also, the throughput of all connections is decreasing as these parametrs increase. 

%(Synthesis) The important point here is that .....

\subsection{End-to-End Delay}

We turn in this section to study the end-to-end delay when the fundamental parameters ($f^{M|A}_{U|D}, h, \theta, \lambda,\omega$) vary.

\begin{figure}
\centering     %%% not \center
\subfloat[End-to End delay of uplink queue Vs Forwarding probability ]{\includegraphics[trim= 0cm 5cm 0cm 9cm, width=0.25\textwidth, keepaspectratio]{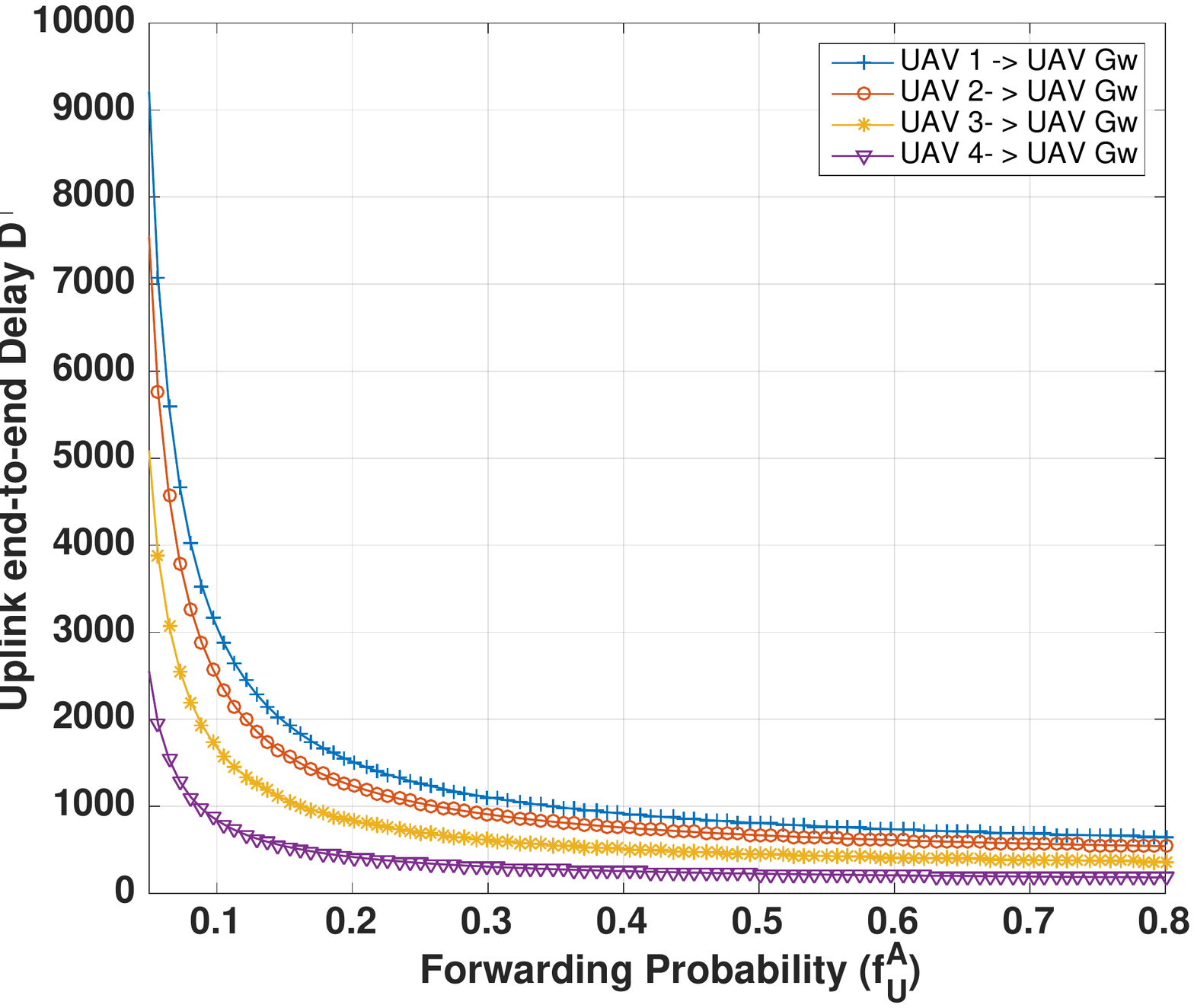}}
\subfloat[End-to End delay of uplink queue Vs Altitude]{\includegraphics[trim= 0cm 5cm 0cm 9cm, width=0.25\textwidth, keepaspectratio]{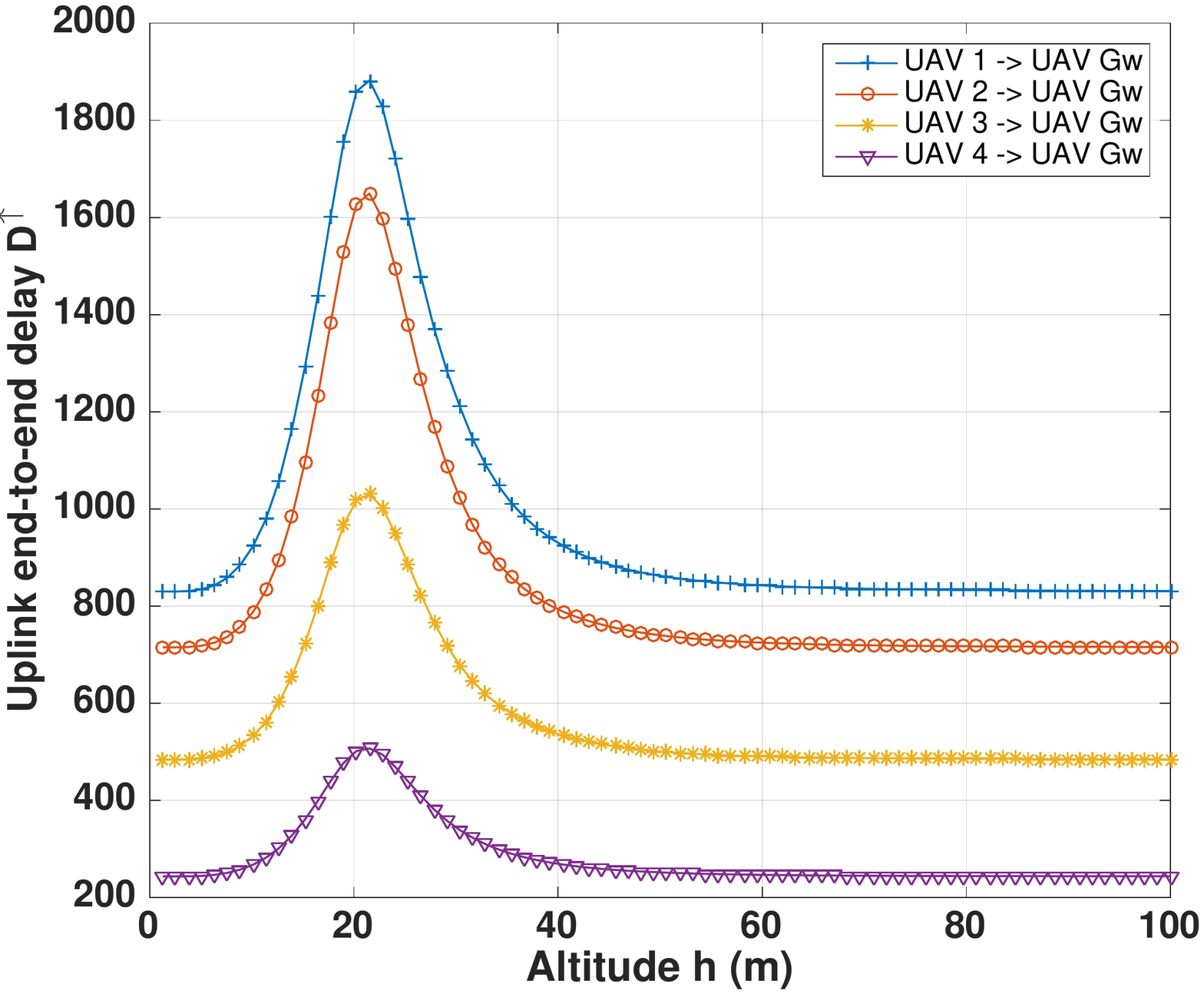}\label{A1}}\\
\subfloat[End-to End delay of uplink queue Vs Aperture Angle]{\includegraphics[trim= 0cm 5cm 0cm 9cm, width=0.25\textwidth, keepaspectratio]{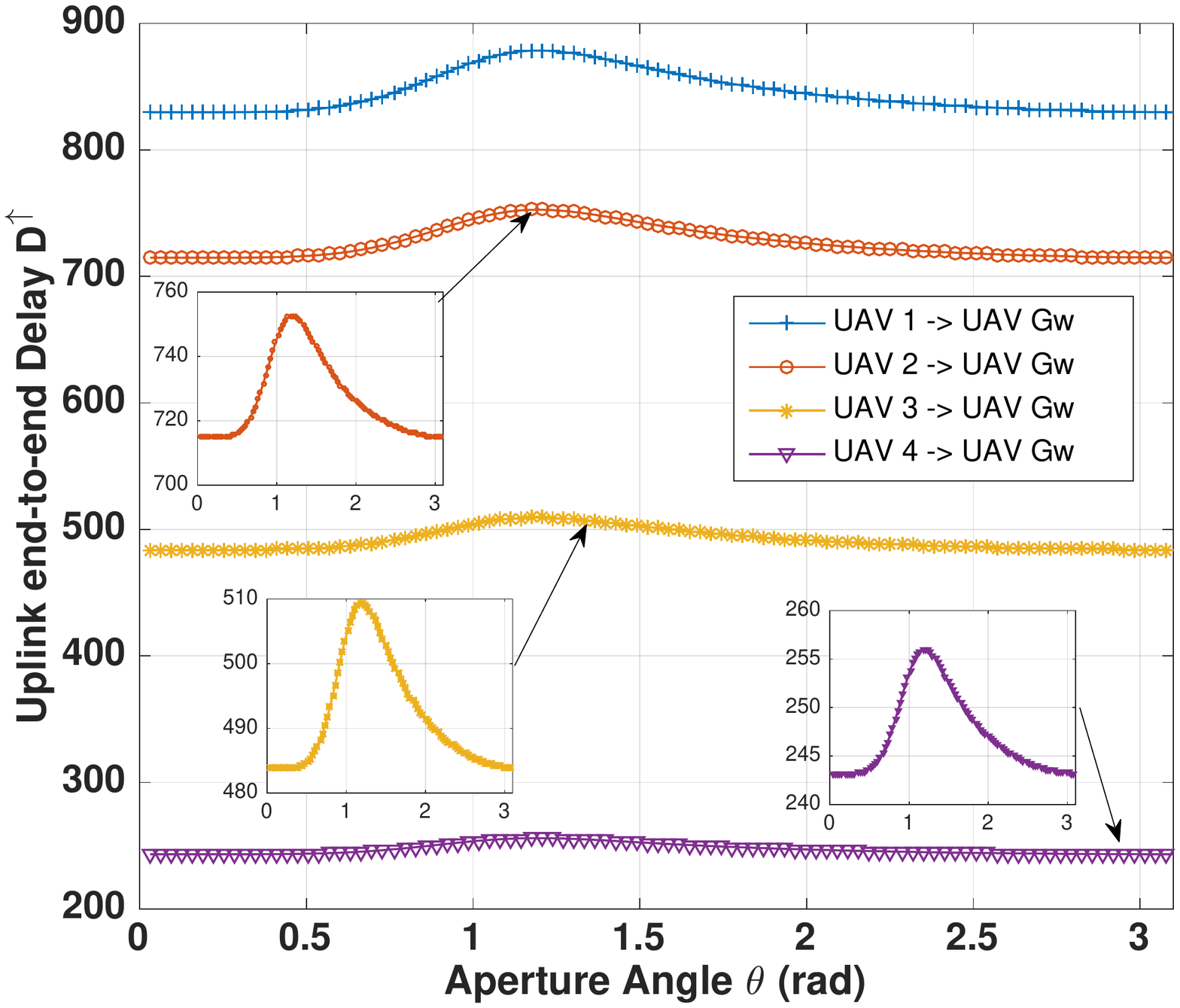}\label{A2}}
\subfloat[End-to End delay of uplink queue Vs Density]{\includegraphics[trim= 0cm 5cm 0cm 7cm, width=0.25\textwidth, keepaspectratio]{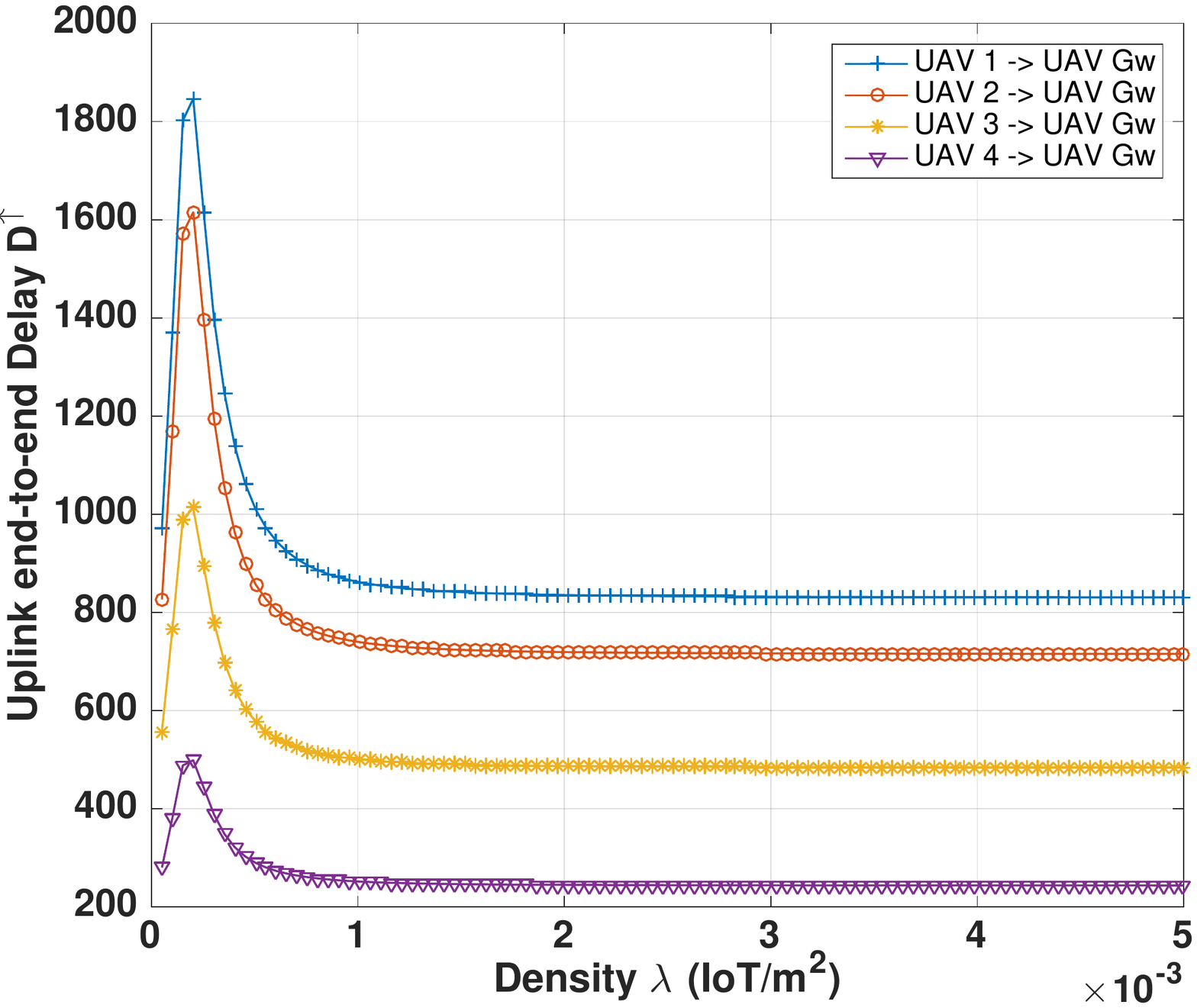}\label{A3}}\\
%\subfloat[$D^{\uparrow}$ Vs Rotation radius]{\includegraphics[trim= 0cm 5cm 0cm 5cm, width=0.3\textwidth, keepaspectratio]{Radius/e2eDU.pdf}\label{A4}}
\begin{comment}
\subfloat[End-to End delay of uplink queue Vs Angular Velocity]{\includegraphics[trim= 0cm 5cm 0cm 5cm, width=0.25\textwidth, keepaspectratio]{Angularvelocity/e2eDU.pdf}\label{A5}}
\end{comment}
\caption{End-to-End delay of uplink Queue Vs parameters: Forwarding probability $f^{A}_{U}$, Altitude $h$, Aperture Angle $\theta$ and Density $\lambda$. }
\label{22}
\end{figure}

In figure. \ref{22} the end-to-end delay of uplink queue is plotted. According to the figures. \ref{aa} and \ref{kk}, it is clear that when the uplink load is filling, it implies that UAV  suffer from huge delay, thus the end-to-end delay per connection is high, explaining that an arriving packet into queues cannot be forwarded immediately due to a busy MAC layer as well as the packets that must be sent before it. 

\begin{figure}
\centering     %%% not \center
\subfloat[End-to End delay of downlink queue Vs Forwarding probability $f^{G}_{D}$]{\includegraphics[trim= 1cm 5cm 0cm 7cm, width=0.25\textwidth, keepaspectratio]{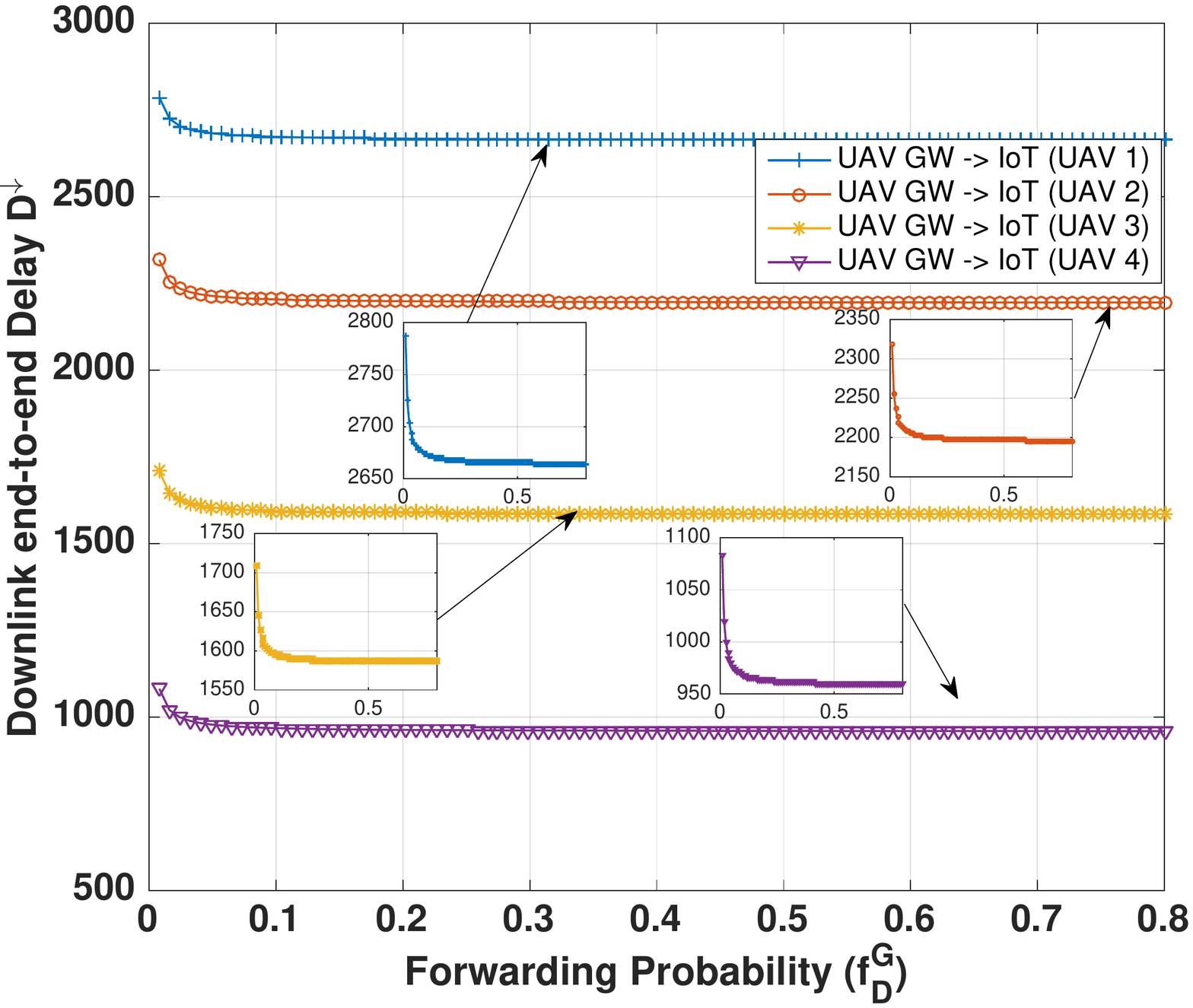}\label{A12}}
\subfloat[End-to End delay of downlink queue Vs Forwarding probability $f^{A}_{D}$]{\includegraphics[trim= 1cm 5cm 0cm 7cm, width=0.25\textwidth, keepaspectratio]{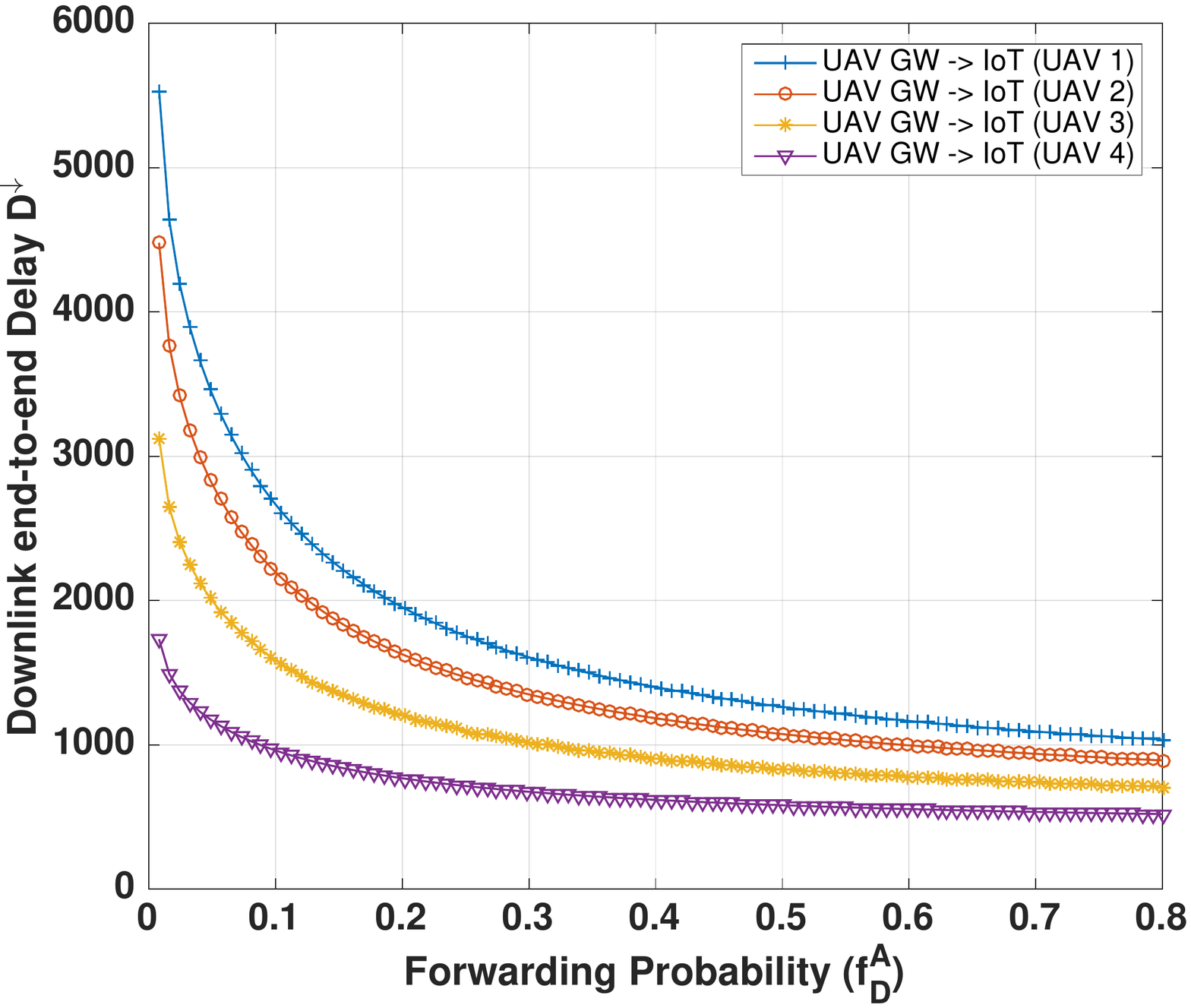}\label{A13}}\\
\begin{comment}
\subfloat[End-to End delay of downlink queue Vs Altitude]{\includegraphics[trim= 0cm 5cm 0cm 6cm, width=0.25\textwidth, keepaspectratio]{Altitude/e2eDD.pdf}\label{A6}}
\subfloat[End-to End delay of downlink queue Vs Aperture Angle]{\includegraphics[trim= 0cm 5cm 0cm 6cm, width=0.25\textwidth, keepaspectratio]{ApertureAngle/e2eDD.pdf}\label{A7}}\\
\subfloat[End-to End delay of downlink queue Vs Density]{\includegraphics[trim= 0cm 5cm 0cm 7cm, width=0.25\textwidth, keepaspectratio]{Density/e2eDD.pdf}\label{A8}}
%\subfloat[$D^{\downarrow}$ Vs Rotation radius]{\includegraphics[trim= 0cm 5cm 0cm 5cm, width=0.3\textwidth, keepaspectratio]{Radius/e2eDD.pdf}\label{A9}}\\
\subfloat[End-to End delay of downlink queue Vs Angular Velocity]{\includegraphics[trim= 0cm 5cm 0cm 5cm, width=0.25\textwidth, keepaspectratio]{Angularvelocity/e2eDD.pdf}\label{A10}}
\end{comment}
\caption{End-to-End delay of downlink Queue Vs Forwarding probability $f^{G}_{D}, f^{A}_{D}$.}
\label{A11}
\end{figure}

Unlike when the uplink queue is relieved the delay decrease in turn as these parameters increase. Indeed, the UAV far to the gateway has the highest delay compared to drone nearest to the gateway. Because packets in the farthest drones which desiring reaches a gateway as a destination should spend waiting time in all the drone along its path.  

The figure.  \ref{A11} shows the downlink end-to-end delay against $h$, $\theta$, $\lambda$, $\omega$ and the forwarding probability. The delay is high when the forwarding probability on Air-to-Air channel is very small and it decreases as $f^{A}_{D}$ takes large values (see figure. \ref{A13}). In figure.\ref{A12} the end-to-end delay smoothly decreases as $f^{G}_{D}$ increases per connection (according to figure. \ref{dd}), till getting fixed.  
%The figures. \ref{A6}, \ref{A7}, \ref{A8} and \ref{A10} are obtained while fixing $f^{A}_{D}$ at 0.1, which justified the greater value of delay. 
Therefore, match up with downlink load queue, when the queue is congested the delay is higher and as the queue becomes stable the delay decreases.
% Evidently, MMAC performs the worst since it induces high latencies and becomes unstable when the throughput is low. After this point, the throughput of MMAC starts to decrease while delay continues to grow. 

%Based on this observation, we conclude that to minimize access delay, G-McMAC should be used. Whereas, if it is important to maximize throughput at the expense of access delays, SYN-MAC should be exploited.

\section{Discussion and Insights}
\subsection{Full Cross-Layering}
The cross-layer design used in this paper shall exchange quality information across layers for better end-to-end performance that would satisfy a demanding application. Such an important information sustains a valuable synergy among OSI layers and permits to implement a dynamic and context-aware setting of control parameters. For example, using the interaction and the information from PHY concerning the channel state  (Noise, Interference level, etc.), the MAC layer can select the suitable value of $K$ or also can schedule the transmission in which the channel is better. As well as using the interaction between Application and physical layers, the PHY layer can adjust its transmission power and/or bit rate and/or modulation-coding scheme according to the application supported and its traffic features. Moreover, in the scenario when an arbitrary network topology is deployed, the protocols in the network layer could perform strategic routing decisions in order to improve the end-to-end performance (throughput, delay, jitter, latency, loss rate, etc.).

\subsection{Worst-Case Performance}
The proposed framework is conceived to understand the impact of the UAV parameters on the e2e throughput, e2e delay, as well as the traffic stability. Moreover, extensive simulation runs provide insights and some hints on how to set the altitude, and the aperture angle to meet the specific QoS level imposed by an application. We point out that the altitude, aperture angle and density are correlated parameters, as the UAV altitude  and the aperture angle (pointing toward the ground) increases, the number of covered ground IoT devices increases also. However, ground nodes cannot be covered for large values of $h$ and $\theta$ due to a limited transmit power and coverage probability between UAV and target ground. Therefore, the simulation exhibits the interval of $h$ and $\theta$ where covered nodes are efficiently served. Otherwise, on the standpoint of offered traffic generated, UAVs are deployed to gather high traffic demand, when all the ground IoT devices always have data to transmit (saturated nodes), which, in turn, provides an upper-bound in terms of traffic and a lower-bound in regards to e2e performance. For sure, the general case of unsaturated nodes will experience better performance. \\
From another perspective, our framework assumes a safe UAV design. Yet, our framework ensures a safe  UAV  flight through the pre-planned semi-static topology, which avoids the collision between drones due to the respected separation distance between them.

\begin{comment}
\subsection{Throughput-Delay trade-off}
You need to talk here about delay-sensitive applications, and how one could meet an efficient trade-off between delay and throughput.
\end{comment}

\subsection{Network Design}
%Planning, dimensioning, battery lifetime, net topology, etc.\\e

As shown above, our study offers an interaction overview of parameters that manages Flying Mesh UAV Network. This study or more precisely a network design is helpful to a network service provider (NSP). It involves studying, evaluating, planning and dimensioning the FMN before deploying. Through the obtained results, an WSP can chooses the optimal parameters (such as altitude $h$, Aperture Angle $\theta$ and Density $\lambda$ and many other parameters) to configure the network according to a specific application service (Video, e-Mail, Online Game, File sharing....), while depending on the desired throughput and delay.
For the desired application, the decision of which values of UAV parameters should be taken to ensure the stability region, as well as the throughput and delay, is our aim. For instance, suppose that the required throughput value to improve the performance of an application is $\Theta_{s,d}^{\uparrow}= 1.1 \, Mbit/s$ (Taking the channel capacity is equal to $11 Mbit/s$), from the graphs, we can see that to achieve this value while ensuring the network stability, it is desirable to take the values correspond to $10^{-1}$ in the end-to-end throughput axis: $f^{A}_{U}= 0.4 , \, h> 18 m, \theta=\pi/4, \, \omega > 5 rad/s  $. Thus, we can choose the values of UAV parameters that ensure a certain value of downlink queue throughput regardless of the stability region, because this queue is always stable.

Indeed, to guarantee the QoS, an average delay between $150 ms - 400 ms$ is generally acceptable for a specific application. According to the obtained graphs (see  \ref{A3}), for example, to guarantee a delay of uplink queue equal to $D^{e2e|}_{s,d} = 2000 T.S$, it is highly recommended to choose these parameters value $f^{A}_{U}>0.2, \,  h > 39 m, \lambda > 0.5\cdot 10^{-3}, \theta= \forall,\omega> 5 \, rad/s$.

It is important to note that the analytic model proposed as well as the obtained results are appropriate to the proposed topology management. This topology plays a fundamental role in maintaining network connectivity. On the other hand, the organization of the drones in the network is managed by the UAV battery lifetime. Along this work, the angular velocity is assumed to be almost the same for all drones. This constraint leads to provide a significant energy consumption for drones having a large rotation radius compared to the drone with a short flying radius. To be effective, the choice of which drone must be positioned in the center, as well as the other drones in the network, should be decided the UAV battery lifetime. It is strongly recommended that the drone with lower battery level be located as the first drone in the network (Center), and by increasing the battery level the rotation radius also expands.
%he video quality level changes according to perceived network quality (which is visible to the network service provider) and buffer status (which is not). So you can look at long-term data rates over 30 seconds, but it's possible that the actual video quality level changed several times over that 30 seco nds.

\section{Conclusion}

Flying Mesh UAV networks promise to efficiently collect data from ground IoT devices deployed in a target area as well as relaying their data to legacy systems. In this paper, we present a complete theoretic framework to analyze and understand the dynamics of such a network. The proposed model aims to help planning and sizing a UAV network while ensuring target/satisfactory performance and efficient deployment. Namely, we construct a queuing-based framework, allowing a cross-layered optimization over APP, NET, MAC and PHY layers. The proposed model was validated through discrete event simulation, and it turns to predict network performance with high accuracy. Despite considering a specific network topology, our model is vbalid for any network topology and allows to measure traffic intensity, end-to-end throughput, and end-to-end delay for active streams. It allows to analyze the effect of routing, drones altitude, aperture angle, angular velocity, transmit poser, MAC parameters, ground IoT nodes density, transmit power, etc. Many trade-offs have been defined and a comprehensive discussion on parameter setting and network design has been provided.\\

This article paves the way for many interesting directions such as network design and on-demand deployment, optimal setting, energy efficiency, wireless energy transfer, flexible infrastructure, etc.

\bibliographystyle{IEEEtran}
\bibliography{Biblio}

%Dr. Reveryrand would like to acknowledge the funding by XLIM, Limoges, France. 
%The authors would like to thank Dr. David Root and Dr. Jean-Pierre Teyssier at Agilent Technologies for the loan of the time-domain nonlinear measurement equipment and TriQuint Semiconductor for the donation of the transistors. 

% if have a single appendix:
%\appendix[Proof of the Zonklar Equations]
% or
%\appendix  % for no appendix heading
% do not use \section anymore after \appendix, only \section*
% is possibly needed

% use appendices with more than one appendix
% then use \section to start each appendix
% you must declare a \section before using any
% \subsection or using \label (\appendices by itself
% starts a section numbered zero.)
%

% ============================================
%\appendices
%\section{Proof of the First Zonklar Equation}
%Appendix one text goes here %\cite{Roberg2010}.

% you can choose not to have a title for an appendix
% if you want by leaving the argument blank
%\section{}
%Appendix two text goes here.

% use section* for acknowledgement
%\section*{Acknowledgment}

%The authors would like to thank D. Root for the loan of the SWAP. The SWAP that can ONLY be usefull in Boulder...

% Can use something like this to put references on a page
% by themselves when using endfloat and the captionsoff option.
\ifCLASSOPTIONcaptionsoff
 
\fi

% trigger a \newpage just before the given reference
% number - used to balance the columns on the last page
% adjust value as needed - may need to be readjusted if
% the document is modified later
%\IEEEtriggeratref{8}
% The "triggered" command can be changed if desired:
%\IEEEtriggercmd{\enlargethispage{-5in}}

% ====== REFERENCE SECTION

%\begin{thebibliography}{1}

% IEEEabrv,

% Can be used to pull up biographies so that the bottom of the last one
% is flush with the other column.
%\enlargethispage{-5in}

% that's all folks

%\appendixtitles{no} %Leave argument "no" if all appendix headings stay EMPTY (then no dot is printed after "Appendix A"). If the appendix sections contain a heading then change the argument to "yes".
\appendix
%\unskip
%\appendixtitles{yes}
\subsection{Solving the rate balance system} \label{appendix:RBE}
This appendix presents the proof of the uplink queueing system load using the rate balance equation.\\
Let:
\begin{equation}
\left\lbrace
\begin{aligned}
\alpha^{U}_{i,s}=x_{s} \,  \, f^{A}_{U_{s}} \, \prod_{k=s}^{i-1}(1- (1 - P_{k,j})^K) \, \, \, \frac{\overline{L^{A}_{i}}}{f^{A}_{U_{i}} \, p^{A}_{i} \, \xi_{i,i+1}} \\
\beta^{U}_{i}=\frac{S}{n_{i}+1} \, \, n_{i} P_{cov}  \, \, \, \frac{\overline{L^{A}_{i}}}{f^{A}_{U_{i}} \, p^{A}_{i} \, \xi_{i,i+1}}\, \, \, \, \, \,  \, \qquad  \qquad \\ 
%y^{U}_{i}=\frac{f^{A}_{U_{i}} \, p^{A}_{i} \, \xi_{i,i+1}}{\overline{L^{A}_{i}}} \quad  \qquad  \qquad \qquad  \, \\
%y^{U}_{GW}= \zeta_{_{GW}} \quad  \qquad  \qquad \qquad  \, \\
%z^{U}_{i,s}= x_{s} \,  \, f^{A}_{U_{s}} \, \prod_{k=s}^{i-1}(1- (1 - P_{k,j})^K)
\end{aligned}
\right.
\end{equation}
  Then:
  \\
$\overbrace{  \begin{pmatrix} 
1 & 0   & 0 & 0 & 0 & 0 &   \ldots  \\
-\alpha_{2,1}^{U}& 1 & 0 & 0 & 0  & 0 & \ldots \\
0 & 0 & 1 & 0 & 0 & 0  & \ldots \\
-\alpha_{3,1}^{U} & 0 & 0 & 1   & 0 & 0 & \ldots  \\
0& 0 & -\alpha_{3,2}^{U} & 0 & 1   & 0 & \ldots \\
0& 0 & 0 & 0 & 0  & 1 & \ldots \\
\vdots & \vdots & \vdots & \vdots & \vdots & \vdots & \vdots\\
\end{pmatrix}}^{X}.
  \begin{pmatrix} 
\pi_{1,1,c}^{U} \\
\pi_{2,1,c}^{U} \\
\pi_{2,2,c}^{U} \\
\pi_{3,1,c}^{U} \\
\pi_{3,2,c}^{U} \\
\pi_{3,3,c}^{U} \\
\vdots\\
\end{pmatrix}
=
 \overbrace{\begin{pmatrix} 
 \beta{1} \\
0 \\
 \beta{2}  \\
0\\
0\\
 \beta{3} \\
\vdots\\
\end{pmatrix}}^{Y}$
  
\begin{equation}
  \boldsymbol{\Pi}^{U}=\textbf{X}^{-1}\cdot \textbf{Y},
\end{equation}
This appendix presents the proof of the traffic intensity of downlink queue using the rate balance equation.\\
Let:   
\begin{equation}
  \left\lbrace
\begin{aligned}
\gamma^{D}_{i}=\frac{f^{A}_{D_{i}}}{\overline{L_{i}^{A}}}\, p^{A}_{i}\, P_{cov}  \quad  \qquad   \qquad \qquad \qquad \qquad \quad  \qquad  \qquad \qquad \quad  \qquad  \qquad \qquad \, \, \\ 
\eta^{D}_{i}= \frac{f^{A}_{D_{i}}}{\overline{L_{i}^{A}}}\,  p^{A}_{i}\, \xi_{i,i-1}   \quad  \qquad  \qquad\qquad  \qquad \qquad \quad  \qquad  \qquad \qquad \quad  \qquad  \qquad \qquad\\
\delta^{D}_{i,d}=( \upsilon_{ack}.\,d^{U}_{Gw,d,c}+\frac{1}{p^{SR}_{pu}} \,\,\, P_{PU,d}) \prod_{k=Gw}^{i+1}(1- (1 - P_{k,j}^K )  \qquad \qquad \qquad \qquad \qquad\\
%z^{D}_{i,d}=  (\upsilon_{ack}.\,d^{U}_{Gw,i,c})+\frac{1}{p^{SR}_{pu}} \,\,\, P_{PU,d} \, \,  \qquad  \qquad \qquad  \qquad  \qquad \qquad \qquad \qquad \qquad \\
\end{aligned}
\right .
\end{equation}
\begin{equation}
\overbrace{  \begin{pmatrix} 
  \gamma_{1}^{D} & 0 & 0 & 0 & 0 & 0 &\ldots\\
  0 & \eta_{2}^{D} & 0 & 0 & 0 & 0 &\ldots\\
  0 & 0 & \gamma_{2}^{D} & 0 &0 & 0&\ldots\\
   0 & 0 & 0 & \eta_{3}^{D} & 0 & 0 &\ldots \\
    0 & 0 & 0 & 0 &\eta_{3}^{D} & 0 &\ldots \\
     0 & 0 & 0 & 0 &0& \gamma_{3}^{D} &\ldots \\
       \vdots  & \vdots & \vdots & \vdots & \vdots & \vdots &  \vdots \\
  \end{pmatrix}}^{Z}.
\begin{pmatrix} 
\pi_{1,c,1}^{D} \\
\pi_{2,c,1}^{D} \\
\pi_{2,c,2}^{D} \\
\pi_{3,c,1}^{D} \\
\pi_{3,c,2}^{D} \\
\pi_{3,c,3}^{D} \\
\vdots\\
\end{pmatrix}
=
\overbrace{   \begin{pmatrix} 
\delta^{D}_{1,1}\\
\delta^{D}_{2,1}\\
\delta^{D}_{2,2} \\
\delta^{D}_{3,1} \\
\delta^{D}_{3,2}\\
\delta^{D}_{3,3}\\
\vdots\\
\end{pmatrix} }^{W}
\nonumber 
\end{equation}

\begin{equation}
  \boldsymbol{\Pi}^{D}=\textbf{W}^{-1}\cdot \textbf{Z},
\end{equation}

\subsection{Uplink Average e2e Delay} \label{appendix:Uplinke2eDelay}
The sojourn time over the air channel is divided into two terms: 1) The waiting time of packet enter to uplink queue, plus 2) average service time of uplink packet  over AC $\tau^{A}_{U_{i}}$. The waiting time also divided into two terms: 1)  The mean residual time $\overline{R}^{A}_{i}$ seen by a newly entered packet, plus 2) the queuing time $Q^{A}_{U_{i}}$ spent in the queue $U_{i}$ which is the time required for all the packets entered earlier to get served.
\begin{equation}
   D^{\uparrow}_{i}=W^{A}_{U_{i}}+\tau^{A}_{U_{i}}=\overline{R}^{A}_{i}+Q^{A}_{U_{i}}+\tau^{A}_{U_{i}}.
   \label{1}
\end{equation}
\begin{description}
\item \textbf{Average residual time at UAV $i$: }
An arriving packet has to wait until the packet in service over the air channel is done. The serving packet could be a beacon packet, a packet from uplink queue $U_{i}$ or a packet from downlink queue $D_{i}$. Hence, the average residual service time seen by a given packet writes:
\begin{multline}
   \overline{R}^{A}_{i}= \sum_{s} \pi^{U}_{i,s,c} \, \, f^{A}_{U_{i}}   \, \, \overline{R}^{A}_{i,j}+ \sum_{\substack{{d}}} \pi_{i,c,d}^{D}\,\, f^{A}_{D_{i}} \,\,\overline{R}^{A}_{i,j} \\
   + (1-\sum_{s} \pi^{U}_{i,s,c} \, \, f^{A}_{U_{i}}- \sum_{d} \pi_{i,c,d}^{D}\,\, f^{A}_{D_{i}}) \, \, \overline{R}^{A}_{i,j}  \qquad     j = i\pm 1 
\end{multline}
Using a discrete renewal process \cite{1025}, $R^{A}_{i,j}$ is obtained as follows:
\begin{equation}
\overline{R}^{A}_{i,j}= \frac{\tau^{(2)_{A}}_{i,j}}{2 \, \,\tau^{A}_{i,j} }+\frac{1}{2}
 \qquad     j = i\pm 1 
\end{equation}
where $\tau^{{A}}_{i,j}$ and $\tau^{(2)_{A}}_{i,j}$ are the first and second moment respectively. Thus, the average service time of a packet from node $i$ to $j$ in MAC layer over Air-to-Ground channel and/or Air-to-Air channel are \cite{1025}:

\begin{multline}
    \tau^{A}_{i,j}=\frac{L^{A}_{i,j}}{p^{A}_{i}} ,
    \quad  
     \tau^{G}_{i,j}=\frac{L^{G}_{i,j}}{p^{G}_{i}},    \quad 
     \tau^{(2)_{A}}_{i,j}=\frac{L^{(2)_{A}}_{i,j}+L^{A}_{i,j}(1-p^{A}_{i})}{p^{2_{A}}_{i}} 
\end{multline}
where
\small
\begin{equation}
L^{(2)_{A}}_{i,j}=L^{A}_{i,j}+\frac{2(1-P_{i,j})}{P^{2}_{i,j}}-\frac{2(1-P_{i,j})^{K} (K-(1-P_{i,j})(K-1))}{P^{2}_{i,j}}
\end{equation}
 \item \textbf{Queuing time: }
A packet enter to uplink Queue has to wait the service of packets stay before it once is at the head of uplink queue, it  has to wait the packets that will be served before it from Beacon Queue and downlink Queue $n^{A}_{(B,D)_{i}}$.
The queuing time in uplink queue $U_{i}$ can be written as:
\begin{equation}
Q^{A}_{U_{i}} = N^{A}_{U_{i}}\, \tau^{A}_{U_{i}}+(N^{A}_{U_{i}}+1)\, n^{A}_{(B,D)_{i}} \, \tau^{A}_{(B,D)_{i}}
\label{2}
\end{equation}
where $N^{A}_{U_{i}}$ is the number of packets  entered  earlier  waiting in uplink queue to  get  served  over Air-to-Air channel, $\tau_{B,D}^{A}$ is the average service time for a beacon packets and  downlink packets. $\tau_{B,D}^{A}=\frac{\tau_{D_{i}}^{A}\,f^{A}_{D_{i}}+\tau_{B_{i}}^{A}\,f^{A}_{B_{i}}}{f^{A}_{D_{i}}+f^{A}_{B_{i}}}$ and $\tau^{A}_{U_{i}}=\frac{L^{A}_{i,i+1}}{p^{A}_{i}} $, $\tau^{A}_{D_{i}}=\frac{L^{A}_{i,i-1}}{p^{A}_{i}} $, $\tau^{A}_{B}=\frac{L^{A}_{i,i\pm 1 }}{p^{A}_{i}} $ represent respectively the service time of uplink, downlink and beacon packet served over AC. A packet at the head of \textit{Uplink} queue ready to be transmitted has to wait a number of cycle \textit{X} (Random Variable) before it can moves to the MAC layer. \textit{X} corresponds to the number of cycles that serving whether the beacons packets or downlink packets. $P[X=k]=(1-f^{A}_{U_{i}}) ^k \, f^{A}_{U_{i}}$. The expected value of a random variable \textit{X} is $E[V] \simeq n^{A}_{(B,D)_{i}} \simeq \frac{1-f^{A}_{U_{i}}}{f^{A}_{U_{i}}}$.\\
According to Little's formula $N^{A}_{U_{i}} = a^{U}_{i} \, .\overline{W^{A}_{U_{i}}}$ , we obtain the waiting time in uplink queue by using the equation ($~ \ref{1}$), ($~ \ref{2}$) as follow:
\begin{equation}
\overline{W}^{A}_{U_{i}}= \frac{ \overline{R}^{A}_{i} + \left(\frac{1-f^{A}_{U_{i}}}{f^{A}_{U_{i}}}\right) \, \tau^{A}_{(B,D)_{i}}}{1- a^{A}_{i}\, \,\left[\tau^{A}_{U_{i}}+ \left(\frac{1-f^{A}_{U_{i}}}{f^{A}_{U_{i}}}\right)\, \tau^{A}_{(B,D)_{i}}\right]}
\end{equation}
\end{description}

\subsection{Downlink Average e2e Delay} \label{appendix:Downlinke2eDelay}
Similar to the uplink queue, a packet enters to queue $D_{i}$ has to wait a period of time which are waiting time and its service time.

\begin{description}[leftmargin=0.7cm]
\item \textbf{Average Residual Time: } : The amount of time for another packet being served over the Air-to-Ground channel, this packet can be a beacon packet or a packet from downlink queue. For any packet on the channel the average residual time is: 

\begin{equation}
  \overline{R}^{G}_{i}= \sum_{\substack{{d}}} \pi_{i,c,d}^{D}\,\,  f^{G}_{D_{i}}\,\,\overline{R}^{G}_{i,j}+ (1- \sum_{d} \pi_{i,c,d}^{D}\,\,   f^{G}_{D_{i}}) \, \, \overline{R}^{G}_{i,j}   \qquad
j \in IoT.
\end{equation}
Where $\overline{R}^{G}_{i,j}= \frac{\tau^{(2)_{G}}_{i,j}}{2 \, \,\tau^{G}_{i,j} }+\frac{1}{2}$, and $\tau^{(2)_{G}}_{i,j}=\frac{L^{(2)_{G}}_{i,j}+L^{G}_{i,j}(1-p^{(G)}_{i})}{p^{2_{(A)}}_{i}}$ and $L^{(2)_{G}}_{i,j }=L^{G}_{i,j}+\frac{2(1-P^{i,j}_{s})}{(P^{i,j}_{s})^{2}}-\frac{2(1-P^{i,j}_{s})^{K} (K-(1-P^{i,j}_{s})(K-1))}{(P^{i,j}_{s})^{2}}$\\
\item \textbf{Queuing time: } for a packet in the downlink queue can be divided into two parts:
\begin{enumerate}
\item A packet in the downlink queue coming from UAV $j$ to  UAV $i$, has to wait: The packets before it in the queue $D$ that will be send over AC($N^{A}_{D_{i}}$). Once it arrives at the head of downlink queue, it has to wait the beacon packets service and downlink packets services. $n^{A}_{(B,U)_{i}}$.% (see figure \ref{Delay}).
\begin{equation}
Q^{A}_{D_{i}} = N^{A}_{D_{i}}\, \tau^{A}_{D_{i}}+(N^{A}_{D_{i}}+1)\, n^{A}_{(B,U)_{i}} \, \tau^{A}_{(B,U)_{i}}
\end{equation}
%\begin{equation}
%P[V=k]=(1-f^{A}_{D_{i}}) ^k \, f^{A}_{D_{i}}
%\end{equation}
Where  $n^{A}_{(B,U)_{i}} \simeq \frac{1-f^{A}_{D_{i}}}{f^{A}_{D_{i}}}$ and $\tau_{B,U}^{A}=\frac{\tau_{B}^{A}\,f^{A}_{B_{i}}+\tau_{U}^{A}\,f^{A}_{U_{i}}}{f^{A}_{B_{i}}+f^{A}_{U_{i}}}$ follows the same procedure as that of \textit{Uplink} queue the waiting time for a packet coming from  UAV $j$  can be written as:
%\begin{equation}
%N^{A}_{D_{i}} = a^{D}_{i} \, .\overline{W^{A}_{D_{i}}}
%\end{equation}

\begin{equation}
\overline{W}^{A}_{D_{i}}= \frac{ \overline{R}^{A}_{i} + \left(\frac{1-f^{A}_{D_{i}}}{f^{A}_{D_{i}}}\right) \, \tau^{A}_{(B,U)_{i}}}{1- a^{A}_{i}\, \,\left[\tau^{A}_{D_{i}}+ \left(\frac{1-f^{A}_{D_{i}}}{f^{A}_{D_{i}}}\right)\, \tau^{A}_{(B,U)_{i}}\right]}
\end{equation}

\item A packet in the downlink queue coming from UAV $j$ to  IoT devices, has to wait the packets before it in the queue $D$ that will be send to an IoT devices ($N^{G}_{D_{i}}$), once the packet is at the head of downlink queue has to wait the beacon packets that will be served before $n^{G}_{(B)_{i}}$.% (see figure \ref{Delay}).
\begin{equation}
Q^{G}_{D_{i}} = N^{G}_{D_{i}}\, \tau^{G}_{D_{i}}+(N^{G}_{D_{i}}+1)\, n^{G}_{(B)_{i}} \, \tau^{G}_{B_{i}}
\end{equation}
\end{enumerate}
 $\tau^{G}_{D_{i}}=\frac{L^{G}_{i}}{p^{G}_{i}} $, $\tau^{G}_{B_{i}}=\frac{L^{G}_{i}}{p^{G}_{i}} $ represent respectively the service time of  downlink and beacon packet served over GC.
According to the same method as before we obtain the waiting time for a packet in the downlink queue of UAV $i$ that will be send to an IoT served by a UAV $i$
\begin{equation}
\overline{W^{G}_{D_{i}}}= \frac{ \overline{R}^{G}_{i} + n^{G}_{(B)_{i}} \, \tau^{G}_{B_{i}}}{1- a^{G}_{i}\, \,(\tau^{G}_{D_{i}}+ n^{G}_{B_{i}}\, \tau^{G}_{B_{i}})}
\end{equation}
Then, the delay in the downlink queue of an single UAV $i$  is 
\begin{equation}
D^{\downarrow}_{s,d} =\sum_{i=s}^{d+1}  D_{i}^{D}= \sum_{i=s}^{d+1}  \overline{W}^{A}_{D_{i}}+ \tau^{A}_{D_{i}}+ \overline{W}^{G}_{D_{d}} +\tau ^{G}_{D_{d}}
\end{equation}

\end{description}
%\begin{figure}[h!]
%\centering
%    \includegraphics[scale=0.23]{Delay.png}
%\caption{......}
%\label{Delay} 
%\end{figure}

\end{document}